\documentclass[prx,twocolumn]{revtex4}

\usepackage{graphicx}
\usepackage{dcolumn}
\usepackage{bm}
\usepackage{multirow}
\usepackage{amsmath,amssymb}
\usepackage{threeparttable}

\begin{document}

\title{Electronic modes induced by spin and charge perturbations in Mott and Kondo insulators}

\author{Masanori Kohno}
\email{KOHNO.Masanori@nims.go.jp}
\affiliation{Research Center for Materials Nanoarchitectonics, National Institute for Materials Science, Tsukuba 305-0003, Japan}

\date{\today}

\begin{abstract}
Electronic band structures usually remain unaffected by doping via a chemical-potential shift or by increasing the temperature in conventional band insulators. 
In contrast, it has been shown that those of Mott and Kondo insulators can be altered by doping or by increasing the temperature: 
electronic modes are induced within the band gap, exhibiting momentum-shifted magnetic dispersion relations from the band edges. 
Here, this study demonstrates that the underlying mechanism of the remarkable strong-correlation effects can be generalized to the emergence of electronic modes 
caused by various spin and charge perturbations, including magnetization of spin-gapped Mott and Kondo insulators. 
These emergent modes can alter the band structure if a macroscopic number of spins or charges are excited by the perturbations at a given moment. 
The origins and dispersion relations of these emergent modes, particularly why and how the dispersion relations depend on the momentum and energy of the perturbations, 
are elucidated by investigating the selection rules and using the Bethe ansatz and the effective theory for weak inter-unit-cell hopping. 
The validity and generality of the theoretical results across different models and spatial dimensions are verified by numerical calculations for the one- and two-dimensional Hubbard models, 
periodic Anderson models, Kondo lattice models, and ladder and bilayer Hubbard models. 
This study provides crucial insights into why and how spin and charge perturbations can alter the band structure in strongly correlated insulators, thereby 
paving the way for band-structure engineering in strong-correlation electronics, which enables previously unexplored functionalities by exploiting the unconventional characteristics revealed in this paper. 
\end{abstract}

\maketitle
\section{Introduction} 
A remarkable feature of strongly correlated insulators such as Mott and Kondo insulators is the emergence of electronic modes within the band gap by doping or by increasing the temperature, 
which exhibit momentum-shifted magnetic dispersion relations from the band edges 
\cite{Kohno1DHub,Kohno2DHub,KohnoDIS,KohnoRPP,KohnoHubLadder,KohnoKLM,KohnoGW,KohnoAF,KohnoSpin,Kohno1DtJ,Kohno2DtJ,KohnoMottT,KohnoTinduced}. 
This feature reflects spin-charge separation, i.e., the existence of spin excitation in the energy regime lower than the charge gap (band gap). 
It is contrasted with the feature of conventional band insulators in which interactions between electrons can be neglected; 
electronic modes do not emerge within the band gap by doping via a chemical-potential shift or by increasing the temperature in band insulators, 
because the lowest spin-excitation energy is equal to the band gap 
(that is, spin-charge separation does not occur). 
\par
The emergence of electronic modes can be generalized to the cases driven by various spin and charge perturbations. 
In this paper, to demonstrate how and why electronic modes are induced by perturbations, 
the band structures of electronic excitations from spin- and charge-perturbed states are studied in models of strongly correlated insulators. 
Spin (charge) perturbed states are states with spin (charge) quantum numbers different from those of the unperturbed state whereas the charge (spin) quantum numbers remain the same. 
Charge-perturbed states include not only doped equilibrium states that can be experimentally realized by chemical doping or application of a gate voltage but also nonequilibrium states that can be caused by photon (light) radiation. 
Spin-perturbed states include magnetized equilibrium states that can be realized by the application of a magnetic field and nonequilibrium states that can be achieved through neutron radiation. 
\par
In previous studies on electronic band structures of strongly correlated insulators in nonequilibrium, 
complicated states have been considered, such as time-dependent states and 
excited states involving featureless states 
\cite{lightPump1dHub,pumpProbe1dEHM,photoSubbands,FKFloquetDMFT,noneqDMFTRMP,DCAkw1dHub}. 
In this study, to clarify the underlying mechanism of the change in the band structure caused by perturbations, 
low-energy or dynamically dominant characteristic spin- and charge-excited states are considered perturbed states, 
and a clear theoretical basis is presented for interpretations of unconventional electronic states induced by perturbations. 
\par
In the conventional view, electronic states induced in the spectral function by perturbations or in nonequilibrium might be considered to be due to electronic excited states affected by perturbations. 
However, this study theoretically shows that electronic states are induced by perturbations primarily because the state from which electronic excitations are made is changed by perturbations. 
The situation is similar to the temperature-induced change in the band structure in Mott and Kondo insulators; 
although temperature effects on electronic excited states would usually be considered a primary cause, 
it has recently been shown that spin-excited states involved in the thermal state, from which electronic excitations are made, primarily cause the temperature-induced change in the band structure \cite{KohnoMottT,KohnoTinduced}. 
\par
In this paper, including the origins and dispersion relations of the emergent electronic modes, the fundamental characteristics of the change in the band structure caused by spin and charge perturbations in strongly correlated insulators are clarified 
by investigating the selection rules and using the Bethe ansatz in the one-dimensional (1D) Hubbard model (HM) along with 
the effective theory for weak inter-unit-cell hopping in spin-gapped Mott and Kondo insulators. 
The theoretical results are validated by numerical calculations 
in the 1D and two-dimensional (2D) HMs, periodic Anderson models (PAMs), Kondo lattice models (KLMs), 
and ladder and bilayer HMs. 
\par
The outline of this paper is as follows: 
In Sec. \ref{sec:modelsMethos}, the models, parameters, and quantities considered in this paper are defined. The details of the methods are presented. 
In Sec. \ref{sec:undperturbed}, the spectral properties of unperturbed systems are reviewed for comparison with those of perturbed systems shown in later sections. 
In Sec. \ref{sec:chargePerturbed}, the spectral properties of charge-perturbed systems are considered. 
After reviewing how and why electronic states are induced by doping \cite{Kohno1DHub,Kohno2DHub,KohnoDIS,KohnoRPP,KohnoHubLadder,KohnoKLM,KohnoGW,KohnoAF,KohnoSpin,Kohno1DtJ,Kohno2DtJ,KohnoMottT,KohnoTinduced}, 
the characteristics are extended to those of the emergent electronic states in undoped charge-perturbed systems with nonzero values of the charge ($\eta$) quantum number, using relations 
based on $\eta$-SU(2) symmetry. 
Remarks on nonzero-$\eta$ states at half filling are made, because photoinduced $\eta$-pairing states have recently attracted considerable attention 
\cite{entropyCoolingDMFT,etaPairingSuper,nonthermalSuper,KanekoEtaPairing,etaPairingHubLad}, which are generally nonsuperconducting states. 
The dispersion relations of electronic modes induced by charge fluctuation are also derived. 
In Sec. \ref{sec:spinPerturbed}, the spectral properties of spin-perturbed systems are considered. The origins and dispersion relations of magnetization-induced electronic modes in spin-gapped Mott and Kondo insulators are explained. 
Based on the results, the emergent electronic modes in unmagnetized spin-perturbed systems with nonzero values of the spin quantum number are explained. 
The dispersion relations of electronic modes induced by spin fluctuation are also derived. 
In Sec. \ref{sec:bandInsulator}, the differences in spectral features between the strongly correlated insulators and conventional band insulators are discussed to elucidate the strong-correlation effects. 
The relevance of spin-charge separation in strongly correlated insulators to the doping-induced modes and magnetization-induced modes is clarified. 
Implications of this study are described. The paper is summarized in Sec. \ref{sec:summary}. 
\par
The theoretical and numerical results indicate that the spectral weights of the electronic modes induced by spin and charge perturbations 
can become comparable to those of the conventional bands, provided that a macroscopic number of spins or charges are excited by the perturbations at a given moment; 
the spectral weights of the induced modes are proportional to the density of the excited spins or charges in the small-density regime. 
The dispersion relations of the emergent electronic modes can be expressed as combinations of the electronic and either the spin or the charge dispersion relations of unperturbed systems. 
These results can be applied to the basic principle of band-structure engineering for strong-correlation electronics, thus guiding the tuning and selection of perturbations and strongly correlated materials. 
\section{Models and methods} 
\label{sec:modelsMethos}
\subsection{Models} 
The HM, PAM, and KLM are defined by the following Hamiltonians: 
\begin{align}
\label{eq:HamHub}
{\cal H}_{\rm HM}=&-\sum_{\langle i,j\rangle,\sigma}t_{i,j}\left(c^{\dagger}_{i,\sigma}c_{j,\sigma}+{\rm H.c.}\right)\nonumber\\
&+U\sum_i\left(n^c_{i,\uparrow}-\frac{1}{2}\right)\left(n^c_{i,\downarrow}-\frac{1}{2}\right)\nonumber\\
&-\mu\sum_{i,\sigma}n^c_{i,\sigma}-H\sum_{i}S^{c,z}_i,\\
\label{eq:HamPAM}
{\cal H}_{\rm PAM}=&-\sum_{\langle i,j\rangle,\sigma}t_{i,j}\left(c^{\dagger}_{i,\sigma}c_{j,\sigma}+{\rm H.c.}\right)\nonumber\\
&-t_{\rm K}\sum_{i,\sigma}\left(c^{\dagger}_{i,\sigma}f_{i,\sigma}+{\rm H.c.}\right)-\Delta\sum_{i,\sigma}n^f_{i,\sigma}\nonumber\\
&+U\sum_i n^f_{i,\uparrow}n^f_{i,\downarrow}-\mu\sum_{i,\sigma}(n^c_{i,\sigma}+n^f_{i,\sigma})\nonumber\\
&-H\sum_{i}(S^{c,z}_i+S^{f,z}_i),\\
\label{eq:HamKLM}
{\cal H}_{\rm KLM}=&-\sum_{\langle i,j\rangle,\sigma}t_{i,j}\left(c^{\dagger}_{i,\sigma}c_{j,\sigma}+{\rm H.c.}\right)\nonumber\\
&+J_{\rm K}\sum_i{\bm S}^c_i\cdot{\bm S}^f_i-\mu\sum_{i,\sigma}(n^c_{i,\sigma}+n^f_{i,\sigma})\nonumber\\
&-H\sum_{i}(S^{c,z}_i+S^{f,z}_i),
\end{align}
where $\langle i,j\rangle$ means that sites $i$ and $j$ are nearest neighbors. 
Here, $c^{\dagger}_{i,\sigma}$ and $n^c_{i,\sigma}$ ($f^{\dagger}_{i,\sigma}$ and $n^f_{i,\sigma}$) denote 
the creation and number operators of a conduction orbital (localized orbital) electron with spin $\sigma$ for $\sigma=\uparrow,\downarrow$ at a site $i$, respectively, 
and ${\bm S}^c_i$ (${\bm S}^f_i$) denotes the spin operator of a conduction orbital (localized orbital) electron at a site $i$; 
${\bm S}^\lambda_i=(S^{\lambda,x}_i,S^{\lambda,y}_i,S^{\lambda,z}_i)$ for $\lambda=c$ and $f$. 
In the KLM, each localized orbital has one electron. 
Hereinafter, we set $t_{i,j}=t$ in chains and square-lattice planes, $t_{i,j}=t_{\perp}$ between chains for the ladder HM and between square-lattice planes for the bilayer HM, 
and $\Delta=\frac{U}{2}$ for the PAM; the PAM with $U=2\Delta$ is called the symmetric PAM. 
The parameters $t, t_{\perp}, t_{\rm K}, U$, $\Delta$, and $J_{\rm K}$ are assumed to be positive unless otherwise mentioned. 
\par
In this paper, the ground state at $\mu=H=0$ is considered the unperturbed state. 
In a doped system (magnetized system), the chemical potential $\mu$ (magnetic field $H$) is adjusted 
such that the ground state has the corresponding doping concentration (magnetization) unless otherwise mentioned. 
\par
The numbers of electrons, sites, and unit cells are denoted by $N_{\rm e}$, $N_{\rm s}$, and $N_{\rm u}$, respectively. 
A unit cell is a site in the 1D and 2D HMs ($N_{\rm u}=N_{\rm s}$), 
a pair of sites connected by $t_{\perp}$ in the ladder and bilayer HMs ($N_{\rm u}=\frac{N_{\rm s}}{2}$), 
and a site with a conduction orbital and localized orbital in the PAM and KLM ($N_{\rm u}=N_{\rm s}$). 
The number of doped holes from half filling is denoted by $N_{\rm h}$: $N_{\rm h}=N_{\rm s}-N_{\rm e}$ in the HM and $2N_{\rm u}-N_{\rm e}$ in the PAM and KLM. 
The hole-doping concentration $\delta$ is defined as $\delta=\frac{N_{\rm h}}{N_{\rm s}}(=1-\frac{N_{\rm e}}{N_{\rm s}})$ in the HM and 
$\delta=\frac{N_{\rm h}}{2N_{\rm u}}(=1-\frac{N_{\rm e}}{2N_{\rm u}})$ in the PAM and KLM. 
At half filling, $\delta=N_{\rm h}=0$. In the hole-doped case, $\delta>0$ and $N_{\rm h}>0$; in the electron-doped case, $\delta<0$ and $N_{\rm h}<0$.
Hereinafter, the number of electrons at half filling is assumed to be even, and the unit of $\hbar=1$ is used. The total spin quantum number and $z$ component of the spin are denoted by $S$ and $S^z$, respectively. 
The maximum value of $S$ at half filling is denoted by $S_{\rm max}$: $S_{\rm max}=\frac{N_{\rm s}}{2}$ in the HM and $N_{\rm u}$ in the PAM and KLM. 
The electronic spin opposite to $\sigma$ is denoted by ${\bar \sigma}$: ${\bar \sigma}=\downarrow$ and $\uparrow$ for $\sigma=\uparrow$ and $\downarrow$, respectively. 
The $z$ component of the spin of an electron is denoted by $s^z$: $s^z=\frac{1}{2}$ and $-\frac{1}{2}$ for $\sigma=\uparrow$ and $\downarrow$, respectively. 
\subsection{Spectral function and dynamical structure factors} 
The spectral function of a state $|X\rangle$ at zero temperature is defined as 
\begin{align}
\label{eq:Akwcf}
A_X({\bm k},\omega)=&\sum_{\lambda=c,f}A^{\lambda}_X({\bm k},\omega)\\
\label{eq:Akwspin}
=&\frac{1}{2}\sum_{\sigma=\uparrow,\downarrow}A^\sigma_X({\bm k},\omega),
\end{align}
where
\begin{align}
\label{eq:Akwcfspin}
A^\sigma_X({\bm k},\omega)&=\sum_{\lambda=c,f}A^{\lambda,\sigma}_X({\bm k},\omega),\\
\label{eq:Akwlambda}
A^\lambda_X({\bm k},\omega)&=\frac{1}{2}\sum_{\sigma=\uparrow,\downarrow}A^{\lambda,\sigma}_X({\bm k},\omega),\\
A^{\lambda,\sigma}({\bm k},\omega)&=-\frac{1}{\pi}{\rm Im}G^{\lambda}_{{\bm k},\sigma}(\omega)
\end{align}
using the following Green's function: 
\begin{align}
\label{eq:GreenFunc}
G^{\lambda}_{{\bm k},\sigma}(\omega)&=-\frac{i}{Z_X}\int_0^\infty dt {\rm e}^{i\omega t-\epsilon t}\langle X|
\{\lambda_{{\bm k},\sigma}(t),\lambda^\dagger_{{\bm k},\sigma}\}|X\rangle,\\
Z_X&=\langle X|X\rangle,\quad\lambda_{{\bm k},\sigma}(t)={\rm e}^{i{\cal H}t}\lambda_{{\bm k},\sigma}{\rm e}^{-i{\cal H}t}\nonumber
\end{align}
in the limit of $\epsilon\rightarrow +0$. Here, ${\cal H}$ represents ${\cal H}_{\rm HM}$, ${\cal H}_{\rm PAM}$, and ${\cal H}_{\rm KLM}$, 
and $\lambda^{\dagger}_{{\bm k},\sigma}$ denotes the Fourier transform of $\lambda^{\dagger}_{i,\sigma}$ for $\lambda=c$ and $f$. 
For an eigenstate of the Hamiltonian $|X\rangle$ with energy $E_X$, 
\begin{align}
\label{eq:Akwdef}
A^{\lambda,\sigma}_X({\bm k},\omega)&=\frac{1}{Z_X}\sum_{n}
\left[\begin{array}{r}
|\langle n|\lambda^{\dagger}_{{\bm k},\sigma}|X\rangle|^2\delta(\omega-E_n+E_X)\\
+|\langle n|\lambda_{{\bm k},\sigma}|X\rangle|^2\delta(\omega+E_n-E_X)\end{array}\right],
\end{align}
where $|n\rangle$ denotes the $n$th eigenstate of the Hamiltonian with energy $E_n$. The eigenstates of the Hamiltonian are assumed to be normalized to unity. 
In the HM and KLM, $A^{f,\sigma}_X({\bm k},\omega)=0$. 
\par
The momentum ${\bm k}$ in the ladder (bilayer) HM has the leg (in-plane) component ${\bm k}_{\parallel}$ and rung (out-of-plane) component $k_{\perp}$: 
${\bm k}_{\parallel}=k_x$ and $k_{\perp}=k_y$ in the ladder HM; 
${\bm k}_{\parallel}=(k_x,k_y)$ and $k_{\perp}=k_z$ in the bilayer HM. In the 1D and 2D HMs, PAMs, and KLMs, ${\bm k}={\bm k}_{\parallel}$. 
\par
The dynamical spin structure factor $S({\bm k},\omega)$ and dynamical charge structure factor $N({\bm k},\omega)$ at zero temperature are defined as 
\begin{align}
\label{eq:SkwHub}
&S({\bm k},\omega)=\sum_{m,\alpha}|\langle m|S^{c,\alpha}_{\bm k}|{\rm GS}\rangle|^2\delta(\omega-e_m),\\
\label{eq:NkwHub}
&N({\bm k},\omega)=\sum_{m}|\langle m|{\tilde n}^{c}_{\bm k}|{\rm GS}\rangle|^2\delta(\omega-e_m)
\end{align}
in the HM and 
\begin{align}
\label{eq:SkwPAMKLM}
&S({\bm k},\omega)=\frac{1}{2}\sum_{m,\alpha}|\langle m|(S^{c,\alpha}_{\bm k}-S^{f,\alpha}_{\bm k})|{\rm GS}\rangle|^2\delta(\omega-e_m),\\
\label{eq:NkwPAMKLM}
&N({\bm k},\omega)=\frac{1}{2}\sum_{m}|\langle m|({\tilde n}^{c}_{\bm k}-{\tilde n}^{f}_{\bm k})|{\rm GS}\rangle|^2\delta(\omega-e_m)
\end{align}
in the PAM and KLM, 
where $e_m$ denotes the excitation energy of $|m\rangle$ from the ground state $|{\rm GS}\rangle$, and 
$S^{\lambda,\alpha}_{\bm k}$ denotes the Fourier transform of $S^{\lambda,\alpha}_i$ for $\lambda=c$ and $f$; 
$S^{\lambda,\alpha}_i$ denotes the $\alpha$ component of ${\bm S}^{\lambda}_i$ for $\alpha=x$, $y$, and $z$. 
The charge fluctuation of a momentum ${\bm k}$ is caused by the following operator: 
\begin{equation}
{\tilde n}^\lambda_{\bm k}=\frac{1}{\sqrt{N_{\rm s}}}\sum_{i}e^{i{\bm k}\cdot{\bm r}_i}{\tilde n}^\lambda_i
\label{eq:nk}
\end{equation}
with ${\tilde n}^\lambda_i=n^\lambda_i-{\bar n}$, where $n^\lambda_i=\sum_{\sigma}n^\lambda_{i,\sigma}$ for $\lambda=c$ and $f$; 
${\bar n}=\frac{N_{\rm e}}{N_{\rm s}}$ in the HM and $\frac{N_{\rm e}}{2N_{\rm u}}$ in the KLM and PAM. At half filling, ${\bar n}=1$. 
\par
The transverse dynamical spin susceptibility of the HM at zero temperature is defined as 
\begin{align}
\label{eq:Chikw}
&\chi({\bm k},\omega)=\frac{1}{2}[\chi^{+-}({\bm k},\omega)+\chi^{-+}({\bm k},\omega)],\\
&\chi^{\pm\mp}({\bm k},\omega)=-\sum_{m}
\left[\frac{|\langle m|S^{c,\mp}_{\bm k}|{\rm GS}\rangle|^2}{\omega-e_m+i\epsilon}
-\frac{|\langle m|S^{c,\pm}_{-{\bm k}}|{\rm GS}\rangle|^2}{\omega+e_m+i\epsilon}\right]\nonumber
\end{align}
for $\epsilon\rightarrow +0$, where $S^{c,\pm}_{\bm k}=S^{c,x}_{\bm k}\pm iS^{c,y}_{\bm k}$. 
\par
The ground state in the subspace of $N_{\rm h}=n$ is denoted by $|{\rm GS}\rangle^{N_{\rm h}=n}$ regardless of the value of $\mu$ for $H=0$. 
In this paper, unless otherwise mentioned, the ground state means the ground state at half filling for $H=0$, $|{\rm GS}\rangle^{N_{\rm h}=0}$, 
which is denoted by $|{\rm GS}\rangle$. 
The energies of $|{\rm GS}\rangle$ and $|{\rm GS}\rangle^{N_{\rm h}=n}$ are denoted by $E_{\rm GS}$ and $E_{\rm GS}^{N_{\rm h}=n}$, respectively. 
\subsection{Methods} 
In the 1D models (1D HM, 1D PAM, 1D KLM, and ladder HM), 
the spectral function of the magnetized ($S^z\ne0$) ground state for the 1D KLM was calculated using the conventional [U(1)] dynamical density-matrix renormalization group (DDMRG) method \cite{Kohno1DHub,DDMRG,DDMRGAkw}, and 
the spectral function and dynamical spin and charge structure factors of the undoped and hole-doped ground states with $S=0$ were calculated 
using the spin-SU(2)-symmetric non-Abelian DDMRG method 
\cite{KohnoDIS,Kohno1DtJ,Kohno2DtJ,KohnoHubLadder,KohnoKLM,KohnoGW,KohnoMottT,KohnoTinduced,KohnoRPP,nonAbelianHub,nonAbeliantJ,nonAbelianThesis}. 
The DDMRG calculations were performed under open boundary conditions on clusters of $N_{\rm u}=60$ 
in the ladder HM, 1D PAM, and 1D KLM, and $N_{\rm s}=120$ in the 1D HM, 
where 120 (240) eigenstates of the density matrix were retained in the 1D and ladder HMs and 1D KLM (1D PAM). 
\par
In the 2D models (2D HM, 2D PAM, 2D KLM, and bilayer HM), 
the spectral function was calculated using the cluster perturbation theory (CPT) \cite{CPTPRL,CPTPRB} 
on $4\times 4$-site clusters in the 2D HM, $3\times 3$-unit-cell clusters in the 2D KLM, and $3\times 2$-unit-cell clusters in the bilayer HM and 2D PAM. 
The symmetrized spectral function is defined as ${\bar A}_X({\bm k},\omega)=\frac{1}{2}\sum_\sigma{\bar A}^\sigma_X({\bm k},\omega)$ 
for ${\bar A}^\sigma_X({\bm k},\omega)=\frac{1}{2}[A^\sigma_X(k_x,k_y,k_{\perp},\omega)+A^\sigma_X(k_y,k_x,k_{\perp},\omega)]$ in the bilayer HM and 
$\frac{1}{2}[A^\sigma_X(k_x,k_y,\omega)+A^\sigma_X(k_y,k_x,\omega)]$ in the 2D PAM. 
\par
The lowest spin-excitation energy at each momentum in the 2D models was calculated under periodic boundary conditions 
using the Lanczos method. 
The transverse dynamical spin susceptibility was calculated 
using the random-phase approximation \cite{KohnoAF, spinbag} 
based on the antiferromagnetic mean-field approximation \cite{SlaterAF, PennAF} in the 2D HM. 
\par
The spectral function of $|{\rm GS}\rangle^{N_{\rm h}=-n}$ for $n>0$ was obtained using that of $|{\rm GS}\rangle^{N_{\rm h}=n}$ via Eq. (\ref{eq:AkwphX}). 
The spectral functions of magnetized states in the HM and PAM were obtained using the Shiba transformation [Eqs. (\ref{eq:AkwuEta}) and (\ref{eq:AkwdEta}); Sec. \ref{sec:AkwM}]. 
The spectral functions of $S^{c,z}_{\bm q}|{\rm GS}\rangle$ and ${\tilde n}^c_{\bm q}|{\rm GS}\rangle$ in the HM 
and those of $\frac{1}{\sqrt{2}}(S^{c,z}_{\bm q}-S^{f,z}_{\bm q})|{\rm GS}\rangle$ and $\frac{1}{\sqrt{2}}({\tilde n}^c_{\bm q}-{\tilde n}^f_{\bm q})|{\rm GS}\rangle$ 
in the PAM and KLM were calculated using CPT on sixteen-site clusters in the 1D, 2D, and ladder HMs, six-unit-cell clusters in the bilayer HM and 2D PAM, eight-unit-cell clusters in the 1D PAM, and nine-unit-cell clusters in the 1D and 2D KLMs. 
These calculations were performed using Eq. (\ref{eq:Akwdef}), 
where $|X\rangle$ is represented by $\sum_i e^{i{\bm q}\cdot{\bm r}_i}O_i|{\rm GS}\rangle$ for $O_i=S^{c,z}_i$, ${\tilde n}^c_i$, $\frac{1}{\sqrt{2}}(S^{c,z}_i-S^{f,z}_i)$, and $\frac{1}{\sqrt{2}}({\tilde n}^c_i-{\tilde n}^f_i)$, 
under the assumption that these states can be effectively identified as the dominant states in the dynamical structure factors. 
\par
The local Green's functions for the intermediate electron density, magnetization ($S^z$) density, spin ($S$) density, and $\eta$ density (Sec. \ref{sec:etaSU2}) were 
obtained as the weighted averages of the local Green's functions of larger and smaller densities on a cluster in CPT \cite{CPTHanke}. 
The numerical calculations were performed for perturbed and unperturbed states with even numbers of electrons. 
\par
The results with Gaussian broadening were obtained through deconvolution from those with Lorentzian broadening 
\cite{Kohno1DHub,Kohno2DHub,KohnoDIS,KohnoRPP,KohnoHubLadder,KohnoKLM,KohnoGW,KohnoAF,KohnoSpin,Kohno1DtJ,Kohno2DtJ,KohnoMottT,KohnoTinduced,KohnottpHub,KohnottpJ}. 
\par
In the figures presented in this paper, the numerical and theoretical results are shown for 
$U/t=6$ in the 1D HM, $U/t=9$ in the 2D HM, $U/t=8$ and $t_{\perp}/t=4$ in the ladder HM (except for Fig. \ref{fig:U0} below), $U/t=16$ and $t_{\perp}/t=8$ in the bilayer HM, 
$U/t=2\Delta/t=14$ and $t_{\rm K}/t=4$ in the 1D PAM, $U/t=2\Delta/t=28$ and $t_{\rm K}/t=8$ in the 2D PAM, $J_{\rm K}/t=4$ in the 1D KLM, and $J_{\rm K}/t=8$ in the 2D KLM. 
For the spectral functions, the dynamical structure factors, and the imaginary part of the dynamical spin susceptibility, Gaussian broadening is used with the standard deviations of 
$0.1t$ for the 1D and 2D HMs, $0.16t$ for the ladder HM and 1D PAM, $0.32t$ for the bilayer HM and 2D PAM, $0.12t$ for the 1D KLM, and $0.24t$ for the 2D KLM. 
\section{Excitations from the unperturbed state} 
\label{sec:undperturbed}
\subsection{Excitations from the unperturbed state in the 1D and 2D HMs} 
\label{sec:undperturbedHub}
\subsubsection{Excitations in the unperturbed 1D HM} 
\label{sec:excitationUnperturbedHub1d}
\begin{figure} 
\includegraphics[width=\linewidth]{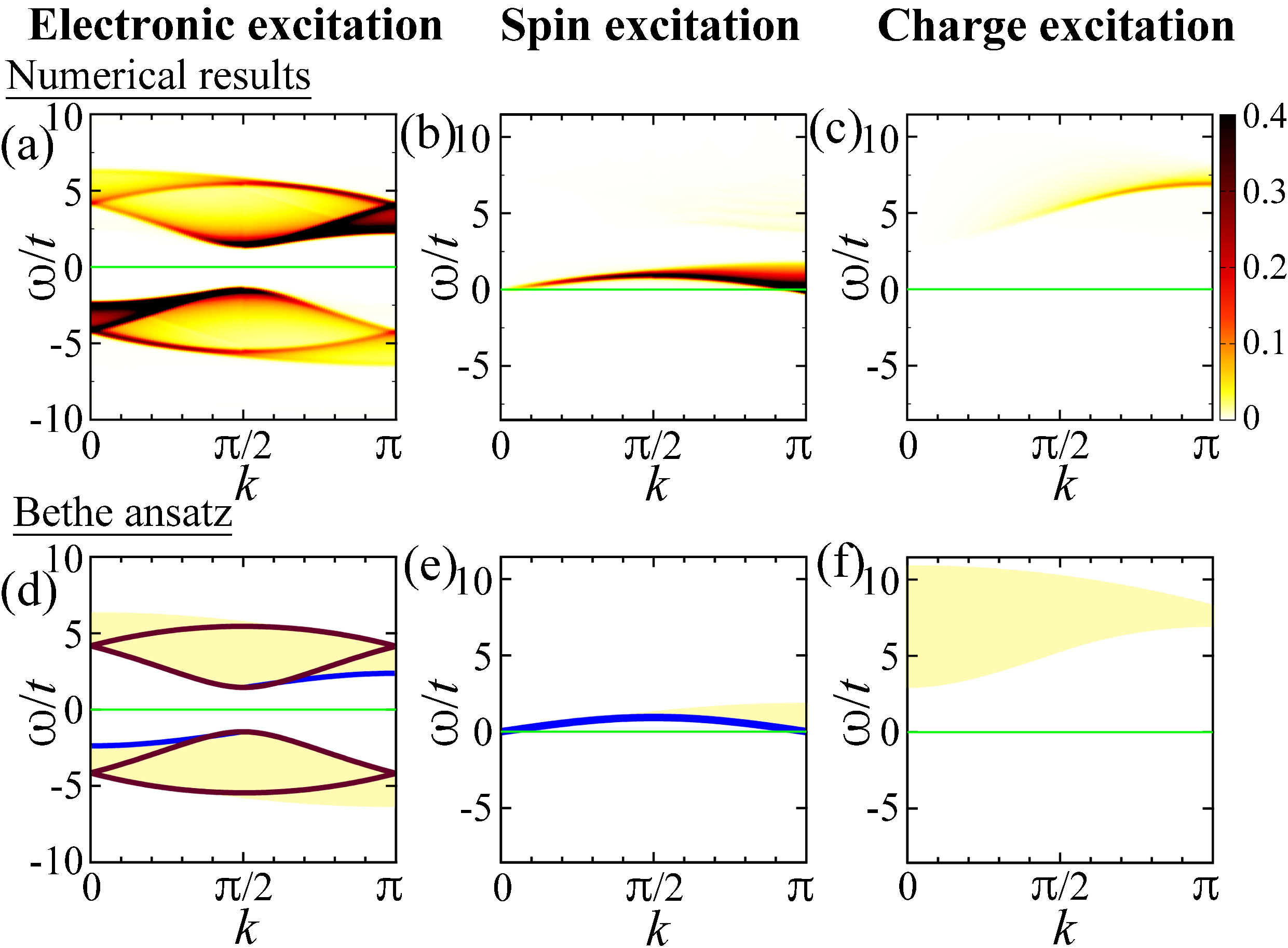}
\caption{Excitation from the ground state at half filling in the 1D HM. 
(a) $A_{\rm GS}(k,\omega)t$, (b) $S(k,\omega)t/3$, (c) $N(k,\omega)t/4$. 
(d)--(f) Dispersion relations obtained using the Bethe ansatz. 
(d) Holon modes (brown curves), spinon modes (blue curves), and holon-spinon continua (light-yellow regions) in electronic excitation. 
(e) Two-spinon continuum (light-yellow region) and lower edge of the continuum (blue curve) in spin excitation. 
(f) Two-holon continuum (light-yellow region) in charge excitation. 
The solid green lines indicate $\omega=0$.}
\label{fig:Hub1d}
\end{figure}
Before discussing the electronic modes induced by perturbations, the excitations from the unperturbed state (ground state at half filling) at $\mu=H=0$ are reviewed. 
\par
In the 1D HM, the band structure primarily consists of the upper ($\omega>0$) and lower ($\omega<0$) Hubbard bands, 
which are separated by a band gap [Fig. \ref{fig:Hub1d}(a)]. 
This band gap is called a Hubbard gap or Mott gap, which becomes larger as the interaction increases \cite{Essler,TakahashiBook,LiebWu,HubbardI}. 
\par
The dominant parts of the electronic, charge, and spin excitations in the 1D HM are identified in terms of spinons and holons using the Bethe ansatz, 
as follows \cite{Kohno1DHub,DDMRGAkw,Essler,TakahashiBook}: 
\par
In the lower band ($\omega<0$) of the electronic excitation, the fan-shaped dispersion relation is identified as the holon modes [brown curves for $\omega<0$ in Fig. \ref{fig:Hub1d}(d)], 
and the low-$|\omega|$ mode for $0\le k<\pi/2$ is identified as the spinon mode [blue curve for $\omega<0$ in Fig. \ref{fig:Hub1d}(d)]. 
The dispersion relations in the upper band ($\omega>0$) are obtained using the particle--hole transformation 
[Appendix \ref{sec:phSym}; Eq. (\ref{eq:AkwphX})], 
\begin{equation}
A_{\rm GS}({\bm k},\omega)=A_{\rm GS}({\bm \pi}-{\bm k},-\omega).
\label{eq:AkwphTransHF}
\end{equation}
\par
The spin excitation [Fig. \ref{fig:Hub1d}(b)] is primarily explained by the two-spinon continuum [light-yellow region in Fig. \ref{fig:Hub1d}(e)]. 
The lower edge is dominant and behaves as a spin mode [blue curve in Fig. \ref{fig:Hub1d}(e)]. 
\par
The continuum in the charge excitation [Fig. \ref{fig:Hub1d}(c)] corresponds to the two-holon states of $N_{\rm h}=2$ ($\eta=1, \eta^z=-1$) 
transformed to $\eta^z=0$ using the ${\hat \eta}^+$ operator under $\eta$-SU(2) symmetry for $\mu=0$ (Sec. \ref{sec:etaSU2}) [light-yellow region in Fig. \ref{fig:Hub1d}(f)] \cite{Essler}. 
It should be noted that although the two-holon states of the $k$-$\Lambda$ string solutions \cite{Woynarovich} exhibit the same dispersion relation as this continuum \cite{Essler}, 
they do not appear in the charge ($\eta>0$) excitation because they are $\eta$-singlet ($\eta=0$) states \cite{Essler}. 
\subsubsection{$\eta$-SU(2) symmetry} 
\label{sec:etaSU2}
The HM, symmetric PAM, and KLM on a bipartite lattice possess not only spin-SU(2) symmetry at $H=0$, but also $\eta$-SU(2) symmetry for charge at $\mu=0$, 
and SO(4)(=[SU(2)$\times$SU(2)]/Z$_2$) symmetry at $\mu=H=0$ \cite{Essler,TakahashiBook,SO4}. 
\par
Each eigenstate of the Hamiltonian, with or without a chemical potential $\mu$ or a magnetic field $H$, 
can have definite values of $S$, $S^z$, $\eta$, and $\eta^z$ \cite{Essler,TakahashiBook,SO4}, 
where $\eta$ and $\eta^z$ denote the total $\eta$ quantum number and eigenvalue of ${ \hat \eta}^z$, respectively. 
The $\eta$ operators are defined as 
\begin{equation}
\label{eq:eta}
\begin{array}{lll}
{\hat \eta}^{+}=\sum_{\lambda,i}\eta^{\lambda,+}_i,&
{\hat \eta}^{-}=\sum_{\lambda,i}\eta^{\lambda,-}_i,&
{\hat \eta}^{z}=\sum_{\lambda,i}\eta^{\lambda,z}_i,
\end{array}
\end{equation}
where $\lambda=c$ in the HM; $\lambda=c$ and $f$ in the PAM and KLM, and 
\begin{align}
\label{eq:etai}
&\begin{array}{ll}
\eta^{c,+}_i=(-)^ic^{\dagger}_{i,\uparrow}c^{\dagger}_{i,\downarrow},&
\eta^{f,+}_i=-(-)^if^{\dagger}_{i,\uparrow}f^{\dagger}_{i,\downarrow},\\
\eta^{c,-}_i=(-)^ic_{i,\downarrow}c_{i,\uparrow},&
\eta^{f,-}_i=-(-)^if_{i,\downarrow}f_{i,\uparrow},\\
\eta^{c,z}_i=\frac{1}{2}(n^c_{i,\uparrow}+n^c_{i,\downarrow}-1),&
\eta^{f,z}_i=\frac{1}{2}(n^f_{i,\uparrow}+n^f_{i,\downarrow}-1),
\end{array}\\
&(-)^i=\left\{\begin{array}{rl}
1&\text{for a site $i$ on the A sublattice},\\
-1&\text{for a site $i$ on the B sublattice}
\end{array}\right.
\end{align}
on a bipartite lattice. The eigenvalue of ${\hat \eta}^z$ is $-\frac{N_{\rm h}}{2}$. 
\subsubsection{Excitations in the unperturbed 2D HM} 
\label{sec:excitationUnperturbedHub2d}
\begin{figure} 
\includegraphics[width=\linewidth]{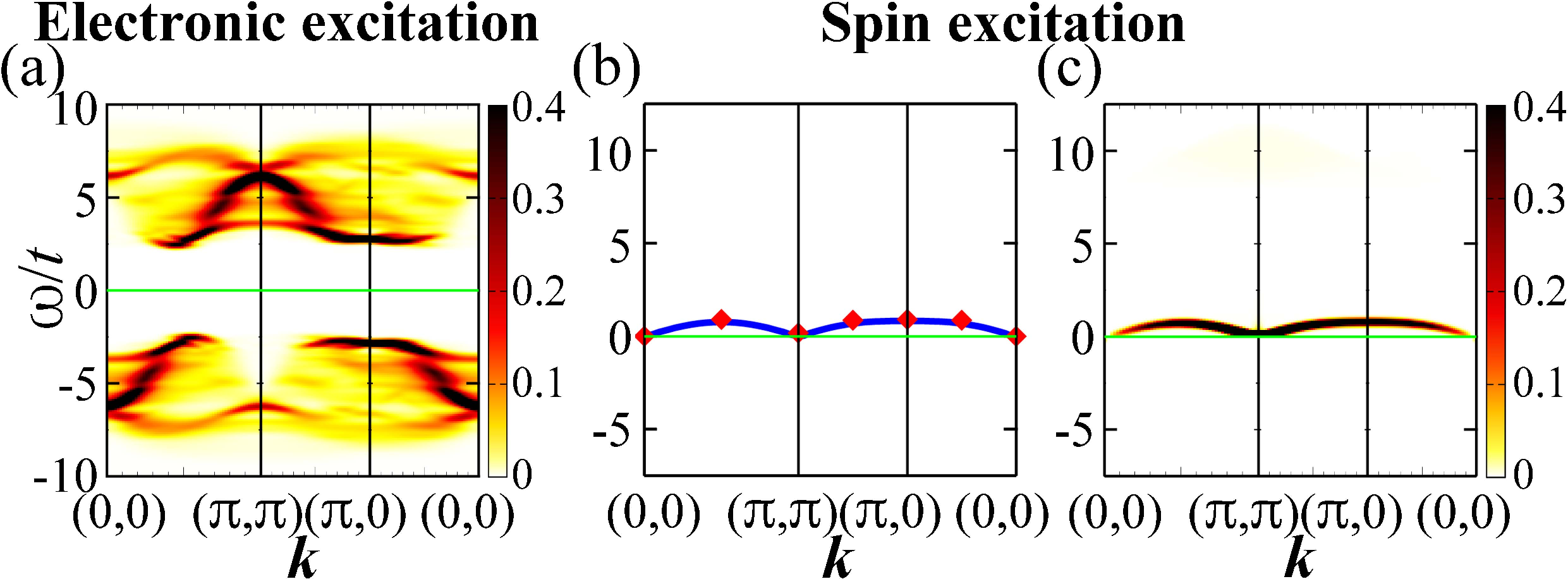}
\caption{Excitation from the ground state at half filling in the 2D HM. 
(a) $A_{\rm GS}({\bm k},\omega)t$. 
(b) Lowest spin-excitation energies in a $4\times 4$-site cluster (red diamonds). 
The blue curve indicates the dispersion relation of the spin-wave mode extracted from (c). 
(c) $\frac{1}{2\pi}{\rm Im}\chi({\bm k},\omega)t$ for $\omega>0$. 
The solid green lines indicate $\omega=0$.}
\label{fig:Hub2d}
\end{figure}
In the 2D HM, the overall spectral features of the electronic excitation [Fig. \ref{fig:Hub2d}(a)] are similar to those of the 1D HM [Fig. \ref{fig:Hub1d}(a)], 
and the dominant electronic excitations of the momentum ${\bm k}$ along $(0,0)$--$(\pi,\pi)$ can basically be interpreted as those of the 1D HM deformed by interchain hopping \cite{Kohno2DHub}. 
\par
The spin excitation in the 2D HM is dominated by a gapless mode, 
which can be identified as the spin-wave mode (magnon) [Figs. \ref{fig:Hub2d}(b) and \ref{fig:Hub2d}(c)] \cite{spinbag,AndersonSW}. 
\subsection{Excitations from the unperturbed state in spin-gapped Mott and Kondo insulators} 
\label{sec:unperturbedSpinGap}
\subsubsection{Overall spectral features of unperturbed spin-gapped Mott and Kondo insulators} 
\label{sec:overallUnperturbedSpinGap}
\begin{figure*} 
\includegraphics[width=\linewidth]{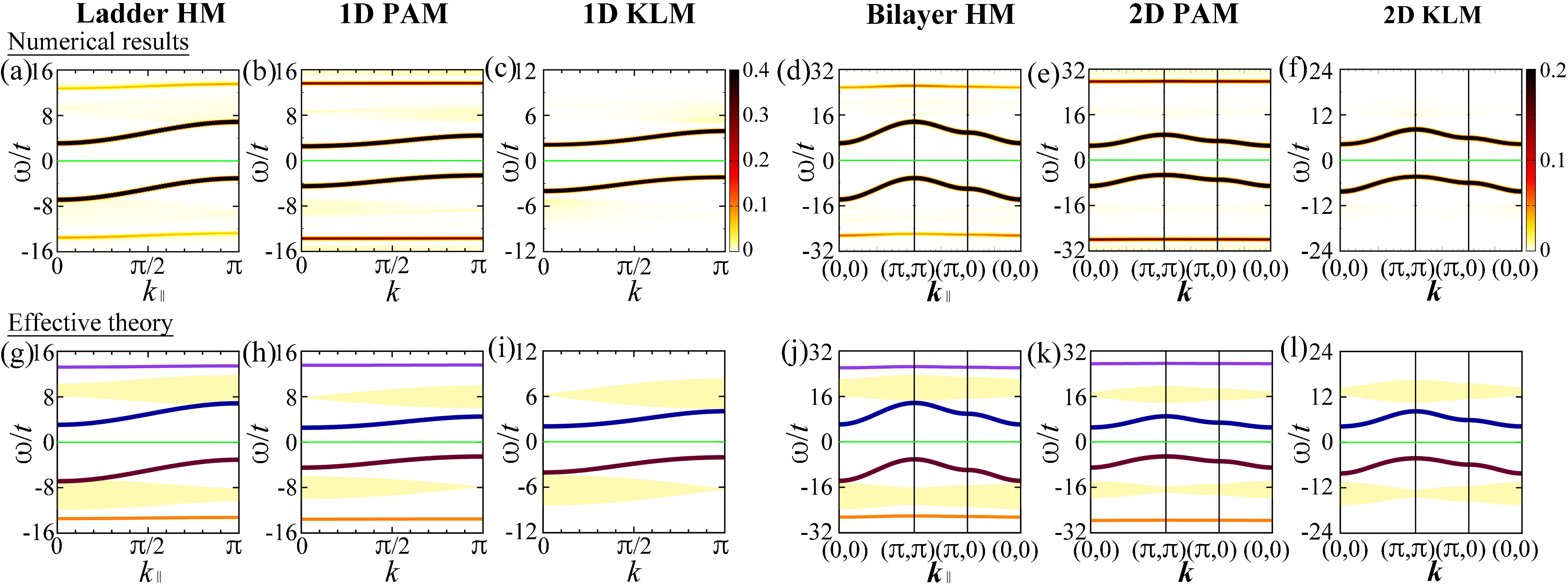}
\caption{Electronic excitation from the ground state at half filling in the ladder HM [(a), (g)], 1D PAM [(b), (h)], 
1D KLM [(c), (i)], bilayer HM [(d), (j)], 2D PAM [(e), (k)], and 2D KLM [(f), (l)]. 
(a)--(f) $A_{\rm GS}(k_\parallel,0,\omega)t+A_{\rm GS}({k_\parallel},\pi,\omega)t$ [(a)], $A_{\rm GS}({\bm k},\omega)t$ [(b), (c), (f)], 
${\bar A}_{\rm GS}({\bm k_\parallel},0,\omega)t+{\bar A}_{\rm GS}({\bm k_\parallel},\pi,\omega)t$ [(d)], and ${\bar A}_{\rm GS}({\bm k},\omega)t$ [(e)]. 
(g)--(l) Dispersion relations of electronic excitations in the effective theory; 
$\omega=\varepsilon^{\cal A}_{{\bm k}_\parallel}$ (solid blue curves), 
$\omega=-\varepsilon^{\cal R}_{-{\bm k}_\parallel}$ (solid brown curves), 
$\omega=\varepsilon^{\bar {\cal A}}_{{\bm k}_\parallel}$ (solid purple curves), 
$\omega=-\varepsilon^{\bar {\cal R}}_{-{\bm k}_\parallel}$ (solid orange curves), 
$\omega=\varepsilon^{{\cal A}{\cal T}}_{{\bm k}_\parallel;{\bm p}_\parallel}$ (light-yellow regions for $\omega>0$), and 
$\omega=-\varepsilon^{{\cal R}{\cal T}}_{-{\bm k}_\parallel;-{\bm p}_\parallel}$ (light-yellow regions for $\omega<0$). 
The solid green lines indicate $\omega=0$.}
\label{fig:AkwLadBil}
\end{figure*}
\begin{figure*} 
\includegraphics[width=\linewidth]{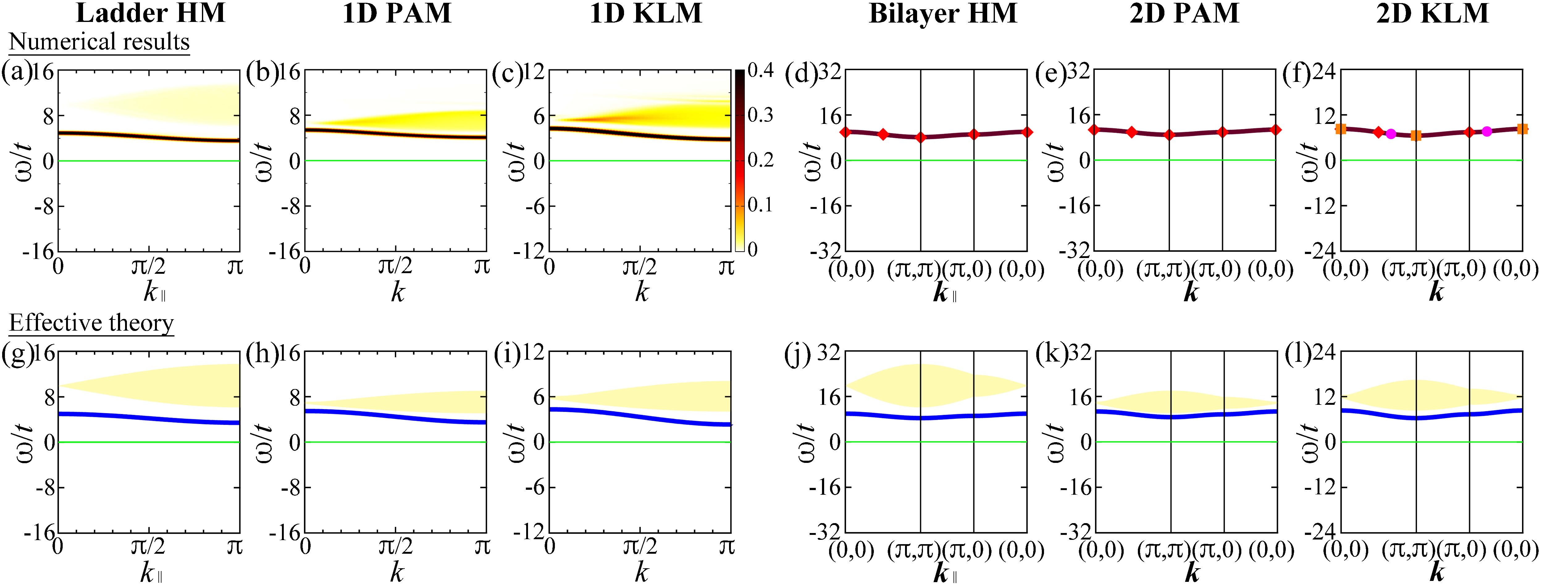}
\caption{Spin excitation from the ground state at half filling in the ladder HM [(a), (g)], 1D PAM [(b), (h)], 1D KLM [(c), (i)], 
bilayer HM [(d), (j)], 2D PAM [(e), (k)], and 2D KLM [(f), (l)]. 
(a)--(c) $S(k_\parallel,\pi,\omega)t/3$ [(a)] and $S(k,\omega)t/3$ [(b), (c)]. 
(d)--(f) Lowest spin-excitation energies for $N_{\rm u}=\sqrt{10}\times\sqrt{10}$ (orange squares) [(f)], $3\times 3$ (magenta circles) [(f)], and $\sqrt{8}\times\sqrt{8}$ (red diamonds) [(d)--(f)]. 
The brown curves indicate the least-squares fitting in the form of $\omega=Jd\gamma_{{\bm k}_{\parallel}}+\Delta E$. 
(g)--(l) Dispersion relations of spin excitations in the effective theory; 
$\omega=e^{\cal T}_{{\bm k}_\parallel}$ (solid blue curves) and 
$\omega=e^{{\cal A}{\cal R}}_{{\bm k}_\parallel;{\bm p}_\parallel}$ (light-yellow regions). 
The solid green lines indicate $\omega=0$.}
\label{fig:SkwLadBil}
\end{figure*}
\begin{figure} 
\includegraphics[width=\linewidth]{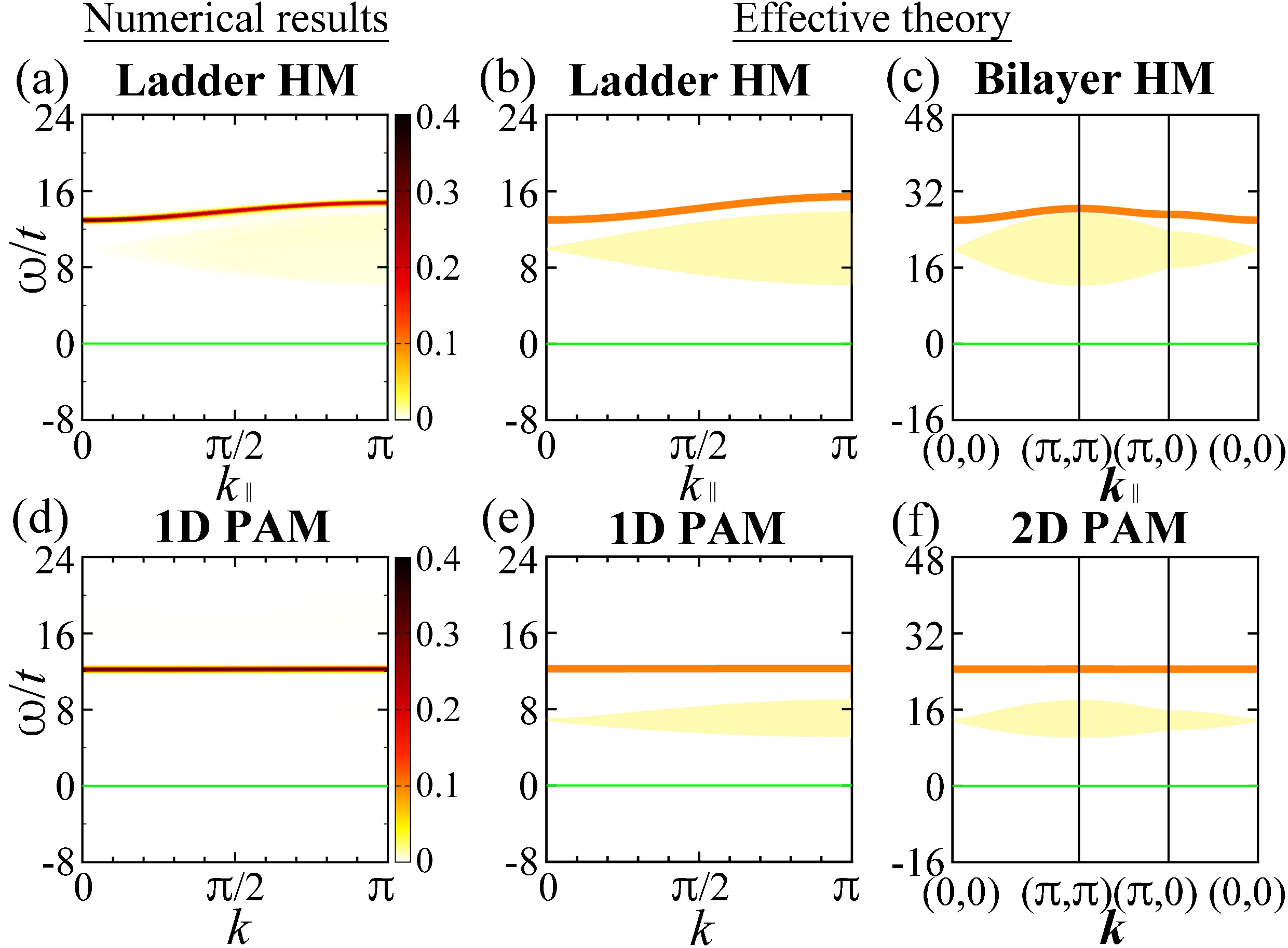}
\caption{Charge excitation from the ground state at half filling in the ladder HM [(a), (b)], 1D PAM [(d), (e)], bilayer HM [(c)], and 2D PAM [(f)]. 
(a), (d) $N(k_\parallel,\pi,\omega)t/4$ [(a)] and $N(k,\omega)t/4$ [(d)]. 
(b), (c), (e), (f) Dispersion relations of charge excitations in the effective theory; 
$\omega=e^{{\cal C}^0}_{{\bm k}_\parallel}$ (solid orange curves) and 
$\omega=e^{{\cal A}{\cal R}}_{{\bm k}_\parallel;{\bm p}_\parallel}$ (light-yellow regions). 
The solid green lines indicate $\omega=0$.}
\label{fig:charge}
\end{figure}
In the ladder and bilayer HMs and 1D and 2D PAMs and KLMs in the regime of large $U/t$, $t_{\perp}/t$, $t_{\rm K}/t$, $\Delta/t$, and $J_{\rm K}/t$, 
there are two dominant modes (one for $\omega>0$ and the other for $\omega<0$) 
separated by a band gap in the electronic excitation from the ground state at half filling [Figs. \ref{fig:AkwLadBil}(a)--\ref{fig:AkwLadBil}(f)]. 
In addition, in the HM and PAM, there are two almost flat modes in the high-$|\omega|$ regime (one for $\omega>0$ and the other for $\omega<0$) 
[Figs. \ref{fig:AkwLadBil}(a), \ref{fig:AkwLadBil}(b), \ref{fig:AkwLadBil}(d), and \ref{fig:AkwLadBil}(e)]. 
\par
In the spin excitation, a dominant mode with an excitation-energy gap exists [Figs. \ref{fig:SkwLadBil}(a)--\ref{fig:SkwLadBil}(f)]. 
The energy gap of the spin excitation is smaller than the band gap (charge gap) 
\cite{KohnoHubLadder,KohnoKLM,KohnoMottT,KohnoTinduced,TsunetsuguRMP}. 
\par
In the charge excitation, a dominant mode exists in the high-$\omega$ regime in the HM and PAM [Figs. \ref{fig:charge}(a) and \ref{fig:charge}(d)]. 
\par
In addition, the electronic, spin, and charge excitations have continua 
[Figs. \ref{fig:AkwLadBil}(a)--\ref{fig:AkwLadBil}(f), \ref{fig:SkwLadBil}(a)--\ref{fig:SkwLadBil}(c), and \ref{fig:charge}(a)].
\subsubsection{Excited states and dispersion relations in unperturbed spin-gapped Mott and Kondo insulators in the effective theory} 
\label{sec:effTheoryUnperturbedSpinGap}
In the effective theory for weak inter-unit-cell hopping ($|t|\ll U, t_{\perp}, t_{\rm K}, \Delta$, and $J_{\rm K}$) 
(Appendix \ref{sec:effTheory}) \cite{KohnoDIS,KohnoHubLadder,KohnoKLM,KohnoTinduced,KohnoUinf,TsunetsuguRMP,OneSiteKondo,localRungHubLadder}, 
the modes and continua can be identified as single-particle and two-particle excitations, respectively. 
\begin{table} 
\caption{Symbols of states in a unit cell for the ladder and bilayer HMs and 1D and 2D PAMs and KLMs.}
\begin{tabular}{lcccc}
\hline\hline
States&Symbols\tnote&HM&PAM&KLM\\\hline
Ground state&${\cal G}$&$\psi^-$&$\psi^-$&${\rm S}$\\ 
States of $N_{\rm e}=1$&${\tilde {\cal R}}_{\sigma}$&${\rm A}_{\sigma}, {\rm B}_{\sigma}$&${\rm R}^{\pm}_{\sigma}$&${\rm R}_{\sigma}$\\ 
Low-energy state of $N_{\rm e}=1$&${\cal R}_{\sigma}$&${\rm B}_{\sigma}$&${\rm R}^-_{\sigma}$&${\rm R}_{\sigma}$\\ 
High-energy state of $N_{\rm e}=1$&${\bar {\cal R}}_{\sigma}$&${\rm A}_{\sigma}$&${\rm R}^{+}_{\sigma}$&\\ 
States of $N_{\rm e}=3$&${\tilde {\cal A}}_{\sigma}$&${\rm F}_{\sigma}, {\rm G}_{\sigma}$&${\rm P}^{\pm}_{\sigma}$&${\rm P}_{\sigma}$\\ 
Low-energy state of $N_{\rm e}=3$&${\cal A}_{\sigma}$&${\rm F}_{\sigma}$&${\rm P}^-_{\sigma}$&${\rm P}_{\sigma}$\\ 
High-energy state of $N_{\rm e}=3$&${\bar {\cal A}}_{\sigma}$&${\rm G}_{\sigma}$&${\rm P}^{+}_{\sigma}$&\\ 
State of $S=1$ and $S^z=\gamma$&${\cal T}^\gamma$&${\rm T}^{\gamma}$&${\rm T}^{\gamma}$&${\rm T}^{\gamma}$\\ 
State of $\eta=1$ and $\eta^z=-1$&${\cal C}^{-1}$&${\rm V}$&${\rm V}$&\\ 
State of $\eta=1$ and $\eta^z=0$&${\cal C}^0$&${\rm D}^-$&${\rm D}^-$&\\
State of $\eta=1$ and $\eta^z=1$&${\cal C}^1$&${\rm Z}$&${\rm Z}$&\\\hline\hline 
\end{tabular}
\begin{tablenotes}
\item{The eigenstates in a unit cell are explicitly shown in Tables \ref{tbl:HM}--\ref{tbl:KLM}. 
$\sigma=\uparrow,\downarrow$ and $\gamma=-1,0,1$.}
\end{tablenotes}
\label{tbl:symbols}
\end{table}
\par
The single-particle excited state with a momentum ${{\bm k}_{\parallel}}$, $|X\rangle_{{\bm k}_{\parallel}}$, is obtained 
by replacing the ground state at an $i$th unit cell, $|{\cal G}\rangle_i$, in the effective ground state 
\begin{equation}
|{\rm GS}\rangle=\prod_j |{\cal G}\rangle_j
\label{eq:GS}
\end{equation}
with an excited state at the unit cell, $|X\rangle_i$, and by making a Fourier transform as 
\begin{equation}
|X\rangle_{{\bm k}_{\parallel}}=\frac{1}{\sqrt{N_{\rm u}}}\sum_i e^{i{\bm k}_{\parallel}\cdot{\bm r}_i}|X\rangle_i\prod_{j(\ne i)} |{\cal G}\rangle_j
\label{eq:Xk}
\end{equation}
(Table \ref{tbl:symbols}; Appendix \ref{sec:effTheory}). 
\par
The energy of the effective ground state is obtained as 
\begin{equation}
E_{\rm GS}=({\cal E}_{\cal G}+d\xi_{{\cal G}{\cal G}})N_{\rm u}
\label{eq:eGS}
\end{equation}
up to $\mathcal{O}(t^2)$ in $d$ dimensions; $d=1$ for the ladder HM, 1D PAM, and 1D KLM, and $d=2$ for the bilayer HM, 2D PAM, and 2D KLM. 
Here, ${\cal E}_X$ denotes the local energy of a state $|X\rangle$ at a unit cell (Tables \ref{tbl:symbols}--\ref{tbl:KLM}), 
and $\xi_{XY}$ denotes the bond energy of $\mathcal{O}(t^2)$ between states $X$ and $Y$ on neighboring unit cells 
[Eqs. (\ref{eq:xiHub})--(\ref{eq:xiKLM}); Appendix \ref{sec:effTheory}] \cite{KohnoHubLadder,KohnoKLM,KohnoTinduced,TsunetsuguRMP}. 
The excitation energies of $|X\rangle_{{\bm k}_{\parallel}}$ for a boson and fermion $X$ are obtained up to $\mathcal{O}(t^2)$ \cite{KohnoTinduced,KohnoHubLadder,KohnoKLM,TsunetsuguRMP} as 
\begin{align}
\label{eq:eXkBoson}
e^X_{{\bm k}_{\parallel}}=&-2t_{\rm eff}^Xd\gamma_{{\bm k}_{\parallel}}+{\cal E}_X-{\cal E}_{\cal G}+2d(\xi_{{\cal G}X}-\xi_{{\cal G}{\cal G}}),\\
\label{eq:eXkFermion}
\varepsilon^X_{{\bm k}_{\parallel}}=&-2t_{\rm eff}^Xd\gamma_{{\bm k}_{\parallel}}+{\cal E}_X-{\cal E}_{\cal G}+2d(\xi_{{\cal G}X}-\xi_{{\cal G}{\cal G}})\nonumber\\
&-2t_{3{\rm uc}}^Xd(2d\gamma_{{\bm k}_{\parallel}}^2-1),
\end{align}
respectively, where 
\begin{equation}
\label{eq:gammak}
\gamma_{{\bm k}_{\parallel}}=\frac{1}{d}\sum_{i=1}^d\cos k_{\parallel i} 
\end{equation}
with $k_{\parallel 1}=k_x$ and $k_{\parallel 2}=k_y$ in $d$ dimensions. 
The effective nearest-neighbor hopping parameters $t_{\rm eff}^X$ of the boson and fermion $X$ are $\mathcal{O}(t^2)$ and $\mathcal{O}(t)$, respectively, 
and the effective three-unit-cell hopping parameter $t_{3{\rm uc}}^X$ of the fermion $X$ is $\mathcal{O}(t^2)$ 
[Eqs. (\ref{eq:teffHub})--(\ref{eq:teffKLM}); Appendix \ref{sec:effTheory}] \cite{KohnoHubLadder,KohnoKLM,KohnoTinduced,TsunetsuguRMP}. 
\par
For shorthand notation, the spin indices are omitted for spin-independent energies, e.g., 
$e^{\cal T}_{{\bm k}_{\parallel}}=e^{{\cal T}^\gamma}_{{\bm k}_{\parallel}}$, 
$\varepsilon^{\cal A}_{{\bm k}_{\parallel}}=\varepsilon^{{\cal A}_{\sigma}}_{{\bm k}_{\parallel}}$, 
$\varepsilon^{\cal AT}_{{\bm k}_{\parallel};{\bm p}_{\parallel}}=\varepsilon^{{\cal A}_{\sigma}{\cal T}^{\gamma}}_{{\bm k}_{\parallel};{\bm p}_{\parallel}}$, and 
$e^{\cal AR}_{{\bm k}_{\parallel};{\bm p}_{\parallel}}=e^{{\cal A}_{\sigma}{\cal R}_{\sigma^\prime}}_{{\bm k}_{\parallel};{\bm p}_{\parallel}}$ for $H=0$. 
\par
The dispersion relations of the dominant mode and high-$\omega$ mode in the electron-addition excitation ($\omega>0$) are obtained as 
\begin{align}
\label{eq:ekA}
&\omega=\varepsilon^{\cal A}_{{\bm k}_{\parallel}},\\
\label{eq:ekAbar}
&\omega=\varepsilon^{\bar {\cal A}}_{{\bm k}_{\parallel}}
\end{align}
by regarding $|X\rangle_i$ in Eq. (\ref{eq:Xk}) as the low-energy and high-energy states of $N_{\rm e}=3$ with spin $\sigma$ at the $i$th unit cell, namely, 
$|{\cal A}_{\sigma}\rangle_i$ and $|{\bar {\cal A}_{\sigma}}\rangle_i$, respectively (Table \ref{tbl:symbols}; Appendix \ref{sec:effTheory}). 
By representing the states of $N_{\rm e}=3$ with spin $\sigma$ at the $i$th unit cell as 
$|{\tilde {\cal A}}_{\sigma}\rangle_i(=|{\cal A}_{\sigma}\rangle_i$ and $|{\bar {\cal A}_{\sigma}}\rangle_i)$, 
Eqs. (\ref{eq:ekA}) and (\ref{eq:ekAbar}) are expressed as 
\begin{equation}
\omega=\varepsilon^{\tilde {\cal A}}_{{\bm k}_{\parallel}}.
\label{eq:ekAtilde}
\end{equation}
\par
Similarly, the dispersion relations of the dominant mode and high-$|\omega|$ mode in the electron-removal excitation ($\omega<0$) are obtained as 
\begin{align}
\label{eq:ekR}
&\omega=-\varepsilon^{\cal R}_{-{\bm k}_{\parallel}},\\
\label{eq:ekRbar}
&\omega=-\varepsilon^{\bar {\cal R}}_{-{\bm k}_{\parallel}}
\end{align}
by regarding $|X\rangle_i$ in Eq. (\ref{eq:Xk}) as the low-energy and high-energy states of $N_{\rm e}=1$ with spin $\sigma$ at the $i$th unit cell, namely, 
$|{\cal R}_{\sigma}\rangle_i$ and $|{\bar {\cal R}_{\sigma}}\rangle_i$, respectively (Table \ref{tbl:symbols}; Appendix \ref{sec:effTheory}), for the momentum of ${-{\bm k}_{\parallel}}$. 
By representing the states of $N_{\rm e}=1$ with spin $\sigma$ at the $i$th unit cell as 
$|{\tilde {\cal R}}_{\sigma}\rangle_i(=|{\cal R}_{\sigma}\rangle_i$ and $|{\bar {\cal R}_{\sigma}}\rangle_i)$, 
Eqs. (\ref{eq:ekR}) and (\ref{eq:ekRbar}) are expressed as 
\begin{equation}
\omega=-\varepsilon^{\tilde {\cal R}}_{-{\bm k}_{\parallel}}.
\label{eq:ekRtilde}
\end{equation}
\par
In the ladder and bilayer HMs and 1D and 2D PAMs and KLMs, 
the dispersion relations of the dominant modes in the electron-addition excitation [Eq. (\ref{eq:ekA})] and electron-removal excitation [Eq. (\ref{eq:ekR})] are shown 
by the solid blue curves and solid brown curves, respectively, in Figs. \ref{fig:AkwLadBil}(g)--\ref{fig:AkwLadBil}(l), 
which are consistent with the numerical results [Figs. \ref{fig:AkwLadBil}(a)--\ref{fig:AkwLadBil}(f)]. 
The dispersion relations of the high-$|\omega|$ modes in the electron-addition excitation ($\omega>0$) [Eq. (\ref{eq:ekAbar})] and electron-removal excitation ($\omega<0$) [Eq. (\ref{eq:ekRbar})] 
of the ladder and bilayer HMs and 1D and 2D PAMs 
are shown by the solid purple curves and solid orange curves, respectively, in Figs. \ref{fig:AkwLadBil}(g), \ref{fig:AkwLadBil}(h), \ref{fig:AkwLadBil}(j), and \ref{fig:AkwLadBil}(k), 
which are consistent with the numerical results [Figs. \ref{fig:AkwLadBil}(a), \ref{fig:AkwLadBil}(b), \ref{fig:AkwLadBil}(d), and \ref{fig:AkwLadBil}(e)]. 
\par
The dispersion relation of the spin mode is obtained as 
\begin{equation}
\omega=e^{\cal T}_{{\bm k}_{\parallel}}
\label{eq:ekT}
\end{equation}
by regarding $|X\rangle_i$ in Eq. (\ref{eq:Xk}) as the spin-triplet ($S=1$) state of $S^z=\gamma(=-1,0,1)$ at the $i$th unit cell, $|{\cal T}^{\gamma}\rangle_i$ (Table \ref{tbl:symbols}; Appendix \ref{sec:effTheory}). 
\par
In the ladder and bilayer HMs and 1D and 2D PAMs and KLMs, 
the dispersion relations of the spin modes [Eq. (\ref{eq:ekT})] are shown 
by the solid blue curves in Figs. \ref{fig:SkwLadBil}(g)--\ref{fig:SkwLadBil}(l), 
which are consistent with the numerical results [Figs. \ref{fig:SkwLadBil}(a)--\ref{fig:SkwLadBil}(f)]. 
\par
The dispersion relation of the charge mode is obtained as 
\begin{equation}
\omega=e^{{\cal C}^0}_{{\bm k}_{\parallel}}
\label{eq:ekD-}
\end{equation}
by regarding $|X\rangle_i$ in Eq. (\ref{eq:Xk}) as the $\eta$-triplet ($\eta=1$) state of $\eta^z=0$ at the $i$th unit cell, $|{\cal C}^0\rangle_i$, in the HM and PAM 
(Table \ref{tbl:symbols}; Appendix \ref{sec:effTheory}). 
\par
In the ladder and bilayer HMs and 1D and 2D PAMs, 
the dispersion relations of the charge modes [Eq. (\ref{eq:ekD-})] are shown 
by the solid orange curves in Figs. \ref{fig:charge}(b), \ref{fig:charge}(c), \ref{fig:charge}(e), and \ref{fig:charge}(f), 
which are consistent with the numerical results [Figs. \ref{fig:charge}(a) and \ref{fig:charge}(d)]. 
\par
The two-particle excited state with a momentum ${\bm k}_{\parallel}$ is effectively obtained as 
\begin{align}
|XY\rangle_{{\bm k}_{\parallel};{\bm p}_{\parallel}}&=\frac{1}{\sqrt{N_{\rm u}(N_{\rm u}-1)}}\nonumber\\
&\times\sum_{i\ne j}e^{i({\bm k}_{\parallel}-{\bm p}_{\parallel})\cdot{\bm r}_i}
e^{i{\bm p}_{\parallel}\cdot{\bm r}_j}|X\rangle_i|Y\rangle_j\prod_{l(\ne i,j)}|{\cal G}\rangle_l;
\label{eq:XYk}
\end{align}
its excitation energy is approximately obtained as 
\begin{align}
\label{eq:eXYkf}
\varepsilon^{XY}_{{\bm k}_{\parallel};{\bm p}_{\parallel}}=
\varepsilon_{{\bm k}_{\parallel}-{\bm p}_{\parallel}}^{X}+e^{Y}_{{\bm p}_{\parallel}}&\text{for a fermion $X$ and boson $Y$},\\
\label{eq:eXYkb}
e^{XY}_{{\bm k}_{\parallel};{\bm p}_{\parallel}}=
\varepsilon_{{\bm k}_{\parallel}-{\bm p}_{\parallel}}^{X}+\varepsilon^{Y}_{{\bm p}_{\parallel}}&\text{ for fermions $X$ and $Y$}.
\end{align}
\par
The continuum of the electronic excitation corresponds to 
$|{\cal A}_{\sigma}{\cal T}^{\gamma}\rangle_{{\bm k}_{\parallel};{\bm p}_{\parallel}}$ for $\omega>0$ and 
$|{\cal R}_{\sigma}{\cal T}^{\gamma}\rangle_{-{\bm k}_{\parallel};-{\bm p}_{\parallel}}$ for $\omega<0$, where $s^z+\gamma=\pm\frac{1}{2}$. 
The continuum of the spin excitation and that of the charge excitation correspond to the spin-triplet states 
and $\eta$-triplet states of $\eta^z=0$, respectively, expressed using $|{\cal A}_{\sigma}{\cal R}_{\sigma^{\prime}}\rangle_{{\bm k}_{\parallel};{\bm p}_{\parallel}}$ 
(Appendix \ref{sec:effTheory}) \cite{KohnoHubLadder,KohnoKLM,KohnoTinduced}. 
\par
In the ladder and bilayer HMs and 1D and 2D PAMs and KLMs, 
the dispersion relations of the continua in the electronic excitation are shown 
by the light-yellow regions in Figs. \ref{fig:AkwLadBil}(g)--\ref{fig:AkwLadBil}(l), 
which are consistent with the numerical results [Figs. \ref{fig:AkwLadBil}(a)--\ref{fig:AkwLadBil}(f)]. 
The dispersion relations of the continua in the spin excitation are shown 
by the light-yellow regions in Figs. \ref{fig:SkwLadBil}(g)--\ref{fig:SkwLadBil}(l) and 
are consistent with the numerical results [Figs. \ref{fig:SkwLadBil}(a)--\ref{fig:SkwLadBil}(c)]. 
The dispersion relations of the continua in the charge excitation of the ladder and bilayer HMs and 1D and 2D PAMs are shown 
by the light-yellow regions in Figs. \ref{fig:charge}(b), \ref{fig:charge}(c), \ref{fig:charge}(e), and \ref{fig:charge}(f) and 
are consistent with the numerical results [Fig. \ref{fig:charge}(a)]. 
\section{Electronic modes induced by charge perturbations} 
\label{sec:chargePerturbed}
\subsection{Outline} 
\label{sec:outlineCharge}
Section \ref{sec:chargePerturbed} discusses electronic modes induced by charge perturbations, such as doping-induced states (Secs. \ref{sec:DIS}--\ref{sec:swDIS}), 
emergent states corresponding to the doping-induced states in undoped charge-perturbed systems (Sec. \ref{sec:etaz0DIS}), and 
electronic modes induced by charge fluctuation (Sec. \ref{sec:chargeFluc}). 
\par
In Secs. \ref{sec:DIS}--\ref{sec:swDIS}, the origins, underlying mechanisms, and properties of the doping-induced states elucidated in Refs. 
\cite{Kohno1DHub,Kohno2DHub,KohnoDIS,KohnoRPP,KohnoHubLadder,KohnoKLM,KohnoGW,KohnoAF,KohnoSpin,Kohno1DtJ,Kohno2DtJ,KohnoMottT,KohnoTinduced} 
are reviewed. 
Specifically, Secs. \ref{sec:holeDIS} and \ref{sec:eleDIS} explain the hole- and electron-doping-induced states in parallel 
from the viewpoint of quantum numbers, which are validated by numerical calculations in the 1D and 2D HMs. 
The relation between the hole- and electron-doping-induced states is mentioned based on particle--hole symmetry in Sec. \ref{sec:relationDIS}. 
Section \ref{sec:DISspingap} explains the doping-induced states in spin-gapped Mott and Kondo insulators, 
using the electronic excited states arising from perturbed states in the effective theory for weak inter-unit-cell hopping presented in Sec. \ref{sec:eleEff}. 
The behavior of the spectral weights and dispersion relations of the doping-induced states in the small-doping regime is explained in Sec. \ref{sec:swDIS}. 
\par
In Sec. \ref{sec:etaz0DIS}, the spectral functions of nonzero-$\eta$ states at half filling are shown to be obtained from those of the doped ground states 
via relations between the spectral functions of $\eta$-SU(2) symmetric states (Sec. \ref{sec:Akwetaz0DIS}). 
Remarks are also made on the experimental realization of the nonzero-$\eta$ states and on the fact that nonzero-$\eta$ states by themselves do not imply superconducting states (Sec. \ref{sec:Remarksetaz0DIS}). 
\par
In Sec. \ref{sec:chargeFluc}, the dispersion relations of electronic modes induced by charge fluctuation are derived using quantum-number analysis (Sec. \ref{sec:selectionRulesCF}), 
and the emergence of these modes is demonstrated with the effective theory and numerical calculations for spin-gapped Mott and Kondo insulators (Sec. \ref{sec:effectiveTheoryCF}). 
\subsection{Selection rules for doping-induced states} 
\label{sec:DIS}
\subsubsection{Hole-doping-induced states} 
\label{sec:holeDIS}
The simplest perturbation for charge is the chemical-potential shift down to the top of the lower band (up to the bottom of the upper band). 
This causes hole (electron) doping in the system. The state obtained by this perturbation is the hole-doped (electron-doped) ground state. 
By doping Mott and Kondo insulators, electronic modes emerge within the band gap, 
exhibiting the spin-mode dispersion relation shifted by the Fermi momentum ${\bm k}_{\rm F}^-$ (${\bm k}_{\rm F}^+$) of the hole-doped (electron-doped) system 
\cite{Kohno1DHub,Kohno2DHub,KohnoDIS,KohnoRPP,KohnoHubLadder,KohnoKLM,KohnoGW,KohnoAF,KohnoSpin,Kohno1DtJ,Kohno2DtJ,KohnoMottT,KohnoTinduced}. 
This characteristic can be explained from the viewpoint of quantum numbers 
\cite{KohnoDIS,KohnoRPP,KohnoHubLadder,KohnoKLM,KohnoGW,Kohno1DtJ,KohnoMottT,KohnoTinduced}, as described below. 
\par
By adding an electron with a momentum ${\bm k}$ and $S^z=m^\prime$ to the one-hole-doped ($N_{\rm h}=1$) ground state with the momentum of $-{\bm k}_{\rm F}^-$ and $S^z=m$, 
a state with the momentum of ${\bm k}-{\bm k}_{\rm F}^-$, $S^z=m^\prime+m$, and $N_{\rm h}=0$ is obtained. 
This state can overlap with the ground state having the momentum of ${\bm 0}$ and $S^z=0$ at half filling ($N_{\rm h}=0$) if ${\bm k}={\bm k}_{\rm F}^-$ and $m^\prime=-m$, 
which corresponds to the top of the lower band, as in the case of a noninteracting band insulator. 
\par
In strongly correlated insulators, the electron-added state can also overlap with the spin-excited state having the momentum of ${\bm k}-{\bm k}_{\rm F}^-$ and $S^z=m^\prime+m$ at half filling ($N_{\rm h}=0$) 
and can emerge in the electronic spectrum within the band gap along 
\begin{equation}
\omega=e^{\rm spin}_{{\bm k}-{\bm k}_{\rm F}^-}+\mu_-,
\label{eq:ekhDISspin}
\end{equation}
where $e^{\rm spin}_{\bm p}$ denotes the excitation energy of the spin excitation with a momentum ${\bm p}$ at half filling, 
and $\mu_-$ denotes the $\omega$ value at the top of the lower band, 
\begin{equation}
\mu_-=E_{\rm GS}-E_{\rm GS}^{N_{\rm h}=1}.
\label{eq:mu-}
\end{equation}
In the one-hole-doped system, $\mu_-=0$. 
\par
Equation (\ref{eq:ekhDISspin}) can be explained as follows: 
in the electron-addition excitation from $|X\rangle=|{\rm GS}\rangle^{N_{\rm h}=1}_{-{\bm k}_{\rm F}^-}$ (one-hole-doped ground state with the momentum of $-{\bm k}_{\rm F}^-$) 
to $|n\rangle=|{\rm Spin}\rangle_{{\bm k}-{\bm k}_{\rm F}^-}$ (spin-excited state with the momentum of ${\bm k}-{\bm k}_{\rm F}^-$ at half filling) in Eq. (\ref{eq:Akwdef}), 
$|_{{\bm k}-{\bm k}_{\rm F}^-}\langle{\rm Spin}|\lambda^{\dagger}_{{\bm k},\sigma}|{\rm GS}\rangle^{N_{\rm h}=1}_{-{\bm k}_{\rm F}^-}|^2$ can be nonzero, 
for which $\omega=E_n-E_X=e^{\rm spin}_{{\bm k}-{\bm k}_{\rm F}^-}+\mu_-$ because $E_n=e^{\rm spin}_{{\bm k}-{\bm k}_{\rm F}^-}+E_{\rm GS}$ and $E_X=E_{\rm GS}^{N_{\rm h}=1}$. 
\par
In addition, the electron-added state can overlap with the charge-excited state having the momentum of ${\bm k}-{\bm k}_{\rm F}^-$, $S^z=0$, and $N_{\rm h}=0$ 
if $m^\prime=-m$ and can emerge in the electronic spectrum in the high-$\omega$ regime for $\omega>0$ along 
\begin{equation}
\omega=e^{{\rm charge},N_{\rm h}=0}_{{\bm k}-{\bm k}_{\rm F}^-}+\mu_-,
\label{eq:ekhDISchargeNh0}
\end{equation}
where $e^{{\rm charge},N_{\rm h}=0}_{\bm p}$ denotes the excitation energy of the charge excitation with a momentum ${\bm p}$ at half filling. 
Equation (\ref{eq:ekhDISchargeNh0}) can be explained by regarding $|n\rangle$ and $|X\rangle$ in the electron-addition excitation of Eq. (\ref{eq:Akwdef}) 
as the charge-excited state of $N_{\rm h}=0$ and $|{\rm GS}\rangle^{N_{\rm h}=1}_{-{\bm k}_{\rm F}^-}$, respectively. 
\par
Furthermore, by removing an electron with a momentum ${\bm k}$ and $S^z=m$ from the one-hole-doped ground state 
with the momentum of $-{\bm k}_{\rm F}^-$ and $S^z=m$, 
a state with the momentum of $-{\bm k}-{\bm k}_{\rm F}^-$, $S^z=0$, and $N_{\rm h}=2$ is obtained. 
This state can overlap with the charge-excited state having the momentum of $-{\bm k}-{\bm k}_{\rm F}^-$, $S^z=0$, and $N_{\rm h}=2$ 
and can emerge in the electronic spectrum in the high-$|\omega|$ regime for $\omega<0$ along 
\begin{equation}
\omega=-e^{{\rm charge},N_{\rm h}=2}_{-{\bm k}-{\bm k}_{\rm F}^-}-\mu_-,
\label{eq:ekhDISchargeNh2}
\end{equation}
where $e^{{\rm charge},N_{\rm h}=2}_{\bm p}$ denotes the excitation energy of the charge excitation with a momentum ${\bm p}$ and $N_{\rm h}=2$ from the ground state at half filling. 
Equation (\ref{eq:ekhDISchargeNh2}) can be explained by regarding $|n\rangle$ and $|X\rangle$ in the electron-removal excitation of Eq. (\ref{eq:Akwdef}) 
as the charge-excited state of $N_{\rm h}=2$ and $|{\rm GS}\rangle^{N_{\rm h}=1}_{-{\bm k}_{\rm F}^-}$, respectively. 
\par
In hole-doped Mott insulators of the HM, because the lower band in the momentum regime inside the Fermi sea is essentially entirely filled with electrons, 
the electronic states emerge in the momentum regime primarily outside the Fermi sea in the electron-addition spectrum. 
\par
In the 1D and 2D HMs, the emergence of electronic modes caused by hole doping 
in line with the above selection rules \cite{Kohno1DHub,Kohno2DHub,KohnoRPP} is confirmed as follows: 
by representing the excitation energy of the spin mode as $e^{\cal T}_{\bm k}$ [Figs. \ref{fig:Hub1d}(b), \ref{fig:Hub1d}(e), \ref{fig:Hub2d}(b), and \ref{fig:Hub2d}(c)] 
and noting that the Fermi momentum of the hole-doped system is ${\bm k}^-_{\rm F}=\frac{\bm \pi}{2}$ in the small-doping limit [Figs. \ref{fig:Hub1d}(a), \ref{fig:Hub1d}(d), and \ref{fig:Hub2d}(a)], 
according to Eq. (\ref{eq:ekhDISspin}), an electronic mode can emerge from the top of the lower band ($\mu_-$), exhibiting the dispersion relation of 
\begin{equation}
\omega=e^{\cal T}_{{\bm k}-{\bm k}_{\rm F}^-}+\mu_-
\label{eq:ekhDIST}
\end{equation}
in the momentum regime outside the Fermi sea (${\bm k}>{\bm k}^-_{\rm F}=\frac{\bm \pi}{2}$). 
\par
\begin{figure} 
\includegraphics[width=\linewidth]{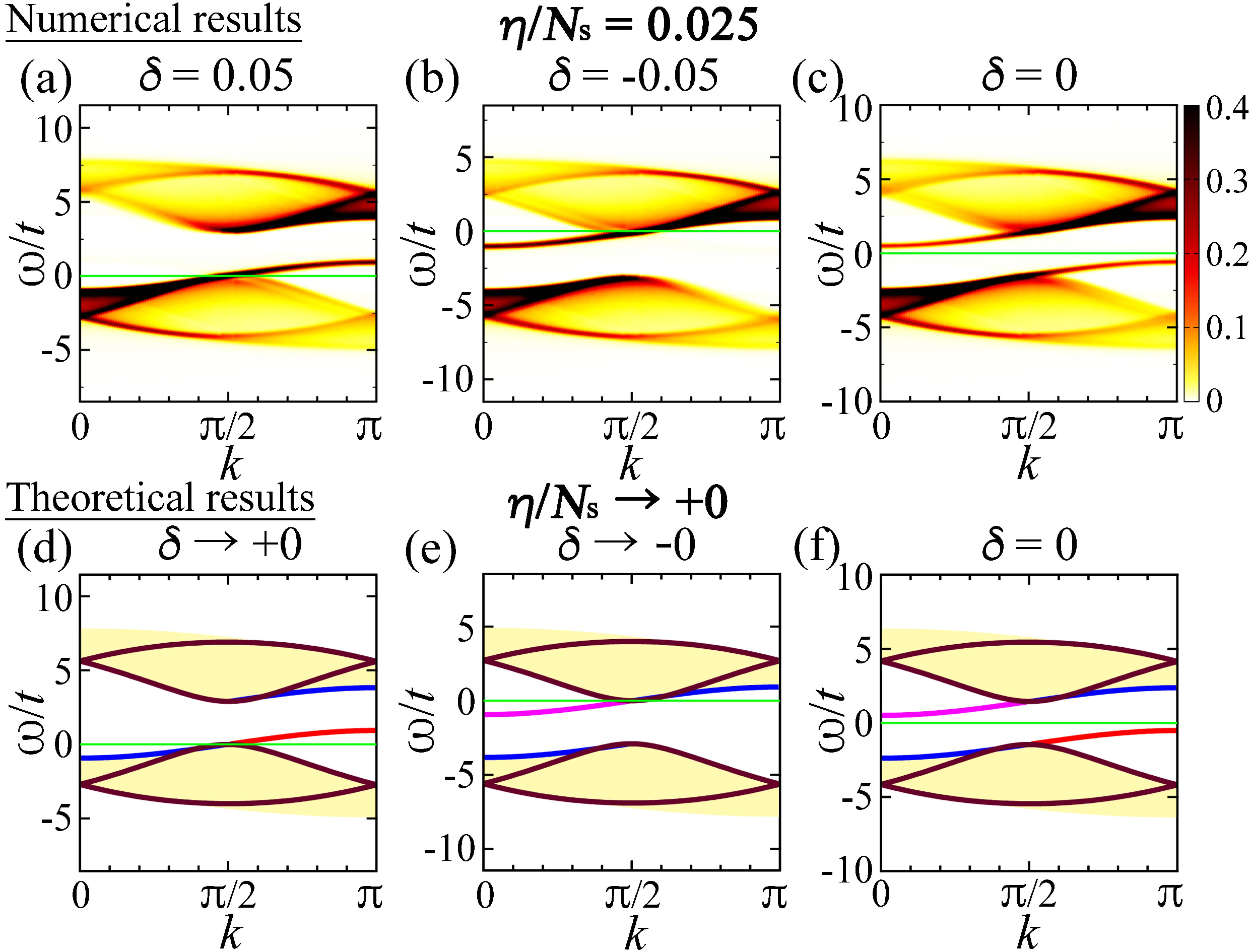}
\caption{Electronic excitation from the lowest-energy charge-perturbed states of small $\eta/N_{\rm s}$ in the 1D HM. 
(a)--(c) $A_X(k,\omega)t$ for $|X\rangle$ of the hole-doped ground state at $\delta=0.05$ ($\eta/N_{\rm s}=0.025$ and $\eta^z=-\eta$) [(a)], 
electron-doped ground state at $\delta=-0.05$ ($\eta/N_{\rm s}=0.025$ and $\eta^z=\eta$) [(b)], 
and lowest-energy state of $\eta/N_{\rm s}=0.025$ at $\delta=0$ ($\eta^z=0$) [(c)]. 
(d)--(f) Dispersion relations obtained using the Bethe ansatz for $\eta/N_{\rm s}\rightarrow+0$
in the small hole-doping limit [(d)], in the small electron-doping limit [(e)], and at half filling [(f)]. 
The dispersion relations of the emergent modes are $\omega=e^{\cal T}_{k-\frac{\pi}{2}}+\mu_-$ for $k>\frac{\pi}{2}$ [red curves in (d) and (f)] 
and $\omega=-e^{\cal T}_{-k+\frac{\pi}{2}}+\mu_+$ for $k<\frac{\pi}{2}$ [magenta curves in (e) and (f)], where $e^{\cal T}_k$ denotes the excitation energy of the spin mode. 
The dispersion relations of the other modes and continua are the same as those in Fig. \ref{fig:Hub1d}(d) 
with $\mu$ adjusted such that $\mu_-=0$ [(d)], $\mu_+=0$ [(e)], and $\mu=0$ [(f)].
The solid green lines indicate $\omega=0$.}
\label{fig:etaHub1d}
\end{figure}
\begin{figure} 
\includegraphics[width=\linewidth]{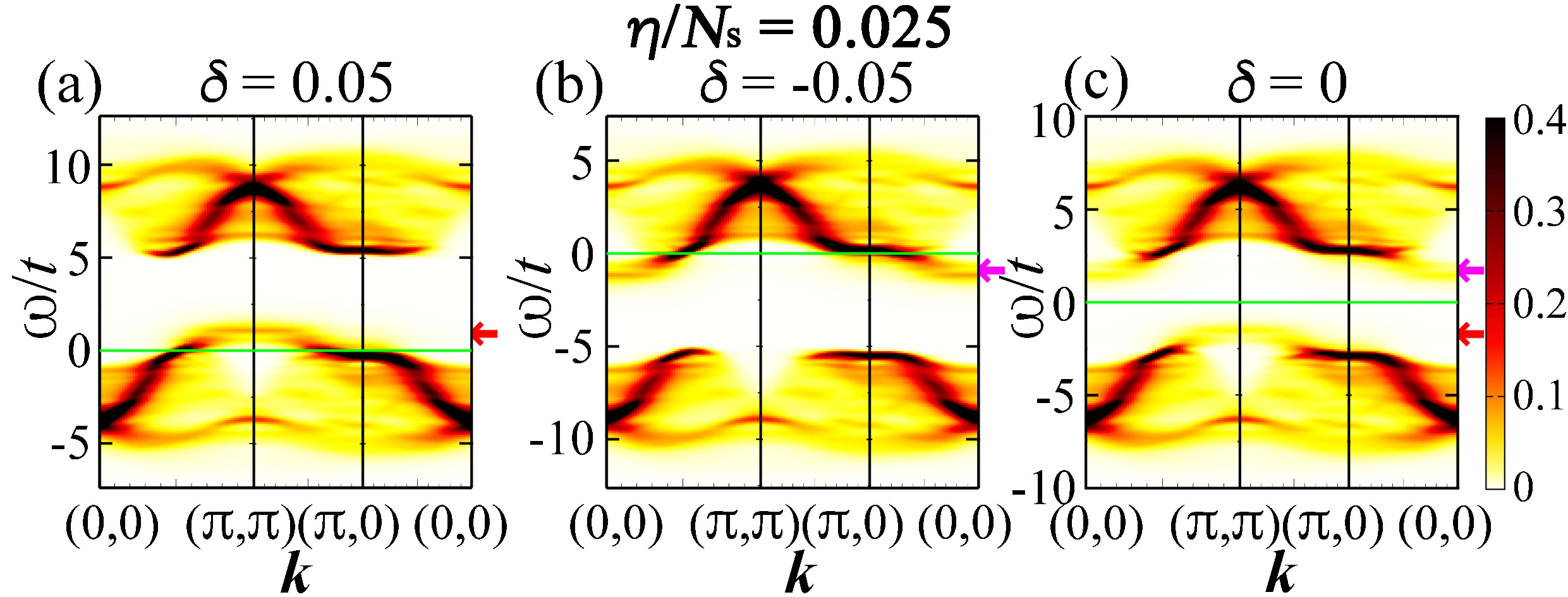}
\caption{Electronic excitation from the lowest-energy charge-perturbed states of $\eta/N_{\rm s}=0.025$ and $\eta^z/N_{\rm s}=-\delta/2$ in the 2D HM. 
$A_X({\bm k},\omega)t$ for $|X\rangle$ of the state corresponding to the hole-doped ground state at $\delta=0.05$ [(a)], electron-doped ground state at $\delta=-0.05$ [(b)], 
and lowest-energy state of $\eta/N_{\rm s}=0.025$ at $\delta=0$ [(c)]. 
The red arrows indicate $\omega=e^{\cal T}_{\frac{\bm \pi}{2}}$ [(a)] and $\omega=e^{\cal T}_{\frac{\bm \pi}{2}}+\mu_{\rm h}$ [(c)]; 
the magenta arrows indicate $\omega=-e^{\cal T}_{\frac{\bm \pi}{2}}$ [(b)] and $\omega=-e^{\cal T}_{\frac{\bm \pi}{2}}+\mu_{\rm e}$ [(c)], 
where $e^{\cal T}_{\frac{\bm \pi}{2}}$ denotes the excitation energy of the spin mode at ${\bm k}=\frac{\bm \pi}{2}$, 
obtained using exact diagonalization in a $4\times 4$-site cluster; 
$\mu_{\rm h}$ and $\mu_{\rm e}$ represent the chemical potential at $\delta=0.05$ and $-0.05$, respectively. 
The solid green lines indicate $\omega=0$.}
\label{fig:etaHub2d}
\end{figure}

In fact, as shown in Figs. \ref{fig:etaHub1d}(a) and \ref{fig:etaHub2d}(a), an electronic mode emerges from the top of the lower band 
outside the Fermi sea in a hole-doped system, 
exhibiting essentially the spin-mode dispersion relation shifted by the Fermi momentum. 
The dispersion relation obtained using the Bethe ansatz \cite{Kohno1DHub} [red curve in Fig. \ref{fig:etaHub1d}(d)] and 
$\omega$ value at the top of the emergent mode expected from Eq. (\ref{eq:ekhDIST}) [red arrow in Fig. \ref{fig:etaHub2d}(a)] are consistent with the numerical results. 
\subsubsection{Electron-doping-induced states} 
\label{sec:eleDIS}
Similarly, by removing an electron with a momentum ${\bm k}$ and $S^z=m^\prime$ from the one-electron-doped ($N_{\rm h}=-1$) ground state 
with the momentum of ${\bm k}_{\rm F}^+$ and $S^z=m$, 
a state with the momentum of $-{\bm k}+{\bm k}_{\rm F}^+$, $S^z=-m^\prime+m$, and $N_{\rm h}=0$ is obtained. 
This state can overlap with the ground state at half filling ($N_{\rm h}=0$) if ${\bm k}={\bm k}_{\rm F}^+$ and $m^\prime=m$, 
which corresponds to the bottom of the upper band, as in the case of a noninteracting band insulator. 
\par
In strongly correlated insulators, the electron-removed state can also overlap with the spin-excited state having the momentum of $-{\bm k}+{\bm k}_{\rm F}^+$ and $S^z=-m^\prime+m$ at half filling ($N_{\rm h}=0$) 
and can emerge in the electronic spectrum within the band gap along 
\begin{equation}
\omega=-e^{\rm spin}_{-{\bm k}+{\bm k}_{\rm F}^+}+\mu_+,
\label{eq:ekeDISspin}
\end{equation}
where $\mu_+$ denotes the $\omega$ value at the bottom of the upper band, 
\begin{equation}
\mu_+=E_{\rm GS}^{N_{\rm h}=-1}-E_{\rm GS}.
\label{eq:mu+}
\end{equation}
In the one-electron-doped system, $\mu_+=0$. 
\par
Equation (\ref{eq:ekeDISspin}) can be explained as follows: 
in the electron-removal excitation from $|X\rangle=|{\rm GS}\rangle^{N_{\rm h}=-1}_{{\bm k}_{\rm F}^+}$ (one-electron-doped ground state with the momentum of ${\bm k}_{\rm F}^+$) 
to $|n\rangle=|{\rm Spin}\rangle_{-{\bm k}+{\bm k}_{\rm F}^+}$ (spin-excited state with the momentum of $-{\bm k}+{\bm k}_{\rm F}^+$ at half filling) in Eq. (\ref{eq:Akwdef}), 
$|_{-{\bm k}+{\bm k}_{\rm F}^+}\langle{\rm Spin}|\lambda_{{\bm k},\sigma}|{\rm GS}\rangle^{N_{\rm h}=-1}_{{\bm k}_{\rm F}^+}|^2$ can be nonzero, 
for which $\omega=-E_n+E_X=-e^{\rm spin}_{-{\bm k}+{\bm k}_{\rm F}^+}+\mu_+$ because $E_n=e^{\rm spin}_{-{\bm k}+{\bm k}_{\rm F}^+}+E_{\rm GS}$ and $E_X=E_{\rm GS}^{N_{\rm h}=-1}$. 
\par
In addition, the electron-removed state can overlap with the charge-excited state having the momentum of $-{\bm k}+{\bm k}_{\rm F}^+$, $S^z=0$, and $N_{\rm h}=0$ 
if $m^\prime=m$ and can emerge in the electronic spectrum in the high-$|\omega|$ regime for $\omega<0$ along 
\begin{equation}
\omega=-e^{{\rm charge},N_{\rm h}=0}_{-{\bm k}+{\bm k}_{\rm F}^+}+\mu_+,
\label{eq:ekeDISchargeNh0}
\end{equation}
which can be explained by regarding $|n\rangle$ and $|X\rangle$ in the electron-removal excitation of Eq. (\ref{eq:Akwdef}) 
as the charge-excited state of $N_{\rm h}=0$ and $|{\rm GS}\rangle^{N_{\rm h}=-1}_{{\bm k}_{\rm F}^+}$, respectively. 
\par
Furthermore, by adding an electron with a momentum ${\bm k}$ and $S^z=-m$ to the one-electron-doped ground state 
with the momentum of ${\bm k}_{\rm F}^+$ and $S^z=m$, 
a state with the momentum of ${\bm k}+{\bm k}_{\rm F}^+$, $S^z=0$, and $N_{\rm h}=-2$ is obtained. 
This state can overlap with the charge-excited state having the momentum of ${\bm k}+{\bm k}_{\rm F}^+$, $S^z=0$, and $N_{\rm h}=-2$ 
and can emerge in the electronic spectrum in the high-$\omega$ regime for $\omega>0$ along 
\begin{equation}
\omega=e^{{\rm charge},N_{\rm h}=-2}_{{\bm k}+{\bm k}_{\rm F}^+}-\mu_+, 
\label{eq:ekeDISchargeNh-2}
\end{equation}
where $e^{{\rm charge},N_{\rm h}=-2}_{\bm p}$ denotes the excitation energy of the charge excitation with a momentum ${\bm p}$ and $N_{\rm h}=-2$ from the ground state at half filling. 
Equation (\ref{eq:ekeDISchargeNh-2}) can be explained by regarding $|n\rangle$ and $|X\rangle$ in the electron-addition excitation of Eq. (\ref{eq:Akwdef}) 
as the charge-excited state of $N_{\rm h}=-2$ and $|{\rm GS}\rangle^{N_{\rm h}=-1}_{{\bm k}_{\rm F}^+}$, respectively. 
\par
In electron-doped Mott insulators of the HM, because the upper band in the momentum regime outside the Fermi sea has essentially no electrons, 
the electronic states emerge in the momentum regime primarily inside the Fermi sea in the electron-removal spectrum. 
\par
In the 1D and 2D HMs, the emergence of electronic modes caused by electron doping 
in line with the above selection rules is confirmed as follows: 
by representing the excitation energy of the spin mode as $e^{\cal T}_{\bm k}$ [Figs. \ref{fig:Hub1d}(b), \ref{fig:Hub1d}(e), \ref{fig:Hub2d}(b), and \ref{fig:Hub2d}(c)] and 
noting that the Fermi momentum of the electron-doped system is ${\bm k}^+_{\rm F}=\frac{\bm \pi}{2}$ in the small-doping limit [Figs. \ref{fig:Hub1d}(a), \ref{fig:Hub1d}(d), and \ref{fig:Hub2d}(a)], 
according to Eq. (\ref{eq:ekeDISspin}), an electronic mode can emerge from the bottom of the upper band ($\mu_+$), exhibiting the dispersion relation of 
\begin{equation}
\omega=-e^{\cal T}_{-{\bm k}+{\bm k}_{\rm F}^+}+\mu_+
\label{eq:ekeDIST}
\end{equation}
in the momentum regime inside the Fermi sea (${\bm k}<{\bm k}^+_{\rm F}=\frac{\bm \pi}{2}$). 
\par
In fact, as shown in Figs. \ref{fig:etaHub1d}(b) and \ref{fig:etaHub2d}(b), an electronic mode emerges from the bottom of the upper band 
inside the Fermi sea in an electron-doped system, 
exhibiting essentially the (inverted) spin-mode dispersion relation shifted by the Fermi momentum. 
The dispersion relation obtained using the Bethe ansatz \cite{Kohno1DHub} [magenta curve in Fig. \ref{fig:etaHub1d}(e)] and 
$\omega$ value at the bottom of the emergent mode expected from Eq. (\ref{eq:ekeDIST}) [magenta arrow in Fig. \ref{fig:etaHub2d}(b)] 
are consistent with the numerical results. 
\subsubsection{Relation between hole- and electron-doping-induced states} 
\label{sec:relationDIS}
According to the above selection rules, by doping strongly correlated insulators with holes or electrons, electronic states can emerge within the band gap, 
which exhibit the spin-excitation dispersion relation ($\pm e^{\rm spin}_{\pm{\bm k}}$) shifted by the Fermi momentum ($\mp{\bm k}_{\rm F}^{\mp}$) 
from the band edges ($\mu_{\mp}$) [Eqs. (\ref{eq:ekhDISspin}) and (\ref{eq:ekeDISspin})], 
and in the high-$|\omega|$ regime, which exhibit the charge-excitation dispersion relation 
($\pm e^{{\rm charge},N_{\rm h}=0}_{\pm{\bm k}}$ and $\mp e^{{\rm charge},N_{\rm h}=\pm 2}_{\mp{\bm k}}$) shifted by the Fermi momentum ($\mp{\bm k}_{\rm F}^{\mp}$) 
from the band edges ($\mu_{\mp}$ and $-\mu_{\mp}=0$ in the doped systems) 
[Eqs. (\ref{eq:ekhDISchargeNh0}), (\ref{eq:ekhDISchargeNh2}), (\ref{eq:ekeDISchargeNh0}), and (\ref{eq:ekeDISchargeNh-2})]. 
\par
The dispersion relations of the electron-doping-induced states exhibit those of the hole-doping-induced states inverted with respect to $(\omega,{\bm k})=(0,\frac{\bm \pi}{2})$ 
as $\omega\rightarrow-\omega$ and ${\bm k}\rightarrow{\bm \pi-{\bm k}}$, reflecting particle--hole symmetry 
[Appendix \ref{sec:phSym}; Eq. (\ref{eq:AkwphX})] 
[Figs. \ref{fig:etaHub1d}(a), \ref{fig:etaHub1d}(b), \ref{fig:etaHub1d}(d), \ref{fig:etaHub1d}(e), \ref{fig:etaHub2d}(a), and \ref{fig:etaHub2d}(b)]. 
\par
The general characteristics of doping-induced states, which are derived based on the quantum-number analysis in Sec. \ref{sec:DIS}, 
can be explicitly demonstrated using the effective theory in spin-gapped Mott and Kondo insulators \cite{KohnoDIS,KohnoHubLadder,KohnoKLM,KohnoTinduced}, 
as shown in Secs. \ref{sec:eleEff}--\ref{sec:swDIS}. 
\subsection{Electronic excited states from a perturbed state in the effective theory} 
\label{sec:eleEff}
In the effective theory for weak inter-unit-cell hopping in the ladder and bilayer HMs and 1D and 2D PAMs and KLMs (Sec. \ref{sec:effTheoryUnperturbedSpinGap}; Appendix \ref{sec:effTheory}), 
by adding and removing an electron with a momentum ${\bm k}$ and spin $\sigma$ on a state $|X\rangle_{{\bm q}_{\parallel}}$ [Eq. (\ref{eq:Xk})], the following states are obtained \cite{KohnoTinduced,KohnoKLM}: 
\begin{align}
\label{eq:adaggerX}
a^{\dagger}_{{\bm k},\sigma}|X\rangle_{{\bm q}_{\parallel}}
&=\frac{1}{N_{\rm u}}\sum_{i\ne j}e^{i{\bm k}_{\parallel}\cdot{\bm r}_i}e^{i{{\bm q}_{\parallel}}\cdot{\bm r}_j}
a_{i,\sigma}^{\dagger}|{\cal G}\rangle_i|X\rangle_j\prod_{l(\ne i,j)}|{\cal G}\rangle_l\nonumber\\
&+\frac{1}{N_{\rm u}}\sum_{i}e^{i({\bm k}_{\parallel}+{{\bm q}_{\parallel}})\cdot{\bm r}_i}
a_{i,\sigma}^{\dagger}|X\rangle_i\prod_{l(\ne i)}|{\cal G}\rangle_l,\\
\label{eq:aX}
a_{{\bm k},\sigma}|X\rangle_{{\bm q}_{\parallel}}
&=\frac{1}{N_{\rm u}}\sum_{i\ne j}e^{-i{\bm k}_{\parallel}\cdot{\bm r}_i}e^{i{{\bm q}_{\parallel}}\cdot{\bm r}_j}
a_{i,\sigma}|{\cal G}\rangle_i|X\rangle_j\prod_{l(\ne i,j)}|{\cal G}\rangle_l\nonumber\\
&+\frac{1}{N_{\rm u}}\sum_{i}e^{i(-{\bm k}_{\parallel}+{{\bm q}_{\parallel}})\cdot{\bm r}_i}
a_{i,\sigma}|X\rangle_i\prod_{l(\ne i)}|{\cal G}\rangle_l,
\end{align}
where
\begin{equation}
a_{{\bm k},\sigma}^{\dagger}=\frac{1}{\sqrt{N_{\rm u}}}\sum_ie^{i{\bm k}_{\parallel}\cdot{\bm r}_i}a_{i,\sigma}^{\dagger}.
\label{eq:akdagger}
\end{equation}
Here, $a_{i,\sigma}^{\dagger}$ denotes the creation operator of an electron with spin ${\sigma}$ at a unit cell $i$: 
\begin{equation}
a_{i,\sigma}^{\dagger}=\left\{\begin{array}{ll}
c_{i,\sigma}^{\dagger}&\text{for the KLM},\\
c_{i,\sigma}^{\dagger}\text{ and }f_{i,\sigma}^{\dagger}&\text{for the PAM},\\
\frac{1}{\sqrt{2}}(c_{i_1,\sigma}^{\dagger}+e^{i k_{\perp}}c_{i_2,\sigma}^{\dagger})&\text{for the HM},
\end{array}\right.
\label{eq:aidagger}
\end{equation}
where $i_m$ represents the site within the unit cell $i$ on the $m$th leg in the ladder HM and on the $m$th plane in the bilayer HM. 
\par
The first terms on the right-hand sides of Eqs. (\ref{eq:adaggerX}) and (\ref{eq:aX}), 
which can be expressed using $|{\tilde {\cal A}}_{\sigma}X\rangle_{{\bm k}_{\parallel}+{\bm q}_{\parallel};{\bm q}_{\parallel}}$ and 
$|{\tilde {\cal R}}_{\bar \sigma}X\rangle_{-{\bm k}_{\parallel}+{\bm q}_{\parallel};{\bm q}_{\parallel}}$, respectively [Eq. (\ref{eq:XYk})], 
exhibit essentially the same band structure as that in the unperturbed case 
because the excitation energy from $|X\rangle$ can be approximately obtained as 
\begin{align}
\label{eq:AbX}
\varepsilon^{{\tilde {\cal A}}_{\sigma}X}_{{\bm k}_{\parallel}+{\bm q}_{\parallel};{\bm q}_{\parallel}}
-e^{X}_{{\bm q}_{\parallel}}&\overset{\text {Eq. (\ref{eq:eXYkf})}}{=}\varepsilon^{{\tilde {\cal A}}_{\sigma}}_{{\bm k}_{\parallel}},\\
\label{eq:RbX}
\varepsilon^{{\tilde {\cal R}}_{{\bar \sigma}}X}_{-{\bm k}_{\parallel}+{\bm q}_{\parallel};{\bm q}_{\parallel}}
-e^{X}_{{\bm q}_{\parallel}}&\overset{\text {Eq. (\ref{eq:eXYkf})}}{=}\varepsilon^{{\tilde {\cal R}}_{\bar \sigma}}_{-{\bm k}_{\parallel}}
\end{align}
for a boson $X$ and 
\begin{align}
\label{eq:AfX}
e^{{\tilde {\cal A}}_{\sigma}X_{\sigma^\prime}}_{{\bm k}_{\parallel}+{{\bm q}_{\parallel}};{{\bm q}_{\parallel}}}
-\varepsilon^{X_{\sigma^\prime}}_{{\bm q}_{\parallel}}&\overset{\text {Eq. (\ref{eq:eXYkb})}}{=}\varepsilon^{{\tilde {\cal A}}_{\sigma}}_{{\bm k}_{\parallel}},\\
\label{eq:RfX}
e^{{\tilde {\cal R}}_{{\bar \sigma}}X_{\sigma^\prime}}_{-{\bm k}_{\parallel}+{{\bm q}_{\parallel}};{{\bm q}_{\parallel}}}
-\varepsilon^{X_{\sigma^\prime}}_{{\bm q}_{\parallel}}&\overset{\text {Eq. (\ref{eq:eXYkb})}}{=}\varepsilon^{{\tilde {\cal R}}_{\bar \sigma}}_{-{\bm k}_{\parallel}}
\end{align}
for a fermion $X_{\sigma^\prime}$. 
Equations (\ref{eq:AbX}) and (\ref{eq:AfX}) correspond to Eq. (\ref{eq:ekAtilde}), and 
Eqs. (\ref{eq:RbX}) and (\ref{eq:RfX}) correspond to Eq. (\ref{eq:ekRtilde}) \cite{KohnoKLM,KohnoTinduced}. 
\par
The emergent modes caused by the perturbation of $|{\rm GS}\rangle\rightarrow|X\rangle_{{\bm q}_{\parallel}}$ 
are due to the second terms on the right-hand sides of Eqs. (\ref{eq:adaggerX}) and (\ref{eq:aX}), 
as shown below. 
\subsection{Doping-induced states in spin-gapped Mott and Kondo insulators} 
\label{sec:DISspingap}
\subsubsection{Conventional bands in doped systems} 
\label{sec:conventionaldoped}
In the effective theory, the one-hole-doped ground state ($N_{\rm h}=1$) and one-electron-doped ground state ($N_{\rm h}=-1$) with spin $\sigma^\prime$ are obtained [Eq. (\ref{eq:Xk})] as 
\begin{align}
\label{eq:GS1h}
&|{\rm GS}\rangle^{N_{\rm h}=1}=|{\cal R}_{\sigma^\prime}\rangle_{-{\bm k}_{\parallel{\rm F}}^-},\\
\label{eq:GS1e}
&|{\rm GS}\rangle^{N_{\rm h}=-1}=|{\cal A}_{\sigma^\prime}\rangle_{{\bm k}_{\parallel{\rm F}}^+},
\end{align}
whose energies are $E^{N_{\rm h}=1}_{\rm GS}=\varepsilon^{\cal R}_{-{\bm k}_{\parallel{\rm F}}^-}+E_{\rm GS}$ 
and $E^{N_{\rm h}=-1}_{\rm GS}=\varepsilon^{\cal A}_{{\bm k}_{\parallel{\rm F}}^+}+E_{\rm GS}$, respectively, at $H=0$ [Eq. (\ref{eq:eXkFermion})]. 
Here, ${\bm k}_{\parallel{\rm F}}^-={\bm \pi}$ at the top of the lower band and 
${\bm k}_{\parallel{\rm F}}^+={\bm 0}$ at the bottom of the upper band in the ladder and bilayer HMs and 1D and 2D PAMs and KLMs (Fig. \ref{fig:AkwLadBil}). 
\par
According to Eqs. (\ref{eq:mu-}) and (\ref{eq:mu+}), 
\begin{align}
\label{eq:effmu-}
\mu_-&=-\varepsilon^{\cal R}_{-{\bm k}_{\parallel{\rm F}}^-},\\
\label{eq:effmu+}
\mu_+&=\varepsilon^{\cal A}_{{\bm k}_{\parallel{\rm F}}^+}.
\end{align}
In the one-hole-doped and one-electron-doped systems, $\mu$ is adjusted such that 
$\mu_-=-\varepsilon^{\cal R}_{-{\bm k}_{\parallel{\rm F}}^-}=0$ and 
$\mu_+=\varepsilon^{\cal A}_{{\bm k}_{\parallel{\rm F}}^+}=0$, respectively. 
\par
The electronic excited states are obtained as Eqs. (\ref{eq:adaggerX}) and (\ref{eq:aX}) 
for $|X\rangle_{{\bm q}_{\parallel}}=|{\rm GS}\rangle^{N_{\rm h}=1}=|{\cal R}_{\sigma^\prime}\rangle_{-{\bm k}_{\parallel{\rm F}}^-}$ in the one-hole-doped system 
and $|X\rangle_{{\bm q}_{\parallel}}=|{\rm GS}\rangle^{N_{\rm h}=-1}=|{\cal A}_{\sigma^\prime}\rangle_{{\bm k}_{\parallel{\rm F}}^+}$ in the one-electron-doped system. 
The first terms on the right-hand sides of Eqs. (\ref{eq:adaggerX}) and (\ref{eq:aX}) exhibit essentially the same band structure as that in the undoped case 
[Eqs. (\ref{eq:ekAtilde}), (\ref{eq:ekRtilde}), (\ref{eq:AfX}), and (\ref{eq:RfX})] \cite{KohnoKLM,KohnoTinduced}. 
\par
By adding an electron with $\sigma={\bar \sigma^\prime}$ and ${\bm k}_{\parallel}={\bm k}_{\parallel{\rm F}}^-$, 
$a^{\dagger}_{{\bm k}_{\parallel{\rm F}}^-,{\bar \sigma^\prime}}$ (of $k_{\perp}=0$ in the HM), to $|{\rm GS}\rangle^{N_{\rm h}=1}$, 
the undoped ground state $|{\rm GS}\rangle$ can appear at the top of the lower band 
($\omega=\mu_-=E_{\rm GS}-E_{\rm GS}^{N_{\rm h}=1}$), 
as in the case of a noninteracting band insulator, 
because $\langle{\rm GS}|a^{\dagger}_{{\bm k}_{\parallel{\rm F}}^-,{\bar \sigma^\prime}}|{\cal R}_{\sigma^\prime}\rangle_{-{\bm k}_{\parallel{\rm F}}^-}\ne 0$ (for $k_{\perp}=0$ in the HM) 
between Eq. (\ref{eq:GS}) and the second term on the right-hand side of Eq. (\ref{eq:adaggerX}) 
for $|X\rangle_{{\bm q}_{\parallel}}=|{\cal R}_{\sigma^\prime}\rangle_{-{\bm k}_{\parallel{\rm F}}^-}$ 
(Tables \ref{tbl:symbols}--\ref{tbl:KLM}; Appendix \ref{sec:effTheory}). 
\par
Similarly, by removing an electron with $\sigma=\sigma^\prime$ and ${\bm k}_{\parallel}={\bm k}_{\parallel{\rm F}}^+$, 
$a_{{\bm k}_{\parallel{\rm F}}^+,\sigma^\prime}$ (of $k_{\perp}=\pi$ in the HM), from $|{\rm GS}\rangle^{N_{\rm h}=-1}$, 
the undoped ground state $|{\rm GS}\rangle$ can appear at the bottom of the upper band 
($\omega=\mu_+=E_{\rm GS}^{N_{\rm h}=-1}-E_{\rm GS}$), 
as in the case of a noninteracting band insulator, 
because $\langle{\rm GS}|a_{{\bm k}_{\parallel{\rm F}}^+,\sigma^\prime}|{\cal A}_{\sigma^\prime}\rangle_{{\bm k}_{\parallel{\rm F}}^+}\ne 0$ (for $k_{\perp}=\pi$ in the HM) 
between Eq. (\ref{eq:GS}) and the second term on the right-hand side of Eq. (\ref{eq:aX}) 
for $|X\rangle_{{\bm q}_{\parallel}}=|{\cal A}_{\sigma^\prime}\rangle_{{\bm k}_{\parallel{\rm F}}^+}$ 
(Tables \ref{tbl:symbols}--\ref{tbl:KLM}; Appendix \ref{sec:effTheory}). 
\subsubsection{In-gap doping-induced states originating from spin-excited states} 
\label{sec:ingapDoped}
In addition, spectral weights emerge in the electron-addition spectrum [Eqs. (\ref{eq:Akwcf})--(\ref{eq:Akwdef})] within the band gap along 
\begin{equation}
\omega=e^{\cal T}_{{\bm k}_{\parallel}-{\bm k}_{\parallel{\rm F}}^-}-\varepsilon^{\cal R}_{-{\bm k}_{\parallel{\rm F}}^-}
\overset{\text{ Eq. (\ref{eq:effmu-})}}{=}e^{\cal T}_{{\bm k}_{\parallel}-{\bm k}_{\parallel{\rm F}}^-}+\mu_-
\label{eq:ekhDISingap}
\end{equation}
(at $k_{\perp}=\pi$ in the HM) in the one-hole-doped system of $|{\rm GS}\rangle^{N_{\rm h}=1}=|{\cal R}_{\sigma^\prime}\rangle_{-{\bm k}_{\parallel{\rm F}}^-}$, because 
$_{{\bm k}_{\parallel}-{\bm k}_{\parallel{\rm F}}^-}\langle{\cal T}^{\gamma}|a^{\dagger}_{{\bm k}_{\parallel},\sigma}|{\cal R}_{\sigma^\prime}\rangle_{-{\bm k}_{\parallel{\rm F}}^-}\ne 0$ 
(for $k_{\perp}=\pi$ in the HM) between $|{\cal T}^{\gamma}\rangle_{{\bm k}_{\parallel}-{\bm k}_{\parallel{\rm F}}^-}$ [Eq. (\ref{eq:Xk})] and the second term on the right-hand side of Eq. (\ref{eq:adaggerX}) 
for $|X\rangle_{{\bm q}_{\parallel}}=|{\cal R}_{\sigma^\prime}\rangle_{-{\bm k}_{\parallel{\rm F}}^-}$ 
if $\gamma=s^z+s^{z\prime}$, where $s^{z\prime}=\frac{1}{2}$ and $-\frac{1}{2}$ for $\sigma^{\prime}=\uparrow$ and $\downarrow$, respectively 
(Tables \ref{tbl:symbols}--\ref{tbl:KLM}; Appendix \ref{sec:effTheory}) \cite{KohnoDIS,KohnoHubLadder,KohnoKLM,KohnoTinduced}. 
Equation (\ref{eq:ekhDISingap}) corresponds to Eq. (\ref{eq:ekhDISspin}). 
\par
Similarly, spectral weights emerge in the electron-removal spectrum [Eqs. (\ref{eq:Akwcf})--(\ref{eq:Akwdef})] within the band gap along 
\begin{equation}
\omega=-e^{\cal T}_{-{\bm k}_{\parallel}+{\bm k}_{\parallel{\rm F}}^+}+\varepsilon^{\cal A}_{{\bm k}_{\parallel{\rm F}}^+}
\overset{\text{Eq. (\ref{eq:effmu+})}}{=}-e^{\cal T}_{-{\bm k}_{\parallel}+{\bm k}_{\parallel{\rm F}}^+}+\mu_+
\label{eq:ekeDISingap}
\end{equation}
(at $k_{\perp}=0$ in the HM) in the one-electron-doped system of $|{\rm GS}\rangle^{N_{\rm h}=-1}=|{\cal A}_{\sigma^\prime}\rangle_{{\bm k}_{\parallel{\rm F}}^+}$, because 
$_{-{\bm k}_{\parallel}+{\bm k}_{\parallel{\rm F}}^+}\langle{\cal T}^{\gamma}|a_{{\bm k}_{\parallel},\sigma}|{\cal A}_{\sigma^\prime}\rangle_{{\bm k}_{\parallel{\rm F}}^+}\ne 0$ 
(for $k_{\perp}=0$ in the HM) between $|{\cal T}^{\gamma}\rangle_{-{\bm k}_{\parallel}+{\bm k}_{\parallel{\rm F}}^+}$ [Eq. (\ref{eq:Xk})] and the second term on the right-hand side of Eq. (\ref{eq:aX}) 
for $|X\rangle_{{\bm q}_{\parallel}}=|{\cal A}_{\sigma^\prime}\rangle_{{\bm k}_{\parallel{\rm F}}^+}$ if $\gamma=-s^z+s^{z\prime}$ 
(Tables \ref{tbl:symbols}--\ref{tbl:KLM}; Appendix \ref{sec:effTheory}) \cite{KohnoDIS,KohnoHubLadder,KohnoKLM,KohnoTinduced}. 
Equation (\ref{eq:ekeDISingap}) corresponds to Eq. (\ref{eq:ekeDISspin}). 
\par
\begin{figure*} 
\includegraphics[width=\linewidth]{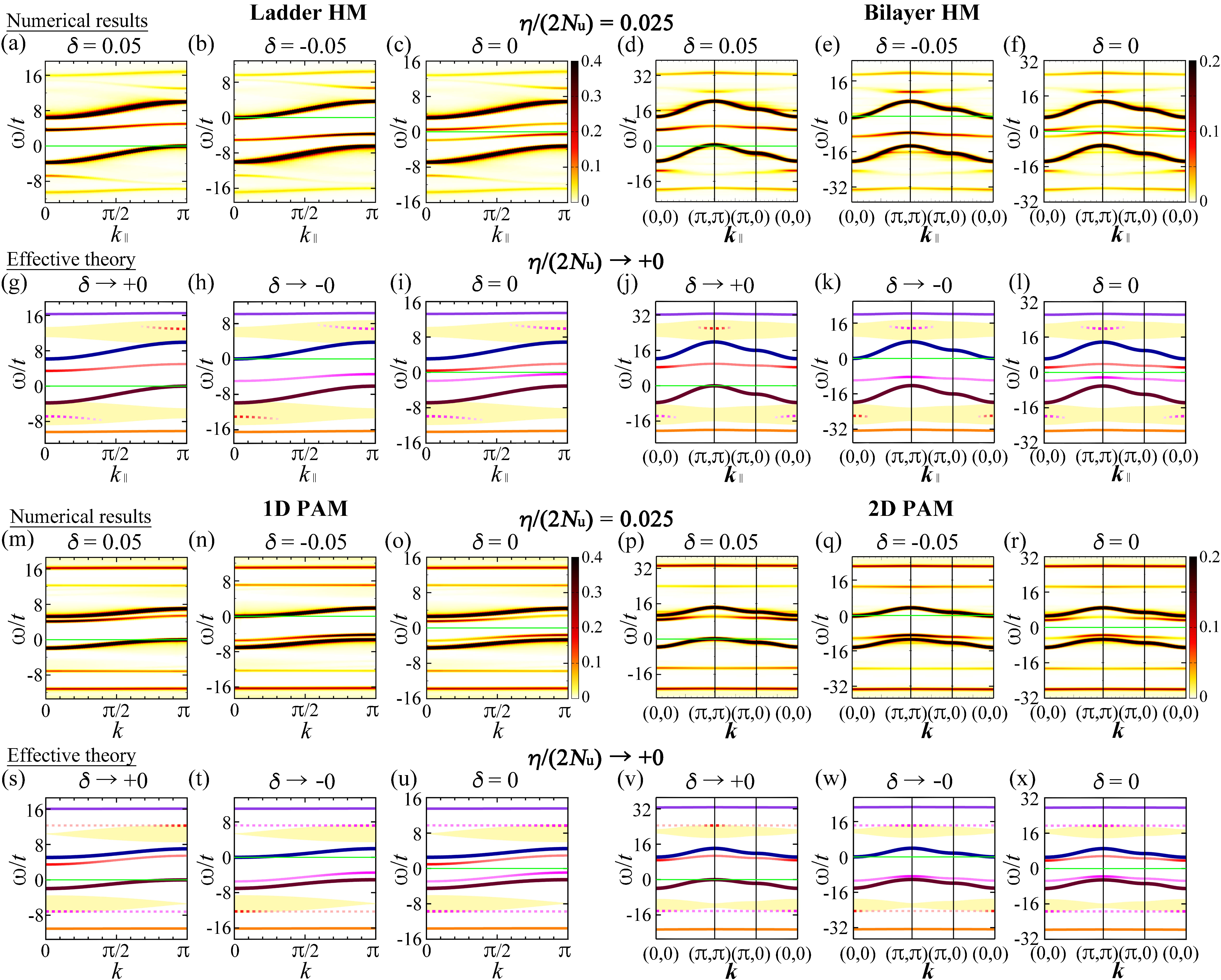}
\caption{Electronic excitation from the lowest-energy charge-perturbed states of small $\eta/(2N_{\rm u})$ in the ladder HM [(a)--(c), (g)--(i)], 
bilayer HM [(d)--(f), (j)--(l)], 1D PAM [(m)--(o), (s)--(u)], and 2D PAM [(p)--(r), (v)--(x)]. 
(a)--(f), (m)--(r) $A_X(k_\parallel,0,\omega)t+A_X(k_\parallel,\pi,\omega)t$ [(a)--(c)], ${\bar A}_X({\bm k}_\parallel,0,\omega)t+{\bar A}_X({\bm k}_\parallel,\pi,\omega)t$ [(d)--(f)], 
$A_X(k,\omega)t$ [(m)--(o)], and ${\bar A}_X({\bm k},\omega)t$ [(p)--(r)] for 
$|X\rangle$ of the state corresponding to the hole-doped ground state at $\delta=0.05$ [$\eta/(2N_{\rm u})=0.025$ and $\eta^z=-\eta$] [(a), (d), (m), (p)], 
electron-doped ground state at $\delta=-0.05$ [$\eta/(2N_{\rm u})=0.025$ and $\eta^z=\eta$] [(b), (e), (n), (q)], 
and lowest-energy state of $\eta/(2N_{\rm u})=0.025$ at $\delta=0$ ($\eta^z=0$) [(c), (f), (o), (r)]. 
(g)--(l), (s)--(x) Dispersion relations from the lowest-energy state of $\eta/(2N_{\rm u})\rightarrow+0$ 
for $\delta\rightarrow+0$ [(g), (j), (s), (v)], $\delta\rightarrow-0$ [(h), (k), (t), (w)], and $\delta=0$ [(i), (l), (u), (x)] in the effective theory. 
The dispersion relations of the emergent modes are 
$\omega=e^{{\cal T}}_{{\bm k}_\parallel-{\bm \pi}}+\mu_-$ (solid red curves), 
$\omega=e^{{\cal C}^0}_{{\bm k}_\parallel-{\bm \pi}}+\mu_-$ for $\omega>0$ (dotted red curves), and 
$\omega=-e^{{\cal C}^{-1}}_{-{\bm k}_\parallel-{\bm \pi}}-\mu_-$ for $\omega<0$ (dotted magenta curves) in (g), (i), (j), (l), (s), (u), (v), and (x); 
$\omega=-e^{{\cal T}}_{-{\bm k}_\parallel}+\mu_+$ (solid magenta curves), 
$\omega=-e^{{\cal C}^0}_{-{\bm k}_\parallel}+\mu_+$ for $\omega<0$ (dotted red curves), and 
$\omega=e^{{\cal C}^{1}}_{{\bm k}_\parallel}-\mu_+$ for $\omega>0$ (dotted magenta curves) in (h), (i), (k), (l), (t), (u), (w), and (x), 
which are shown in the relevant momentum regimes for comparison with the numerical results. 
The dispersion relations of the other modes and continua in (g)--(i), (j)--(l), (s)--(u), and (v)--(x) are the same as 
those in Figs. \ref{fig:AkwLadBil}(g), \ref{fig:AkwLadBil}(j), \ref{fig:AkwLadBil}(h), and \ref{fig:AkwLadBil}(k), respectively, 
with $\mu$ adjusted such that $\mu_-=0$ [(g), (j), (s), (v)], $\mu_+=0$ [(h), (k), (t), (w)], and $\mu=0$ [(i), (l), (u), (x)]. 
The solid green lines indicate $\omega=0$.}
\label{fig:etaLadBilPAM}
\end{figure*}
\begin{figure*} 
\includegraphics[width=\linewidth]{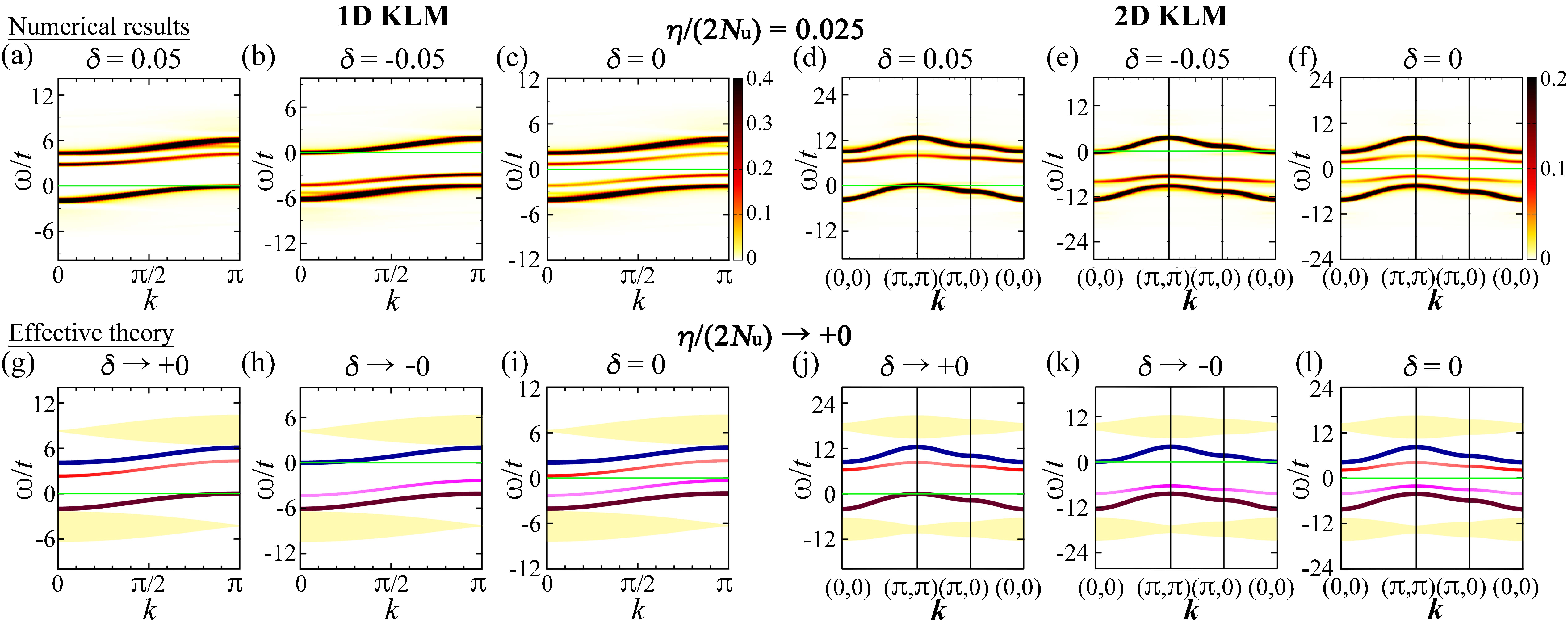}
\caption{Electronic excitation from the lowest-energy charge-perturbed states of small $\eta/(2N_{\rm u})$ in the 1D KLM [(a)--(c), (g)--(i)] and 2D KLM [(d)--(f), (j)--(l)]. 
(a)--(f) $A_X({\bm k},\omega)t$ for 
$|X\rangle$ of the state corresponding to the hole-doped ground state at $\delta=0.05$ [$\eta/(2N_{\rm u})=0.025$ and $\eta^z=-\eta$] [(a), (d)], 
electron-doped ground state at $\delta=-0.05$ [$\eta/(2N_{\rm u})=0.025$ and $\eta^z=\eta$] [(b), (e)], 
and lowest-energy state of $\eta/(2N_{\rm u})=0.025$ at $\delta=0$ ($\eta^z=0$) [(c), (f)]. 
(g)--(l) Dispersion relations from the lowest-energy state of $\eta/(2N_{\rm u})\rightarrow+0$ 
for $\delta\rightarrow+0$ [(g), (j)], $\delta\rightarrow-0$ [(h), (k)], and $\delta=0$ [(i), (l)] in the effective theory. 
The dispersion relations of the emergent modes are 
$\omega=e^{{\cal T}}_{{\bm k}_\parallel-{\bm \pi}}+\mu_-$ (solid red curves) [(g), (i), (j), (l)] and 
$\omega=-e^{{\cal T}}_{-{\bm k}_\parallel}+\mu_+$ (solid magenta curves) [(h), (i), (k), (l)], 
which are shown in the relevant momentum regimes for comparison with the numerical results. 
The dispersion relations of the other modes and continua in (g)--(i) and (j)--(l) are the same as those in Figs. \ref{fig:AkwLadBil}(i) and \ref{fig:AkwLadBil}(l), respectively, 
with $\mu$ adjusted such that $\mu_-=0$ [(g), (j)], $\mu_+=0$ [(h), (k)], and $\mu=0$ [(i), (l)]. 
The solid green lines indicate $\omega=0$.}
\label{fig:etaKLM}
\end{figure*}

The dispersion relations of the in-gap hole-doping-induced mode [Eq. (\ref{eq:ekhDISingap})] are shown 
by the solid red curves in Figs. \ref{fig:etaLadBilPAM}(g), \ref{fig:etaLadBilPAM}(j), \ref{fig:etaLadBilPAM}(s), \ref{fig:etaLadBilPAM}(v), \ref{fig:etaKLM}(g), and \ref{fig:etaKLM}(j), 
and those of the in-gap electron-doping-induced mode [Eq. (\ref{eq:ekeDISingap})] are shown 
by the solid magenta curves in Figs. \ref{fig:etaLadBilPAM}(h), \ref{fig:etaLadBilPAM}(k), \ref{fig:etaLadBilPAM}(t), \ref{fig:etaLadBilPAM}(w), \ref{fig:etaKLM}(h), and \ref{fig:etaKLM}(k), 
which are consistent with the numerical results [Figs. \ref{fig:etaLadBilPAM}(a), \ref{fig:etaLadBilPAM}(b), \ref{fig:etaLadBilPAM}(d), \ref{fig:etaLadBilPAM}(e), 
\ref{fig:etaLadBilPAM}(m), \ref{fig:etaLadBilPAM}(n), \ref{fig:etaLadBilPAM}(p), \ref{fig:etaLadBilPAM}(q), 
\ref{fig:etaKLM}(a), \ref{fig:etaKLM}(b), \ref{fig:etaKLM}(d), and \ref{fig:etaKLM}(e)]. 
\subsubsection{High-$|\omega|$ doping-induced states originating from charge-excited states} 
\label{sec:highwDoped}
In the HM and PAM, electronic modes are induced by doping even in the high-$|\omega|$ regime \cite{KohnoKLM,KohnoTinduced}. 
In the one-hole-doped system of $|{\rm GS}\rangle^{N_{\rm h}=1}=|{\cal R}_{\sigma^\prime}\rangle_{-{\bm k}_{\parallel{\rm F}}^-}$, spectral weights emerge along 
\begin{equation}
\omega=e^{{\cal C}^0}_{{\bm k}_{\parallel}-{\bm k}_{\parallel{\rm F}}^-}-\varepsilon^{\cal R}_{-{\bm k}_{\parallel{\rm F}}^-}
\overset{\text{Eq. (\ref{eq:effmu-})}}{=}e^{{\cal C}^0}_{{\bm k}_{\parallel}-{\bm k}_{\parallel{\rm F}}^-}+\mu_-
\label{eq:ekhDIShighwNh0}
\end{equation}
(at $k_{\perp}=\pi$ in the HM) in the high-$\omega$ regime for $\omega>0$, because 
$_{{\bm k}_{\parallel}-{\bm k}_{\parallel{\rm F}}^-}\langle{\cal C}^0|a^{\dagger}_{{\bm k}_{\parallel},{\bar \sigma^\prime}}|{\cal R}_{\sigma^\prime}\rangle_{-{\bm k}_{\parallel{\rm F}}^-}\ne 0$ 
(for $k_{\perp}=\pi$ in the HM) between $|{\cal C}^0\rangle_{{\bm k}_{\parallel}-{\bm k}_{\parallel{\rm F}}^-}$ [Eq. (\ref{eq:Xk})] and the second term on the right-hand side of Eq. (\ref{eq:adaggerX}) 
for $|X\rangle_{{\bm q}_{\parallel}}=|{\cal R}_{\sigma^\prime}\rangle_{-{\bm k}_{\parallel{\rm F}}^-}$ 
(Tables \ref{tbl:symbols}--\ref{tbl:PAM}; Appendix \ref{sec:effTheory}) \cite{KohnoKLM,KohnoTinduced}. 
Equation (\ref{eq:ekhDIShighwNh0}) corresponds to Eq. (\ref{eq:ekhDISchargeNh0}). 
\par
In addition, spectral weights emerge along 
\begin{equation}
\omega=-e^{{\cal C}^{-1}}_{-{\bm k}_{\parallel}-{\bm k}_{\parallel{\rm F}}^-}+\varepsilon^{\cal R}_{-{\bm k}_{\parallel{\rm F}}^-}
\overset{\text{Eq. (\ref{eq:effmu-})}}{=}-e^{{\cal C}^{-1}}_{-{\bm k}_{\parallel}-{\bm k}_{\parallel{\rm F}}^-}-\mu_-
\label{eq:ekhDIShighwNh2}
\end{equation}
(at $k_{\perp}=0$ in the HM) in the high-$|\omega|$ regime for $\omega<0$, because 
$_{-{\bm k}_{\parallel}-{\bm k}_{\parallel{\rm F}}^-}\langle{\cal C}^{-1}|a_{{\bm k}_{\parallel},\sigma^\prime}|{\cal R}_{\sigma^\prime}\rangle_{-{\bm k}_{\parallel{\rm F}}^-}\ne 0$ 
(for $k_{\perp}=0$ in the HM) between $|{\cal C}^{-1}\rangle_{-{\bm k}_{\parallel}-{\bm k}_{\parallel{\rm F}}^-}$ [Eq. (\ref{eq:Xk})] and the second term on the right-hand side of Eq. (\ref{eq:aX}) 
for $|X\rangle_{{\bm q}_{\parallel}}=|{\cal R}_{\sigma^\prime}\rangle_{-{\bm k}_{\parallel{\rm F}}^-}$ 
(Tables \ref{tbl:symbols}--\ref{tbl:PAM}; Appendix \ref{sec:effTheory}) \cite{KohnoKLM,KohnoTinduced}. 
Equation (\ref{eq:ekhDIShighwNh2}) corresponds to Eq. (\ref{eq:ekhDISchargeNh2}). 
\par
Similarly, in the one-electron-doped system of $|{\rm GS}\rangle^{N_{\rm h}=-1}=|{\cal A}_{\sigma^\prime}\rangle_{{\bm k}_{\parallel{\rm F}}^+}$, spectral weights emerge along 
\begin{equation}
\omega=-e^{{\cal C}^0}_{-{\bm k}_{\parallel}+{\bm k}_{\parallel{\rm F}}^+}+\varepsilon^{\cal A}_{{\bm k}_{\parallel{\rm F}}^+}
\overset{\text{Eq. (\ref{eq:effmu+})}}{=}-e^{{\cal C}^0}_{-{\bm k}_{\parallel}+{\bm k}_{\parallel{\rm F}}^+}+\mu_+
\label{eq:ekeDIShighwNh0}
\end{equation}
(at $k_{\perp}=0$ in the HM) in the high-$|\omega|$ regime for $\omega<0$, because 
$_{-{\bm k}_{\parallel}+{\bm k}_{\parallel{\rm F}}^+}\langle{\cal C}^0|a_{{\bm k}_{\parallel},\sigma^\prime}|{\cal A}_{\sigma^\prime}\rangle_{{\bm k}_{\parallel{\rm F}}^+}\ne 0$ 
(for $k_{\perp}=0$ in the HM) between $|{\cal C}^0\rangle_{-{\bm k}_{\parallel}+{\bm k}_{\parallel{\rm F}}^+}$ [Eq. (\ref{eq:Xk})] and the second term on the right-hand side of Eq. (\ref{eq:aX}) 
for $|X\rangle_{{\bm q}_{\parallel}}=|{\cal A}_{\sigma^\prime}\rangle_{{\bm k}_{\parallel{\rm F}}^+}$ 
(Tables \ref{tbl:symbols}--\ref{tbl:PAM}; Appendix \ref{sec:effTheory}) \cite{KohnoTinduced}. 
Equation (\ref{eq:ekeDIShighwNh0}) corresponds to Eq. (\ref{eq:ekeDISchargeNh0}). 
\par
In addition, spectral weights emerge along 
\begin{equation}
\omega=e^{{\cal C}^1}_{{\bm k}_{\parallel}+{\bm k}_{\parallel{\rm F}}^+}-\varepsilon^{\cal A}_{{\bm k}_{\parallel{\rm F}}^+}
\overset{\text{Eq. (\ref{eq:effmu+})}}{=}e^{{\cal C}^1}_{{\bm k}_{\parallel}+{\bm k}_{\parallel{\rm F}}^+}-\mu_+
\label{eq:ekeDIShighwNh-2}
\end{equation}
(at $k_{\perp}=\pi$ in the HM) in the high-$\omega$ regime for $\omega>0$, because 
$_{{\bm k}_{\parallel}+{\bm k}_{\parallel{\rm F}}^+}\langle{\cal C}^{1}|a^{\dagger}_{{\bm k}_{\parallel},{\bar \sigma^\prime}}|{\cal A}_{\sigma^\prime}\rangle_{{\bm k}_{\parallel{\rm F}}^+}\ne 0$ 
(for $k_{\perp}=\pi$ in the HM) between $|{\cal C}^{1}\rangle_{{\bm k}_{\parallel}+{\bm k}_{\parallel{\rm F}}^+}$ [Eq. (\ref{eq:Xk})] and the second term on the right-hand side of Eq. (\ref{eq:adaggerX}) 
for $|X\rangle_{{\bm q}_{\parallel}}=|{\cal A}_{\sigma^\prime}\rangle_{{\bm k}_{\parallel{\rm F}}^+}$ 
(Tables \ref{tbl:symbols}--\ref{tbl:PAM}; Appendix \ref{sec:effTheory}) \cite{KohnoTinduced}. 
Equation (\ref{eq:ekeDIShighwNh-2}) corresponds to Eq. (\ref{eq:ekeDISchargeNh-2}). 
\par
The dispersion relations of the high-$|\omega|$ hole-doping-induced modes 
for $\omega>0$ [Eq. (\ref{eq:ekhDIShighwNh0})] and $\omega<0$ [Eq. (\ref{eq:ekhDIShighwNh2})] 
are shown by the dotted red curves and dotted magenta curves, respectively, 
in Figs. \ref{fig:etaLadBilPAM}(g), \ref{fig:etaLadBilPAM}(j), \ref{fig:etaLadBilPAM}(s), and \ref{fig:etaLadBilPAM}(v), 
which are consistent with the numerical results 
[Figs. \ref{fig:etaLadBilPAM}(a), \ref{fig:etaLadBilPAM}(d), \ref{fig:etaLadBilPAM}(m), and \ref{fig:etaLadBilPAM}(p)]. 
The dispersion relations of the high-$|\omega|$ electron-doping-induced modes 
for $\omega<0$ [Eq. (\ref{eq:ekeDIShighwNh0})] and $\omega>0$ [Eq. (\ref{eq:ekeDIShighwNh-2})] 
are shown by the dotted red curves and dotted magenta curves, respectively, 
in Figs. \ref{fig:etaLadBilPAM}(h), \ref{fig:etaLadBilPAM}(k), \ref{fig:etaLadBilPAM}(t), and \ref{fig:etaLadBilPAM}(w), 
which are consistent with the numerical results 
[Figs. \ref{fig:etaLadBilPAM}(b), \ref{fig:etaLadBilPAM}(e), \ref{fig:etaLadBilPAM}(n), and \ref{fig:etaLadBilPAM}(q)]. 
\subsection{Spectral weights and dispersion relations of doping-induced states in the small-doping regime} 
\label{sec:swDIS}
The spectral weights of the doping-induced states at a momentum ${\bm k}_{\parallel}$ in the one-hole-doped system are $\mathcal{O}(\frac{1}{N_{\rm u}})$, 
because 
$|_{{\bm k}_{\parallel}-{\bm k}_{\parallel{\rm F}}^-}\langle{\cal T}^{\gamma}|a^{\dagger}_{{\bm k}_{\parallel},\sigma}|{\cal R}_{\sigma^\prime}\rangle_{-{\bm k}_{\parallel{\rm F}}^-}|^2$ for $\gamma=s^z+s^{z\prime}$, 
$|_{{\bm k}_{\parallel}-{\bm k}_{\parallel{\rm F}}^-}\langle{\cal C}^0|a^{\dagger}_{{\bm k}_{\parallel},{\bar \sigma^\prime}}|{\cal R}_{\sigma^\prime}\rangle_{-{\bm k}_{\parallel{\rm F}}^-}|^2$, and 
$|_{-{\bm k}_{\parallel}-{\bm k}_{\parallel{\rm F}}^-}\langle{\cal C}^{-1}|a_{{\bm k}_{\parallel},\sigma^\prime}|{\cal R}_{\sigma^\prime}\rangle_{-{\bm k}_{\parallel{\rm F}}^-}|^2$ 
are $\mathcal{O}(\frac{1}{N_{\rm u}})$ according to 
Eq. (\ref{eq:Xk}) for $|X\rangle_{{\bm k}_{\parallel}}=|{\cal T}^{\gamma}\rangle_{{\bm k}_{\parallel}-{\bm k}_{\parallel{\rm F}}^-}$, 
$|{\cal C}^0\rangle_{{\bm k}_{\parallel}-{\bm k}_{\parallel{\rm F}}^-}$, and $|{\cal C}^{-1}\rangle_{-{\bm k}_{\parallel}-{\bm k}_{\parallel{\rm F}}^-}$ 
and the second terms on the right-hand sides of Eqs. (\ref{eq:adaggerX}) and (\ref{eq:aX}) 
for $|X\rangle_{{\bm q}_{\parallel}}=|{\cal R}_{\sigma^\prime}\rangle_{-{\bm k}_{\parallel{\rm F}}^-}=|{\rm GS}\rangle^{N_{\rm h}=1}$. 
\par
Similarly, the spectral weights of the doping-induced states at a momentum ${\bm k}_{\parallel}$ in the one-electron-doped system are $\mathcal{O}(\frac{1}{N_{\rm u}})$, 
because 
$|_{-{\bm k}_{\parallel}+{\bm k}_{\parallel{\rm F}}^+}\langle{\cal T}^{\gamma}|a_{{\bm k}_{\parallel},\sigma}|{\cal A}_{\sigma^\prime}\rangle_{{\bm k}_{\parallel{\rm F}}^+}|^2$ for $\gamma=-s^z+s^{z\prime}$, 
$|_{-{\bm k}_{\parallel}+{\bm k}_{\parallel{\rm F}}^+}\langle{\cal C}^0|a_{{\bm k}_{\parallel},\sigma^\prime}|{\cal A}_{\sigma^\prime}\rangle_{{\bm k}_{\parallel{\rm F}}^+}|^2$, and 
$|_{{\bm k}_{\parallel}+{\bm k}_{\parallel{\rm F}}^+}\langle{\cal C}^{1}|a^{\dagger}_{{\bm k}_{\parallel},{\bar \sigma^\prime}}|{\cal A}_{\sigma^\prime}\rangle_{{\bm k}_{\parallel{\rm F}}^+}|^2$ 
are $\mathcal{O}(\frac{1}{N_{\rm u}})$ according to 
Eq. (\ref{eq:Xk}) for $|X\rangle_{{\bm k}_{\parallel}}=|{\cal T}^{\gamma}\rangle_{-{\bm k}_{\parallel}+{\bm k}_{\parallel{\rm F}}^+}$, 
$|{\cal C}^0\rangle_{-{\bm k}_{\parallel}+{\bm k}_{\parallel{\rm F}}^+}$, and $|{\cal C}^{1}\rangle_{{\bm k}_{\parallel}+{\bm k}_{\parallel{\rm F}}^+}$ 
and the second terms on the right-hand sides of Eqs. (\ref{eq:adaggerX}) and (\ref{eq:aX}) 
for $|X\rangle_{{\bm q}_{\parallel}}=|{\cal A}_{\sigma^\prime}\rangle_{{\bm k}_{\parallel{\rm F}}^+}=|{\rm GS}\rangle^{N_{\rm h}=-1}$. 
\par
In the $n$-hole-doped ($n$-electron-doped) system for $n\ll N_{\rm u}$, because there are essentially $n$ doped unit cells $|X\rangle_i=|{\cal R}_{\sigma^\prime}\rangle_i$ ($|{\cal A}_{\sigma^\prime}\rangle_i$) 
in the second terms on the right-hand sides of Eqs. (\ref{eq:adaggerX}) and (\ref{eq:aX}), 
the probability of adding (removing) an electron on the doped unit cells is essentially $n$ times as large as that for the one-hole-doped (one-electron-doped) system; 
the spectral weights of the doping-induced states essentially become $n$ times as large as those of the one-hole-doped (one-electron-doped) system. 
This means that the spectral weights of the doping-induced states increase proportionally to the doping concentration $|\delta|$ in the small-$|\delta|$ regime 
and that the doping-induced modes and changes in the band structure can be observed, provided that a macroscopic number $\mathcal{O}(N_{\rm u})$ of holes or electrons can be introduced via doping 
\cite{KohnoDIS,KohnoKLM,KohnoTinduced,KohnoRPP,Eskes,DagottoDOS}. 
\par
The dispersion relations of the doping-induced modes in the small-doping regime are almost the same as those of the small-doping limit 
because the momentum of the $n$-hole-doped ($n$-electron-doped) ground state generally differs 
from that of the $(n-1)$-hole-doped ($(n-1)$-electron-doped) ground state by the Fermi momentum, 
which is almost the same as that of the small-doping limit in the small-doping regime. 
The dispersion relations of the spin and charge excitations in the small-doping regime are also almost the same 
as those of the small-doping limit \cite{KohnoMottT,KohnoTinduced}. 
\subsection{Emergent states corresponding to doping-induced states in undoped charge-perturbed systems} 
\label{sec:etaz0DIS}
\subsubsection{Spectral function of the lowest-energy charge-perturbed state of a nonzero value of $\eta$ at half filling} 
\label{sec:Akwetaz0DIS}
The doping-induced states discussed in Secs. \ref{sec:DIS}, \ref{sec:DISspingap}, and \ref{sec:swDIS} are electronic excited states from the ground states of $N_{\rm h}\ne 0$. 
Reflecting particle--hole symmetry (Appendix \ref{sec:phSym}), 
the spectral function of the electron-doped ground state $|{\rm GS}\rangle^{N_{\rm h}=-n}$ 
is symmetric with that of the hole-doped ground state $|{\rm GS}\rangle^{N_{\rm h}=n}$ as 
\begin{equation}
A_{{\rm GS}^{N_{\rm h}=-n}}({\bm k},\omega)=A_{{\rm GS}^{N_{\rm h}=n}}({\bm \pi}-{\bm k},-\omega),
\label{eq:AkwphTransGS}
\end{equation}
regardless of whether $\mu$ is adjusted for the electron density, where $\mu_{N_{\rm h}=n}=-\mu_{N_{\rm h}=-n}$, or $\mu=0$ 
[Appendix \ref{sec:phSym}; Eq. (\ref{eq:AkwphX})]. Here, $\mu_{N_{\rm h}=n}$ denotes the value of $\mu$ for the $n$-hole-doped system. 
\par
Furthermore, owing to $\eta$-SU(2) symmetry, the spectral function of the undoped state ($\eta^z=0$) of the $\eta$-SU(2)-multiplet with the doped ground states ($|\eta^z|=\eta$) of $\eta=n$, 
$|\eta=n,\eta^z=0\rangle\equiv({\hat \eta}^+)^n|{\rm GS}\rangle^{N_{\rm h}=2n}=({\hat \eta}^-)^n|{\rm GS}\rangle^{N_{\rm h}=-2n}$, for $\mu=0$ can be obtained as 
\begin{align}
&A_{\eta=n,\eta^z=0}({\bm k},{\omega})\nonumber\\
&=\frac{1}{2}[A_{{\rm GS}^{N_{\rm h}=\pm 2n}}({\bm k},\omega)+A_{{\rm GS}^{N_{\rm h}=\pm 2n}}({\bm \pi}-{\bm k},-\omega)]\nonumber\\
&\overset{\text{Eq. (\ref{eq:AkwphTransGS})}}{=}\frac{1}{2}[A_{{\rm GS}^{N_{\rm h}=2n}}({\bm k},\omega)+A_{{\rm GS}^{N_{\rm h}=-2n}}({\bm k},\omega)]
\label{eq:AkwetaGS}
\end{align}
using the spectral functions of the $2n$-hole-doped and $2n$-electron-doped ground states for $\mu=0$ [Appendix \ref{sec:Akweta}; Eq. (\ref{eq:Akwetaz0})]. 
This state ($|\eta=n,\eta^z=0\rangle$) is the lowest-energy state of $\eta=n$ at half filling ($\eta^z=0$), because the doped ground states at $N_{\rm h}=\pm 2n$ are the lowest-energy states of $\eta=n$ and $\eta^z=\mp n$, 
provided that phase separation in the electron density does not occur at $N_{\rm h}=\pm 2n$ for $H=S^z=0$. 
\par
Thus, the charge perturbation that excites particle-hole pairs each having an excitation energy as large as the band gap $(\mu_+-\mu_-)$ 
induces electronic modes exhibiting the dispersion relations of Eqs. (\ref{eq:ekhDISspin}), (\ref{eq:ekhDISchargeNh0}), (\ref{eq:ekhDISchargeNh2}), (\ref{eq:ekeDISspin}), (\ref{eq:ekeDISchargeNh0}), and (\ref{eq:ekeDISchargeNh-2}) for $\mu=0$. 
\par
The dispersion relations of electronic excitations from the lowest-energy charge-perturbed state in the small $\eta$-density limit at half filling obtained via Eq. (\ref{eq:AkwetaGS}) using the Bethe ansatz and the effective theory are shown 
in Figs. \ref{fig:etaHub1d}(f), \ref{fig:etaLadBilPAM}(i), \ref{fig:etaLadBilPAM}(l), 
\ref{fig:etaLadBilPAM}(u), \ref{fig:etaLadBilPAM}(x), \ref{fig:etaKLM}(i), and \ref{fig:etaKLM}(l). 
The spectral functions of the state corresponding to the lowest-energy charge-perturbed state of small $\eta$-density at half filling obtained via Eq. (\ref{eq:AkwetaGS}) using numerical calculations are shown 
in Figs. \ref{fig:etaHub1d}(c), \ref{fig:etaHub2d}(c), \ref{fig:etaLadBilPAM}(c), \ref{fig:etaLadBilPAM}(f), 
\ref{fig:etaLadBilPAM}(o), \ref{fig:etaLadBilPAM}(r), \ref{fig:etaKLM}(c), and \ref{fig:etaKLM}(f). 
\par
If the emergent modes cross the Fermi level, the band structure can be regarded as metallic at half filling \cite{KohnoMottT}. 
\subsubsection{Remarks on nonzero-$\eta$ states at half filling} 
\label{sec:Remarksetaz0DIS}
By the definition of ${\hat \eta}^z[=\frac{1}{2}\sum_{\lambda,i}(n^\lambda_{i,\uparrow}+n^\lambda_{i,\downarrow}-1)]$ [Eqs. (\ref{eq:eta}) and (\ref{eq:etai})], 
the eigenvalue of ${\hat \eta}^z$ is $-\frac{N_{\rm h}}{2}$. 
Hence, $\eta^z$ is zero only at half filling, and any doped state has $\eta\ge|\eta^z|>0$. 
In addition, at half filling, all the excited states from the ground state of $\eta=0$ have $\eta>0$ 
except those caused by $\eta$-singlet operators (rank-0 spherical tensor operators for ${\hat {\bm \eta}}$ [Appendix \ref{sec:Akweta}; Eq. (\ref{eq:tensorOp})]); 
charge ($\eta>0$) excited states generally have $\eta>0$. 
States of $\eta>0$ at half filling ($\eta^z=0$) by themselves do not imply superconducting states \cite{entropyCoolingDMFT,etaPairingSuper,nonthermalSuper,KanekoEtaPairing,etaPairingHubLad}, 
as doped states ($\eta\ge|\eta^z|>0$) by themselves do not mean superconducting states. 
\par
The ground state in a doped system can be realized by a chemical-potential shift that moves the Fermi level into the lower or upper band. 
In this case, a macroscopic number $\mathcal{O}(N_{\rm u})$ of charges (holes or electrons) can be introduced. 
The change in the band structure caused by doping can be observed (Sec. \ref{sec:swDIS}) if the chemical potential is kept in a band, for example, by the application of a gate voltage or chemical doping. 
\par
On the other hand, to realize the lowest-energy charge-perturbed state of a nonzero value of $\eta$ at half filling, an equal number of particles and holes should be introduced 
around the bottom of the upper band and top of the lower band, respectively. 
To observe the change in the band structure, a macroscopic number $\mathcal{O}(N_{\rm u})$ of particle-hole pairs should be kept excited at the time of measurement. 
\par
In the case of photon (light) radiation, because a particle-hole pair can be excited by a photon, 
a macroscopic number $\mathcal{O}(N_{\rm u})$ of photons that excite the particle-hole pairs (excluding reflected and transmitted photons that are irrelevant to the excitations) 
should be injected every lifetime (relaxation time) of the particle-hole excited states 
to keep the number of excited particle-hole pairs macroscopic, as in a steady state, during the measurement. 
\subsection{Electronic modes induced by charge fluctuation} 
\label{sec:chargeFluc}
\subsubsection{Selection rules for electronic excitation from a state perturbed by charge fluctuation} 
\label{sec:selectionRulesCF}
Electronic modes can also be induced by charge fluctuation. 
As a charge-perturbed state, we consider ${\tilde n}^{\lambda}_{\bm q}|{\rm GS}\rangle$ at half filling for $\lambda=c$ and $f$ [Eq. (\ref{eq:nk})]. 
The charge-perturbed state ${\tilde n}^{\lambda}_{\bm q}|{\rm GS}\rangle$ has the momentum of ${\bm q}$, $N_{\rm h}=0$, and $S^z=0$ 
for the unperturbed ground state $|{\rm GS}\rangle$ with the momentum of ${\bm 0}$, $N_{\rm h}=0$, and $S^z=0$. 
The excitation energy of each eigenstate in ${\tilde n}^{\lambda}_{\bm q}|{\rm GS}\rangle$ from $|{\rm GS}\rangle$ is denoted by $e^{{\cal C}^0}_{\bm q}$. 
\par
The electron-addition ($N_{\rm h}=-1$) and electron-removal ($N_{\rm h}=1$) excited states with a momentum ${\bm k}$ in the conventional bands from $|{\rm GS}\rangle$ 
are represented by $|{\tilde {\cal A}}\rangle_{\bm k}$ and $|{\tilde {\cal R}}\rangle_{\bm k}$; their excitation energies from $|{\rm GS}\rangle$ are denoted by $\varepsilon^{\tilde {\cal A}}_{\bm k}$ and $\varepsilon^{\tilde {\cal R}}_{\bm k}$, respectively. 
The dispersion relations of the electronic modes from $|{\rm GS}\rangle$ are $\omega=\varepsilon^{\tilde {\cal A}}_{\bm k}$ for $\omega>0$ and 
$\omega=-\varepsilon^{\tilde {\cal R}}_{-{\bm k}}$ for $\omega<0$ [Figs. \ref{fig:Hub1d}(a), \ref{fig:Hub1d}(d), \ref{fig:Hub2d}(a), and \ref{fig:AkwLadBil}]. 
\par
The state obtained by adding an electron with a momentum ${\bm k}$ to ${\tilde n}^{\lambda}_{\bm q}|{\rm GS}\rangle$ has the momentum of ${\bm k}+{\bm q}$ and $N_{\rm h}=-1$. 
Since this state can overlap with $|{\tilde {\cal A}}\rangle_{{\bm k}+{\bm q}}$, electronic states can emerge in the electron-addition spectrum, exhibiting 
\begin{equation}
\omega=\varepsilon^{\tilde {\cal A}}_{{\bm k}+{\bm q}}-e^{{\cal C}^0}_{\bm q}.
\label{eq:ekAN}
\end{equation}
\par
Similarly, the state obtained by removing an electron with a momentum ${\bm k}$ from ${\tilde n}^{\lambda}_{\bm q}|{\rm GS}\rangle$ has the momentum of $-{\bm k}+{\bm q}$ and $N_{\rm h}=1$. 
Since this state can overlap with $|{\tilde {\cal R}}\rangle_{-{\bm k}+{\bm q}}$, electronic states can emerge in the electron-removal spectrum, exhibiting 
\begin{equation}
\omega=-\varepsilon^{\tilde {\cal R}}_{-{\bm k}+{\bm q}}+e^{{\cal C}^0}_{\bm q}.
\label{eq:ekRN}
\end{equation}
\par
Thus, electronic modes exhibiting the dispersion relations of the conventional bands ($\varepsilon^{\tilde {\cal A}}_{\bm k}$ and $-\varepsilon^{\tilde {\cal R}}_{-{\bm k}}$) 
shifted by the charge-excitation energy ($-e^{{\cal C}^0}_{\bm q}$ and $e^{{\cal C}^0}_{\bm q}$) and momentum (${\bm q}$) 
[Eqs. (\ref{eq:ekAN}) and (\ref{eq:ekRN})] can emerge in the electronic spectrum from the charge-perturbed state ${\tilde n}^{\lambda}_{\bm q}|{\rm GS}\rangle$. 
\par
The charge-perturbed state ${\tilde n}^{\lambda}_{\bm q}|{\rm GS}\rangle$ includes all the eigenstates at the momentum of ${\bm q}$ at half filling caused by ${\tilde n}^{\lambda}_{\bm q}$, 
i.e., $\forall|m\rangle$ which satisfies $\langle m|{\tilde n}^{\lambda}_{\bm q}|{\rm GS}\rangle\ne 0$ [Eqs. (\ref{eq:NkwHub}) and (\ref{eq:NkwPAMKLM})]. 
However, if a single state dominates at the momentum of ${\bm q}$ in the charge excitation, 
${\tilde n}^{\lambda}_{\bm q}|{\rm GS}\rangle$ can effectively be regarded as a single state, and the emergent states can be well identified as the emergent modes. 
\par
To observe the emergent modes, a macroscopic number $\mathcal{O}(N_{\rm u})$ of charges ($\prod_i {\tilde n}^{\lambda}_{{\bm q}_i}$) should be 
kept excited at the time of measurement, as mentioned in Sec. \ref{sec:Remarksetaz0DIS}. 
\par
If the emergent modes cross the Fermi level, the band structure can be regarded as metallic. 
\par
\subsubsection{Electronic modes induced by charge fluctuation in spin-gapped Mott and Kondo insulators} 
\label{sec:effectiveTheoryCF}
\begin{figure*} 
\includegraphics[width=\linewidth]{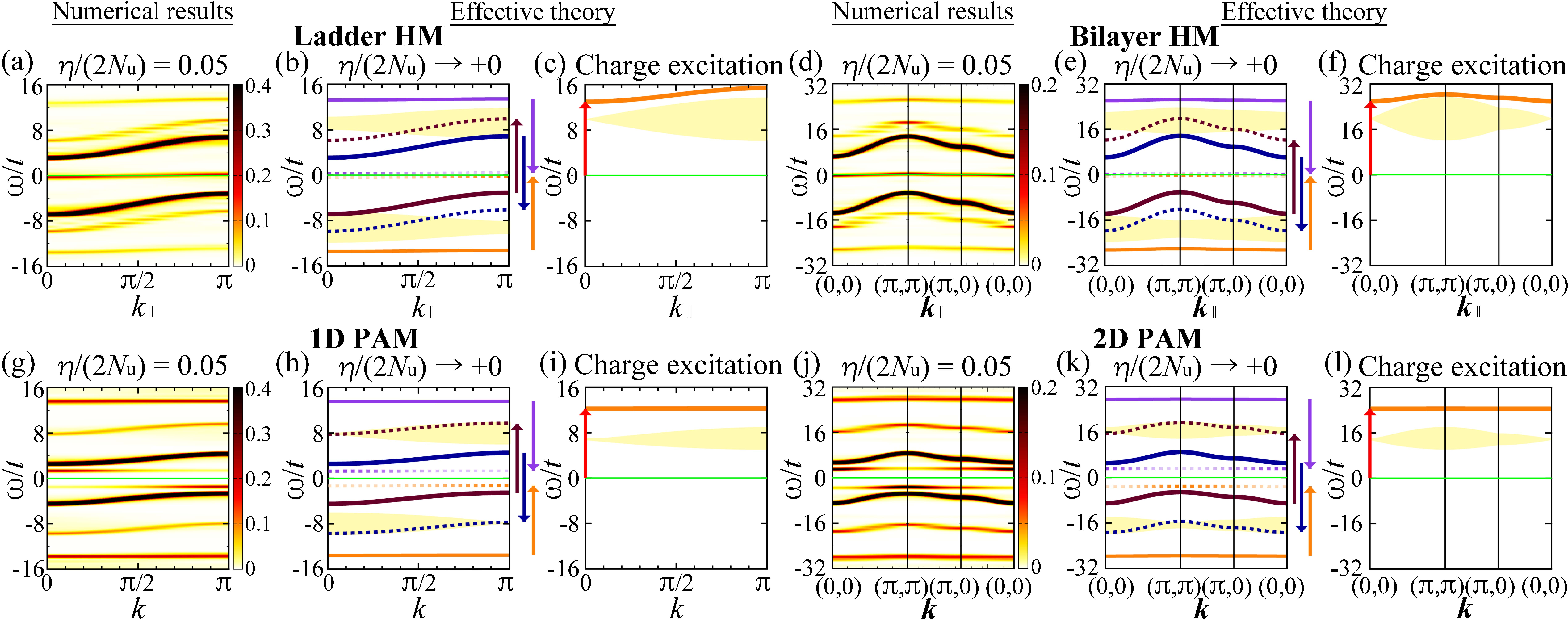}
\caption{Electronic excitation from the charge-perturbed states of small $\eta/(2N_{\rm u})$ caused by charge fluctuation 
in the ladder HM [(a), (b)], bilayer HM [(d), (e)], 1D PAM [(g), (h)], and 2D PAM [(j), (k)]. 
(a), (d), (g), (j) $A_X(k_\parallel,0,\omega)t+A_X(k_\parallel,\pi,\omega)t$ [(a)], ${\bar A}_X({\bm k}_\parallel,0,\omega)t+{\bar A}_X({\bm k}_\parallel,\pi,\omega)t$ [(d)], 
$A_X(k,\omega)t$ [(g)], and ${\bar A}_X({\bm k},\omega)t$ [(j)] for 
$|X\rangle$ of the state corresponding to $\prod_i{\tilde n}^c_{{\bm q}_i}|{\rm GS}\rangle$ at ${\bm q}_i=({\bm 0},\pi)$ in (a) and (d); 
$\prod_i({\tilde n}^c_{{\bm q}_i}-{\tilde n}^f_{{\bm q}_i})|{\rm GS}\rangle$ at ${\bm q}_i={\bm 0}$ in (g) and (j) for $\eta/(2N_{\rm u})=0.05$. 
(b), (e), (h), (k) Dispersion relations of electronic excitations from ${\tilde n}^c_{({\bm 0},\pi)}|{\rm GS}\rangle$ [(b), (e)] and 
$({\tilde n}^c_{\bm 0}-{\tilde n}^f_{\bm 0})|{\rm GS}\rangle$ [(h), (k)] in the effective theory. 
The dispersion relations of the emergent modes are 
$\omega=-\varepsilon^{\cal R}_{-{\bm k}_\parallel}+e^{{\cal C}^0}_{\bm 0}$ (dotted brown curves), 
$\omega=\varepsilon^{\cal A}_{{\bm k}_\parallel}-e^{{\cal C}^0}_{\bm 0}$ (dotted blue curves), 
$\omega=-\varepsilon^{\bar {\cal R}}_{-{\bm k}_\parallel}+e^{{\cal C}^0}_{\bm 0}$ (dotted orange curves), and 
$\omega=\varepsilon^{\bar {\cal A}}_{{\bm k}_\parallel}-e^{{\cal C}^0}_{\bm 0}$ (dotted purple curves), 
which are shown in the relevant momentum regimes for comparison with the numerical results. 
Here, $e^{{\cal C}^0}_{\bm 0}$ denotes the excitation energy of the charge mode at ${\bm k}_\parallel={\bm 0}$ [red arrows in (c), (f), (i), and (l)]. 
The dispersion relations of the other modes and continua in (b), (e), (h), and (k) are the same as 
those in Figs. \ref{fig:AkwLadBil}(g), \ref{fig:AkwLadBil}(j), \ref{fig:AkwLadBil}(h), and \ref{fig:AkwLadBil}(k), respectively. 
The up and down arrows on the right sides of (b), (e), (h), and (k) indicate $\omega$-shifts from the unperturbed modes 
by $e^{{\cal C}^0}_{\bm 0}$ and $-e^{{\cal C}^0}_{\bm 0}$, respectively. 
(c), (f), (i), (l) Dispersion relations of charge excitations in the effective theory in the ladder HM [(c)], bilayer HM [(f)], 1D PAM [(i)], and 2D PAM [(l)]. 
The red arrows indicate the excitation energy of the charge mode at ${\bm k}_\parallel={\bm 0}$, i.e., $e^{{\cal C}^0}_{\bm 0}$. 
The dispersion relations of the modes and continua in (c), (f), (i), and (l) are the same as 
those in Figs. \ref{fig:charge}(b), \ref{fig:charge}(c), \ref{fig:charge}(e), and \ref{fig:charge}(f), respectively. 
The solid green lines indicate $\omega=0$.}
\label{fig:Kn0PLadBil}
\end{figure*}
In the ladder and bilayer HMs and 1D and 2D PAMs, 
\begin{equation}
\label{eq:C0k}
\frac{{\hat {\tilde n}}_{{\bm q}_{\parallel}}|{\rm GS}\rangle}
{\sqrt{\langle{\rm GS}|{\hat {\tilde n}}_{-{\bm q}_{\parallel}}{\hat {\tilde n}}_{{\bm q}_{\parallel}}|{\rm GS}\rangle}}
=|{\cal C}^0\rangle_{{\bm q}_{\parallel}}
\end{equation}
in the effective theory, 
where ${\hat {\tilde n}}_{{\bm q}_{\parallel}}={\tilde n}^{c}_{({\bm q}_{\parallel},\pi)}$ for the ladder and bilayer HMs and 
${\tilde n}^c_{{\bm q}_{\parallel}}-{\tilde n}^f_{{\bm q}_{\parallel}}$ for the 1D and 2D PAMs 
[Eqs. (\ref{eq:NkwHub}), (\ref{eq:NkwPAMKLM}), (\ref{eq:nk}), (\ref{eq:GS}), and (\ref{eq:Xk}); Tables \ref{tbl:symbols}--\ref{tbl:PAM}]. 
The electron-addition and electron-removal excited states from this state are obtained as Eqs. (\ref{eq:adaggerX}) and (\ref{eq:aX}) for $|X\rangle_{{\bm q}_{\parallel}}={|\cal C}^0\rangle_{{\bm q}_{\parallel}}$. 
The first terms on the right-hand sides of Eqs. (\ref{eq:adaggerX}) and (\ref{eq:aX}) exhibit essentially the same band structure as that in the unperturbed case 
[Eqs. (\ref{eq:ekAtilde}), (\ref{eq:ekRtilde}), (\ref{eq:AbX}), and (\ref{eq:RbX})] \cite{KohnoKLM,KohnoTinduced}. 
\par
In addition, spectral weights emerge along 
\begin{align}
\label{eq:ekAC0}
\omega&=\varepsilon^{\tilde {\cal A}}_{{\bm k}_{\parallel}+{\bm q}_{\parallel}}-e^{{\cal C}^0}_{{\bm q}_{\parallel}},\\
\label{eq:ekRC0}
\omega&=-\varepsilon^{\tilde {\cal R}}_{-{\bm k}_{\parallel}+{\bm q}_{\parallel}}+e^{{\cal C}^0}_{{\bm q}_{\parallel}}
\end{align}
(for ${\tilde {\cal A}}_{\sigma}={\rm F}_{\sigma}$ and ${\tilde {\cal R}}_{\bar \sigma}={\rm A}_{\bar \sigma}$ at $k_{\perp}=0$; 
${\tilde {\cal A}}_{\sigma}={\rm G}_{\sigma}$ and ${\tilde {\cal R}}_{\bar \sigma}={\rm B}_{\bar \sigma}$ at $k_{\perp}=\pi$ in the HM), because 
$_{{\bm k}_{\parallel}+{\bm q}_{\parallel}}\langle{\tilde {\cal A}}_{\sigma}|a^{\dagger}_{{\bm k}_{\parallel},\sigma}|{{\cal C}^0}\rangle_{{\bm q}_{\parallel}}\ne 0$ and 
$_{-{\bm k}_{\parallel}+{\bm q}_{\parallel}}\langle{\tilde {\cal R}}_{\bar \sigma}|a_{{\bm k}_{\parallel},\sigma}|{\cal C}^0\rangle_{{\bm q}_{\parallel}}\ne 0$ 
(for ${\tilde {\cal A}}_{\sigma}={\rm F}_{\sigma}$ and ${\tilde {\cal R}}_{\bar \sigma}={\rm A}_{\bar \sigma}$ at $k_{\perp}=0$; 
${\tilde {\cal A}}_{\sigma}={\rm G}_{\sigma}$ and ${\tilde {\cal R}}_{\bar \sigma}={\rm B}_{\bar \sigma}$ at $k_{\perp}=\pi$ in the HM) 
between $|{\tilde {\cal A}}_{\sigma}\rangle_{{\bm k}_{\parallel}+{\bm q}_{\parallel}}$ and $|{\tilde {\cal R}}_{\bar \sigma}\rangle_{-{\bm k}_{\parallel}+{\bm q}_{\parallel}}$ [Eq. (\ref{eq:Xk})] and 
the second terms on the right-hand sides of Eqs. (\ref{eq:adaggerX}) and (\ref{eq:aX}) for $|X\rangle_{{\bm q}_{\parallel}}=|{\cal C}^0\rangle_{{\bm q}_{\parallel}}$ 
(Tables \ref{tbl:symbols}--\ref{tbl:PAM}; Appendix \ref{sec:effTheory}). 
Equations (\ref{eq:ekAC0}) and (\ref{eq:ekRC0}) correspond to Eqs. (\ref{eq:ekAN}) and (\ref{eq:ekRN}), respectively. 
\par
The dispersion relations of the emergent modes are those of the unperturbed modes ($\varepsilon^{\tilde {\cal A}}_{{\bm k}_{\parallel}}$ and $ -\varepsilon^{\tilde {\cal R}}_{-{\bm k}_{\parallel}}$) 
shifted by the charge-excitation energy ($-e^{{\cal C}^0}_{{\bm q}_{\parallel}}$ and $e^{{\cal C}^0}_{{\bm q}_{\parallel}}$) and momentum (${\bm q}_{\parallel}$)
[red arrows at ${\bm k}_{\parallel}={\bm 0}(={\bm q}_{\parallel})$ in Figs. \ref{fig:Kn0PLadBil}(c), \ref{fig:Kn0PLadBil}(f), \ref{fig:Kn0PLadBil}(i), and \ref{fig:Kn0PLadBil}(l)] 
as Eqs. (\ref{eq:ekAC0}) and (\ref{eq:ekRC0}) [dotted curves in Figs. \ref{fig:Kn0PLadBil}(b), \ref{fig:Kn0PLadBil}(e), \ref{fig:Kn0PLadBil}(h), and \ref{fig:Kn0PLadBil}(k)], 
which are consistent with the numerical results [Figs. \ref{fig:Kn0PLadBil}(a), \ref{fig:Kn0PLadBil}(d), \ref{fig:Kn0PLadBil}(g), and \ref{fig:Kn0PLadBil}(j)]. 
\section{Electronic modes induced by spin perturbations} 
\label{sec:spinPerturbed}
\subsection{Outline} 
\label{sec:outlineSpin}
Section \ref{sec:spinPerturbed} discusses electronic modes induced by spin perturbations, such as magnetization-induced states (Secs. \ref{sec:MIS}--\ref{sec:AkwM}), 
emergent states corresponding to the magnetization-induced states in unmagnetized spin-perturbed systems (Secs. \ref{sec:Sz0MIS} and \ref{sec:RemarksMIS}), and 
electronic modes induced by spin fluctuation (Sec. \ref{sec:spinFluc}). 
\par
In Secs. \ref{sec:MIS}--\ref{sec:AkwM}, the origins, underlying mechanisms, and properties of the magnetization-induced states are elucidated. 
Section \ref{sec:MIS} explains the magnetization-induced states from the viewpoint of quantum numbers. 
Their emergence is demonstrated in spin-gapped Mott and Kondo insulators using the effective theory for weak inter-unit-cell hopping and numerical calculations in Sec. \ref{sec:MISspingap}. 
Section \ref{sec:noMISgaplessSpin} shows that electronic modes are not induced inside the band gap by magnetizing insulators with gapless spin excitation. 
Section \ref{sec:AkwM} shows that the spectral functions of the magnetized states can also be obtained from those of the corresponding doped states in the models with $-U$ via the Shiba transformation. 
\par
In Sec. \ref{sec:Sz0MIS}, the spectral functions of unmagnetized nonzero-$S$ states are shown to be obtained from those of the magnetized ground states. 
In Sec. \ref{sec:RemarksMIS}, remarks on nonzero-$S$ states are made, particularly regarding experimental realizations. 
\par
In Sec. \ref{sec:spinFluc}, the dispersion relations of electronic modes induced by spin fluctuation are derived using quantum-number analysis (Sec. \ref{sec:selectionRulesSF}), 
and the emergence of these modes is demonstrated with the effective theory and numerical calculations for spin-gapped Mott and Kondo insulators (Sec. \ref{sec:effectiveTheorySF}), 
as well as in the 1D and 2D HMs (Sec. \ref{sec:Hub1d2dSF}). 
\subsection{Selection rules for magnetization-induced states} 
\label{sec:MIS}
Electronic modes are also induced by magnetizing spin-gapped Mott and Kondo insulators. 
By adding a ${\bar \sigma}$-spin electron with a momentum ${\bm k}$ to the ground state of $S^z=2s^z$, which has the momentum of ${\bm q}$, at half filling, 
a state with the momentum of ${\bm k}+{\bm q}$, $S^z=s^z$, and $N_{\rm h}=-1$ is obtained. 
\par
This state can overlap with the state $|{\tilde {\cal A}}_{\sigma}\rangle_{{\bm k}+{\bm q}}$ having the momentum of ${\bm k}+{\bm q}$, $S^z=s^z$, and $N_{\rm h}=-1$ in the conventional electron-addition bands 
and can emerge in the electron-addition spectrum for ${\bar \sigma}$-spin electrons along 
\begin{equation}
\omega=\varepsilon^{{\tilde {\cal A}}_{\sigma}}_{{\bm k}+{\bm q}}-e^{{\cal T}^{2s^z}}_{\bm q},
\label{eq:ekeMIS}
\end{equation}
where $\varepsilon^{{\tilde {\cal A}}_{\sigma}}_{{\bm k}+{\bm q}}$ denotes the electron-addition excitation energy of $|{\tilde {\cal A}}_{\sigma}\rangle_{{\bm k}+{\bm q}}$ under the magnetic field 
from the unperturbed ground state $|{\rm GS}\rangle$ of $S^z=0$ at half filling, and 
$e^{{\cal T}^{2s^z}}_{\bm q}$ denotes the lowest spin-excitation energy for $S^z=2s^z$ under the magnetic field from $|{\rm GS}\rangle$; the magnetic field $H$ is adjusted for the magnetized ground state of $S^z=2s^z$. 
\par
Similarly, by removing a $\sigma$-spin electron with a momentum ${\bm k}$ from the ground state of $S^z=2s^z$, which has the momentum of ${\bm q}$, at half filling, 
a state with the momentum of $-{\bm k}+{\bm q}$, $S^z=s^z$, and $N_{\rm h}=1$ is obtained. 
\par
This state can overlap with the state $|{\tilde {\cal R}}_{\sigma}\rangle_{-{\bm k}+{\bm q}}$ having the momentum of $-{\bm k}+{\bm q}$, $S^z=s^z$, and $N_{\rm h}=1$ in the conventional electron-removal bands 
and can emerge in the electron-removal spectrum for $\sigma$-spin electrons along 
\begin{equation}
\omega=-\varepsilon^{{\tilde {\cal R}}_{\sigma}}_{-{\bm k}+{\bm q}}+e^{{\cal T}^{2s^z}}_{\bm q},
\label{eq:ekhMIS}
\end{equation}
where $\varepsilon^{{\tilde {\cal R}}_{\sigma}}_{-{\bm k}+{\bm q}}$ denotes the electron-removal excitation energy of $|{\tilde {\cal R}}_{\sigma}\rangle_{-{\bm k}+{\bm q}}$ under the magnetic field from $|{\rm GS}\rangle$. 
\par
In the case of gapless spin excitation ($e^{{\cal T}^{2s^z}}_{\bm q}\rightarrow0$ at $H=0$), because an infinitesimal magnetic field ($H\rightarrow0$) can cause magnetization, 
the emergent states are essentially degenerate in $\omega$ with the conventional bands (Sec. \ref{sec:noMISgaplessSpin}). 
However, in spin-gapped Mott and Kondo insulators, because a magnetic field as large as the spin gap $\Delta_{\rm spin}$ is required for magnetization, 
electronic states different from those of the conventional bands can generally emerge, exhibiting the dispersion relations of Eqs. (\ref{eq:ekeMIS}) and (\ref{eq:ekhMIS}) (Sec. \ref{sec:MISspingap}). 
\subsection{Magnetization-induced electronic modes in spin-gapped Mott and Kondo insulators} 
\label{sec:MISspingap}
\subsubsection{Dispersion relations of magnetization-induced modes in spin-gapped Mott and Kondo insulators} 
\label{sec:dispMISspingap}
\begin{figure*} 
\includegraphics[width=\linewidth]{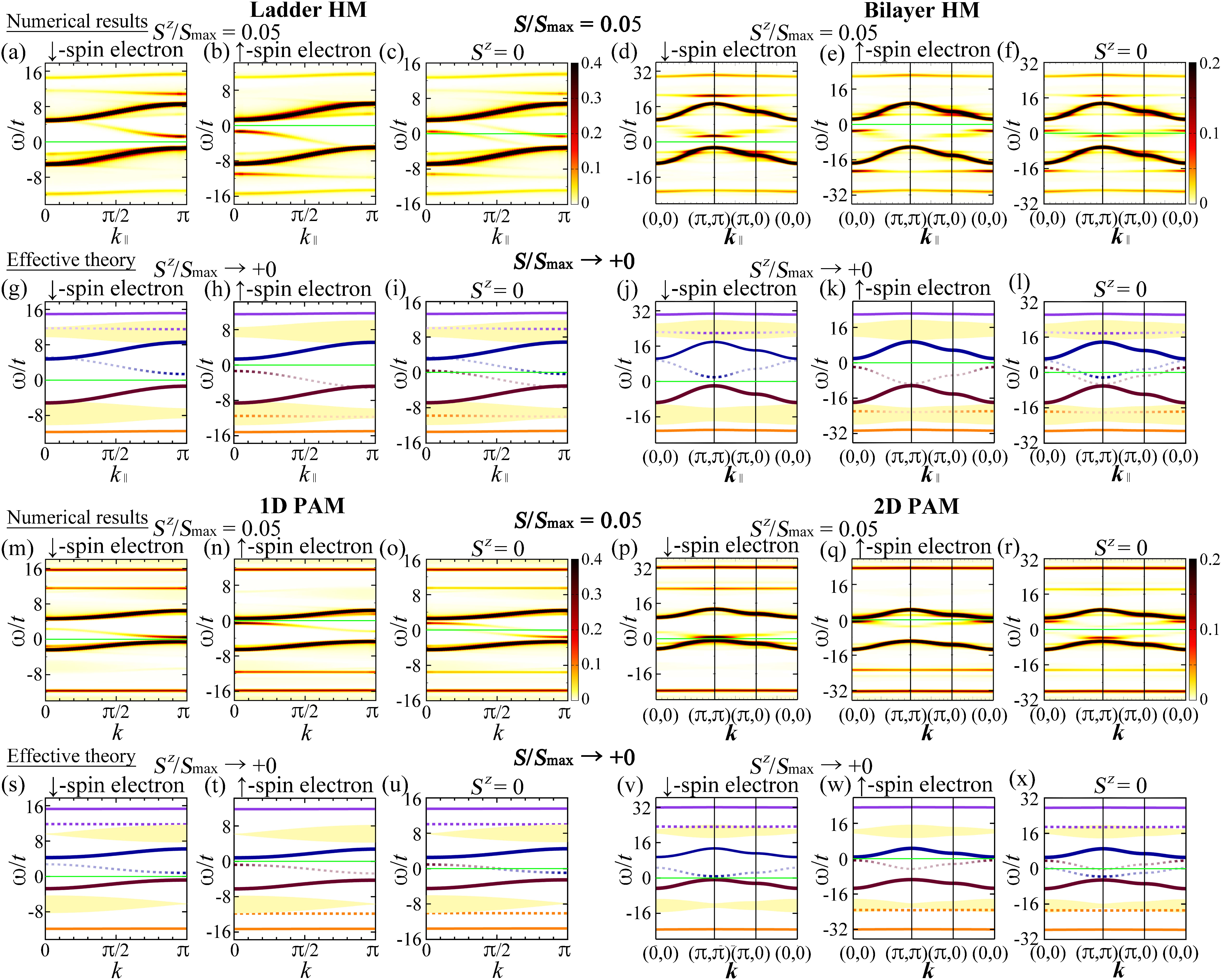}
\caption{Electronic excitation from the lowest-energy spin-perturbed states of small $S/S_{\rm max}$ 
in the ladder HM [(a)--(c), (g)--(i)], bilayer HM [(d)--(f), (j)--(l)], 1D PAM [(m)--(o), (s)--(u)], and 2D PAM [(p)--(r), (v)--(x)]. 
(a), (b), (d), (e), (m), (n), (p), (q) $A^\sigma_X(k_\parallel,0,\omega)t+A^\sigma_X(k_\parallel,\pi,\omega)t$ [(a), (b)], 
${\bar A}^\sigma_X({\bm k}_\parallel,0,\omega)t+{\bar A}^\sigma_X({\bm k}_\parallel,\pi,\omega)t$ [(d), (e)], 
$A^\sigma_X(k,\omega)t$ [(m), (n)], and ${\bar A}^\sigma_X({\bm k},\omega)t$ [(p), (q)] for 
$|X\rangle$ of the state corresponding to the magnetized ground state of $S/S_{\rm max}=0.05$ and $S^z=S$ for $\sigma=\downarrow$ [(a), (d), (m), (p)] and $\uparrow$ [(b), (e), (n), (q)]. 
(c), (f), (o), (r) $A_X(k_\parallel,0,\omega)t+A_X(k_\parallel,\pi,\omega)t$ [(c)], 
${\bar A}_X({\bm k}_\parallel,0,\omega)t+{\bar A}_X({\bm k}_\parallel,\pi,\omega)t$ [(f)], 
$A_X(k,\omega)t$ [(o)], and ${\bar A}_X({\bm k},\omega)t$ [(r)] for 
$|X\rangle$ of the state corresponding to the lowest-energy state of $S/S_{\rm max}=0.05$ and $S^z=0$. 
(g)--(l), (s)--(x) Dispersion relations from the ground state of $S/S_{\rm max}\rightarrow +0$ 
for $\downarrow$-spin electrons [(g), (j), (s), (v)] and $\uparrow$-spin electrons [(h), (k), (t), (w)] 
under the magnetic field of $H=\Delta_{\rm spin}$ ($S^z/S_{\rm max}\rightarrow+0$) [(g), (h), (j), (k), (s), (t), (v), (w)] and without a magnetic field ($S^z=0$) [(i), (l), (u), (x)] 
in the effective theory. 
The dispersion relations of the emergent modes are 
$\omega=\varepsilon^{{\cal A}_{\uparrow}}_{{\bm k}_\parallel+{\bm \pi}}-e^{{\cal T}^1}_{\bm \pi}$ (dotted blue curves) and 
$\omega=\varepsilon^{\bar {\cal A}_{\uparrow}}_{{\bm k}_\parallel+{\bm \pi}}-e^{{\cal T}^1}_{\bm \pi}$ (dotted purple curves) in (g), (i), (j), (l), (s), (u), (v), and (x); 
$\omega=-\varepsilon^{{\cal R}_{\uparrow}}_{-{\bm k}_\parallel+{\bm \pi}}+e^{{\cal T}^1}_{\bm \pi}$ (dotted brown curves) and 
$\omega=-\varepsilon^{\bar {\cal R}_{\uparrow}}_{-{\bm k}_\parallel+{\bm \pi}}+e^{{\cal T}^1}_{\bm \pi}$ (dotted orange curves) in (h), (i), (k), (l), (t), (u), (w), and (x), 
which are shown in the relevant momentum regimes for comparison with the numerical results. 
The dispersion relations of the other modes and continua in (g)--(i), (j)--(l), (s)--(u), and (v)--(x) are the same as 
those in Figs. \ref{fig:AkwLadBil}(g), \ref{fig:AkwLadBil}(j), \ref{fig:AkwLadBil}(h), and \ref{fig:AkwLadBil}(k), respectively, 
with the effective chemical potential of $\mu_{\rm eff}=-\frac{\Delta_{\rm spin}}{2}$ for $\downarrow$-spin electrons [(g), (j), (s), (v)] and 
$\frac{\Delta_{\rm spin}}{2}$ for $\uparrow$-spin electrons [(h), (k), (t), (w)] under the magnetic field of $H=\Delta_{\rm spin}$ and $\mu_{\rm eff}=0$ [(i), (l), (u), (x)] without a magnetic field. 
The solid green lines indicate $\omega=0$.}
\label{fig:SzLadBilPAM}
\end{figure*}
\begin{figure*} 
\includegraphics[width=\linewidth]{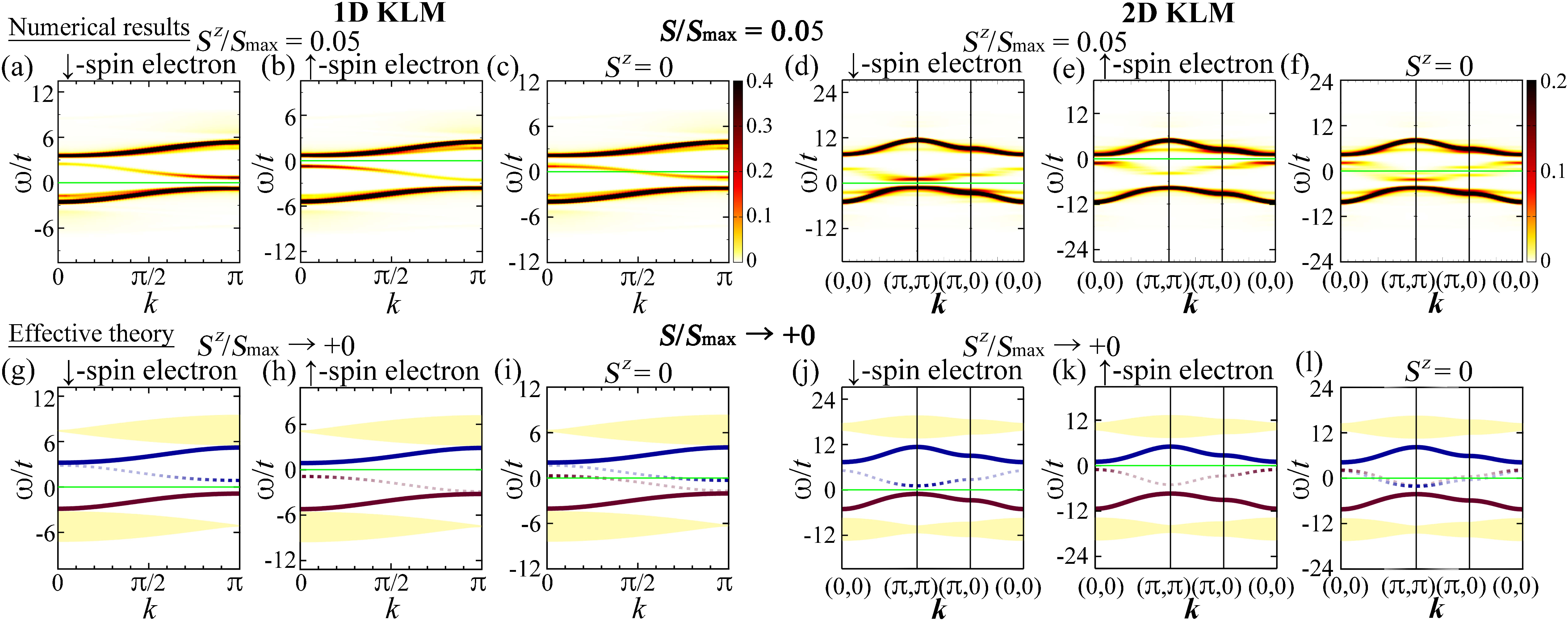}
\caption{Electronic excitation from the lowest-energy spin-perturbed states of small $S/S_{\rm max}$ in the 1D KLM [(a)--(c), (g)--(i)] and 2D KLM [(d)--(f), (j)--(l)]. 
(a), (b), (d), (e) $A^\sigma_X({\bm k},\omega)t$ for 
$|X\rangle$ of the state corresponding to the magnetized ground state of $S/S_{\rm max}=0.05$ and $S^z=S$ for $\sigma=\downarrow$ [(a), (d)] and $\uparrow$ [(b), (e)]. 
(c), (f) $A_X({\bm k},\omega)t$ for 
$|X\rangle$ of the state corresponding to the lowest-energy state of $S/S_{\rm max}=0.05$ and $S^z=0$. 
(g)--(l) Dispersion relations from the ground state of $S/S_{\rm max}\rightarrow +0$ for $\downarrow$-spin electrons [(g), (j)] and $\uparrow$-spin electrons [(h), (k)] 
under the magnetic field of $H=\Delta_{\rm spin}$ ($S^z/S_{\rm max}\rightarrow+0$) [(g), (h), (j), (k)] and without a magnetic field ($S^z=0$) [(i), (l)] in the effective theory. 
The dispersion relations of the emergent modes are 
$\omega=\varepsilon^{{\cal A}_{\uparrow}}_{{\bm k}_\parallel+{\bm \pi}}-e^{{\cal T}^1}_{\bm \pi}$ (dotted blue curves) [(g), (i), (j), (l)] and 
$\omega=-\varepsilon^{{\cal R}_{\uparrow}}_{-{\bm k}_\parallel+{\bm \pi}}+e^{{\cal T}^1}_{\bm \pi}$ (dotted brown curves) [(h), (i), (k), (l)], 
which are shown in the relevant momentum regimes for comparison with the numerical results. 
The dispersion relations of the other modes and continua in (g)--(i) and (j)--(l) are the same as those in Figs. \ref{fig:AkwLadBil}(i) and \ref{fig:AkwLadBil}(l), respectively, 
with the effective chemical potential of $\mu_{\rm eff}=-\frac{\Delta_{\rm spin}}{2}$ for $\downarrow$-spin electrons [(g), (j)] and 
$\frac{\Delta_{\rm spin}}{2}$ for $\uparrow$-spin electrons [(h), (k)] under the magnetic field of $H=\Delta_{\rm spin}$ and $\mu_{\rm eff}=0$ [(i), (l)] without a magnetic field. 
The solid green lines indicate $\omega=0$.}
\label{fig:SzKLM}
\end{figure*}
To demonstrate the emergence of electronic modes caused by magnetizing spin-gapped Mott and Kondo insulators, 
we consider the lowest-energy states of $S^z=\pm1$ at half filling in the ladder and bilayer HMs and 1D and 2D PAMs and KLMs as perturbed states, 
which are obtained as $|{\cal T}^{\pm 1}\rangle_{{\bm q}_{\parallel}}$ [Eq. (\ref{eq:Xk})] in the effective theory. 
In these models, ${\bm q}_{\parallel}={\bm \pi}$, and the spin gap is obtained as $\Delta_{\rm spin}=e^{\cal T}_{\bm \pi}$ at $H=0$ (Fig. \ref{fig:SkwLadBil}). 
\par
The electronic excited states from $|{\cal T}^{\pm 1}\rangle_{{\bm q}_{\parallel}}$ are obtained as Eqs. (\ref{eq:adaggerX}) and (\ref{eq:aX}) for $|X\rangle_{{\bm q}_{\parallel}}=|{\cal T}^{\pm 1}\rangle_{{\bm q}_{\parallel}}$. 
The first terms on the right-hand sides of Eqs. (\ref{eq:adaggerX}) and (\ref{eq:aX}) exhibit essentially the same band structure as that in the unperturbed case 
[Eqs. (\ref{eq:ekAtilde}), (\ref{eq:ekRtilde}), (\ref{eq:AbX}), and (\ref{eq:RbX})]. 
\par
In addition, spectral weights emerge along 
\begin{align}
\label{eq:ekAdTu}
&\omega=\varepsilon^{{\tilde {\cal A}}_{\sigma}}_{{\bm k}_{\parallel}+{\bm q}_{\parallel}}-e^{{\cal T}^{2s^z}}_{{\bm q}_{\parallel}}
\quad\text{in}\quad A^{\bar \sigma}_{{\cal T}^{2s^z}_{{\bm q}_{\parallel}}}({\bm k},\omega),\\
\label{eq:ekRuTu}
&\omega=-\varepsilon^{{\tilde {\cal R}}_{\sigma}}_{-{\bm k}_{\parallel}+{\bm q}_{\parallel}}+e^{{\cal T}^{2s^z}}_{{\bm q}_{\parallel}}
\quad\text{in}\quad A^{\sigma}_{{\cal T}^{2s^z}_{{\bm q}_{\parallel}}}({\bm k},\omega)
\end{align}
(for ${\tilde {\cal A}}_{\sigma}={\rm F}_{\sigma}$ and ${\tilde {\cal R}}_{\sigma}={\rm A}_{\sigma}$ at $k_{\perp}=0$; 
${\tilde {\cal A}}_{\sigma}={\rm G}_{\sigma}$ and ${\tilde {\cal R}}_{\sigma}={\rm B}_{\sigma}$ at $k_{\perp}=\pi$ in the HM), because 
$_{{\bm k}_{\parallel}+{\bm q}_{\parallel}}\langle{\tilde {\cal A}}_{\sigma}|a^{\dagger}_{{\bm k}_{\parallel},{\bar \sigma}}|{{\cal T}^{2s^z}}\rangle_{{\bm q}_{\parallel}}\ne 0$ and
$_{-{\bm k}_{\parallel}+{\bm q}_{\parallel}}\langle{\tilde {\cal R}}_{\sigma}|a_{{\bm k}_{\parallel},\sigma}|{{\cal T}^{2s^z}}\rangle_{{\bm q}_{\parallel}}\ne 0$
(for ${\tilde {\cal A}}_{\sigma}={\rm F}_{\sigma}$ and ${\tilde {\cal R}}_{\sigma}={\rm A}_{\sigma}$ at $k_{\perp}=0$; 
${\tilde {\cal A}}_{\sigma}={\rm G}_{\sigma}$ and ${\tilde {\cal R}}_{\sigma}={\rm B}_{\sigma}$ at $k_{\perp}=\pi$ in the HM) 
between $|{\tilde {\cal A}}_{\sigma}\rangle_{{\bm k}_{\parallel}+{\bm q}_{\parallel}}$ and $|{\tilde {\cal R}}_{\sigma}\rangle_{-{\bm k}_{\parallel}+{\bm q}_{\parallel}}$ [Eq. (\ref{eq:Xk})] and 
the second terms on the right-hand sides of Eqs. (\ref{eq:adaggerX}) and (\ref{eq:aX}) for $|X\rangle_{{\bm q}_{\parallel}}=|{\cal T}^{2s^z}\rangle_{{\bm q}_{\parallel}}$ 
(Tables \ref{tbl:symbols}--\ref{tbl:KLM}; Appendix \ref{sec:effTheory}). 
Equations (\ref{eq:ekAdTu}) and (\ref{eq:ekRuTu}) correspond to Eqs. (\ref{eq:ekeMIS}) and (\ref{eq:ekhMIS}), respectively. 
\par
In the ladder and bilayer HMs and 1D and 2D PAMs and KLMs, 
electronic modes originating from the ${\cal A}_{\uparrow}$ mode emerge within the band gap in the electron-addition spectrum for $\downarrow$-spin electrons, 
exhibiting the dispersion relation of Eq. (\ref{eq:ekAdTu}) for ${\tilde {\cal A}}_{\sigma}={\cal A}_{\uparrow}$ (Tables \ref{tbl:symbols}--\ref{tbl:KLM}; Appendix \ref{sec:effTheory}) 
[dotted blue curves in Figs. \ref{fig:SzLadBilPAM}(g), \ref{fig:SzLadBilPAM}(j), \ref{fig:SzLadBilPAM}(s), \ref{fig:SzLadBilPAM}(v), \ref{fig:SzKLM}(g), and \ref{fig:SzKLM}(j)] 
from the magnetized ground state of $S^z=1$ ($S^z/S_{\rm max}\rightarrow +0$), 
which are consistent with the numerical results 
[Figs. \ref{fig:SzLadBilPAM}(a), \ref{fig:SzLadBilPAM}(d), \ref{fig:SzLadBilPAM}(m), \ref{fig:SzLadBilPAM}(p), \ref{fig:SzKLM}(a), and \ref{fig:SzKLM}(d)]. 
\par
In the electron-removal spectrum for $\uparrow$-spin electrons, electronic modes originating from the ${\cal R}_{\uparrow}$ mode emerge within the band gap, 
exhibiting the dispersion relation of Eq. (\ref{eq:ekRuTu}) for ${\tilde {\cal R}}_{\sigma}={\cal R}_{\uparrow}$ (Tables \ref{tbl:symbols}--\ref{tbl:KLM}; Appendix \ref{sec:effTheory}) 
[dotted brown curves in Figs. \ref{fig:SzLadBilPAM}(h), \ref{fig:SzLadBilPAM}(k), \ref{fig:SzLadBilPAM}(t), \ref{fig:SzLadBilPAM}(w), \ref{fig:SzKLM}(h), and \ref{fig:SzKLM}(k)] 
from the magnetized ground state of $S^z=1$ ($S^z/S_{\rm max}\rightarrow +0$), 
which are consistent with the numerical results 
[Figs. \ref{fig:SzLadBilPAM}(b), \ref{fig:SzLadBilPAM}(e), \ref{fig:SzLadBilPAM}(n), \ref{fig:SzLadBilPAM}(q), \ref{fig:SzKLM}(b), and \ref{fig:SzKLM}(e)]. 
\par
In addition, in the ladder and bilayer HMs and 1D and 2D PAMs, 
electronic modes originating from the ${\bar {\cal A}}_{\uparrow}$ mode emerge in the high-$\omega$ regime for $\omega>0$ in the electron-addition spectrum for $\downarrow$-spin electrons, 
exhibiting the dispersion relation of Eq. (\ref{eq:ekAdTu}) for ${\tilde {\cal A}}_{\sigma}={\bar {\cal A}}_{\uparrow}$ (Tables \ref{tbl:symbols}--\ref{tbl:KLM}; Appendix \ref{sec:effTheory}) 
[dotted purple curves in Figs. \ref{fig:SzLadBilPAM}(g), \ref{fig:SzLadBilPAM}(j), \ref{fig:SzLadBilPAM}(s), and \ref{fig:SzLadBilPAM}(v)] 
from the magnetized ground state of $S^z=1$ ($S^z/S_{\rm max}\rightarrow +0$), 
which are consistent with the numerical results 
[Figs. \ref{fig:SzLadBilPAM}(a), \ref{fig:SzLadBilPAM}(d), \ref{fig:SzLadBilPAM}(m), and \ref{fig:SzLadBilPAM}(p)]. 
\par
In the electron-removal spectrum for $\uparrow$-spin electrons, electronic modes originating from the ${\bar {\cal R}}_{\uparrow}$ mode emerge in the high-$|\omega|$ regime for $\omega<0$, 
exhibiting the dispersion relation of Eq. (\ref{eq:ekRuTu}) for ${\tilde {\cal R}}_{\sigma}={\bar {\cal R}}_{\uparrow}$ (Tables \ref{tbl:symbols}--\ref{tbl:KLM}; Appendix \ref{sec:effTheory}) 
[dotted orange curves in Figs. \ref{fig:SzLadBilPAM}(h), \ref{fig:SzLadBilPAM}(k), \ref{fig:SzLadBilPAM}(t), and \ref{fig:SzLadBilPAM}(w)] 
from the magnetized ground state of $S^z=1$ ($S^z/S_{\rm max}\rightarrow +0$), 
which are consistent with the numerical results 
[Figs. \ref{fig:SzLadBilPAM}(b), \ref{fig:SzLadBilPAM}(e), \ref{fig:SzLadBilPAM}(n), and \ref{fig:SzLadBilPAM}(q)]. 
\subsubsection{Absence of insulator--metal transition caused by magnetization-induced modes in spin-gapped Mott and Kondo insulators} 
\label{sec:insulatingMISspingap}
Even though the magnetization-induced modes emerge under the magnetic field, the system remains insulating in spin-gapped Mott and Kondo insulators at $\mu=0$. 
The chemical potential required for insulator--metal transition under the magnetic field does not change between the cases with and without the magnetization-induced modes. 
These characteristics are explained as follows: 
\par
By rewriting Eqs. (\ref{eq:ekAdTu}) and (\ref{eq:ekRuTu}) for $\sigma=\uparrow$ using the excitation energies at $H=0$ 
(${\bar \varepsilon}^{\tilde {\cal A}}_{{\bm k}_{\parallel}+{\bm q}_{\parallel}}$, ${\bar \varepsilon}^{\tilde {\cal R}}_{-{\bm k}_{\parallel}+{\bm q}_{\parallel}}$, and $\Delta_{\rm spin}$ [Eqs. (\ref{eq:ebarkA})--(\ref{eq:HDelta})]), 
the dispersion relations of the emergent modes [dotted curves in 
Figs. \ref{fig:SzLadBilPAM}(g), \ref{fig:SzLadBilPAM}(h), \ref{fig:SzLadBilPAM}(j), \ref{fig:SzLadBilPAM}(k), \ref{fig:SzLadBilPAM}(s), \ref{fig:SzLadBilPAM}(t), 
\ref{fig:SzLadBilPAM}(v), \ref{fig:SzLadBilPAM}(w), \ref{fig:SzKLM}(g), \ref{fig:SzKLM}(h), \ref{fig:SzKLM}(j), and \ref{fig:SzKLM}(k)] can be expressed as 
\begin{align}
\label{eq:ekAdTuH}
&\omega={\bar \varepsilon}^{\tilde {\cal A}}_{{\bm k}_{\parallel}+{\bm q}_{\parallel}}-\frac{\Delta_{\rm spin}}{2}
\quad\text{in}\quad A^{\downarrow}_{{\cal T}^{1}_{{\bm q}_{\parallel}}}({\bm k},\omega),\\
\label{eq:ekRuTuH}
&\omega=-{\bar \varepsilon}^{\tilde {\cal R}}_{-{\bm k}_{\parallel}+{\bm q}_{\parallel}}+\frac{\Delta_{\rm spin}}{2}
\quad\text{in}\quad A^{\uparrow}_{{\cal T}^{1}_{{\bm q}_{\parallel}}}({\bm k},\omega),
\end{align}
where 
\begin{align}
\label{eq:ebarkA}
&\varepsilon^{{\tilde {\cal A}}_{\sigma}}_{{\bm k}_{\parallel}+{\bm q}_{\parallel}}={\bar \varepsilon}^{\tilde {\cal A}}_{{\bm k}_{\parallel}+{\bm q}_{\parallel}}-s^zH,\\
\label{eq:ebarkR}
&\varepsilon^{{\tilde {\cal R}}_{\sigma}}_{-{\bm k}_{\parallel}+{\bm q}_{\parallel}}={\bar \varepsilon}^{\tilde {\cal R}}_{-{\bm k}_{\parallel}+{\bm q}_{\parallel}}-s^zH,\\
\label{eq:ebarkT}
&e^{{\cal T}^{2s^z}}_{{\bm q}_{\parallel}}=\Delta_{\rm spin}-2s^z H,\\
\label{eq:HDelta}
&H\rightarrow\Delta_{\rm spin}\quad\text{for}\quad S^z/S_{\rm max}\rightarrow +0.
\end{align}
\par
The dispersion relations of the conventional electron-addition and electron-removal modes for $\sigma$-spin electrons, 
which correspond to those of the first terms on the right-hand sides of Eqs. (\ref{eq:adaggerX}) and (\ref{eq:aX}) for $|X\rangle_{{\bm q}_{\parallel}}=|{\cal T}^{1}\rangle_{{\bm q}_{\parallel}}$, under the magnetic field of $H=\Delta_{\rm spin}$ 
[Eqs. (\ref{eq:ekAtilde}), (\ref{eq:ekRtilde}), (\ref{eq:AbX}), and (\ref{eq:RbX}); solid curves in Figs. \ref{fig:SzLadBilPAM}(g), \ref{fig:SzLadBilPAM}(h), \ref{fig:SzLadBilPAM}(j), \ref{fig:SzLadBilPAM}(k), 
\ref{fig:SzLadBilPAM}(s), \ref{fig:SzLadBilPAM}(t), \ref{fig:SzLadBilPAM}(v), \ref{fig:SzLadBilPAM}(w), \ref{fig:SzKLM}(g), \ref{fig:SzKLM}(h), \ref{fig:SzKLM}(j), and \ref{fig:SzKLM}(k)] are expressed as 
\begin{align}
\label{eq:ekAuHcv}
&\omega={\bar \varepsilon}^{\tilde {\cal A}}_{{\bm k}_{\parallel}}-s^z\Delta_{\rm spin},
\quad\text{in}\quad A^{\sigma}_{{\cal T}^{1}_{{\bm q}_{\parallel}}}({\bm k},\omega),\\
\label{eq:ekRuHcv}
&\omega=-{\bar \varepsilon}^{\tilde {\cal R}}_{-{\bm k}_{\parallel}}-s^z\Delta_{\rm spin}
\quad\text{in}\quad A^{\sigma}_{{\cal T}^{1}_{{\bm q}_{\parallel}}}({\bm k},\omega).
\end{align}
\par
Because the charge gap $\Delta_{\rm charge }$, which is equal to the band gap between the bottom of the upper band and top of the lower band ($\mu_+-\mu_-$), 
is larger than the spin gap in spin-gapped Mott and Kondo insulators at $H=0$, i.e., 
\begin{equation}
\label{eq:CSgaps}
\Delta_{\rm charge}>\Delta_{\rm spin},
\end{equation}
and 
\begin{align}
\label{eq:ekAmin}
&{\bar \varepsilon}^{\tilde {\cal A}}_{{\bm k}_{\parallel}}\ge{\bar \varepsilon}^{\cal A}_{{\bm k}_{\parallel{\rm F}}^+}=\frac{\Delta_{\rm charge}}{2},\\
\label{eq:ekRmin}
&{\bar \varepsilon}^{\tilde {\cal R}}_{-{\bm k}_{\parallel}}\ge{\bar \varepsilon}^{\cal R}_{-{\bm k}_{\parallel{\rm F}}^-}=\frac{\Delta_{\rm charge}}{2}
\end{align} 
at $\mu=0$, where $\mu_+=-\mu_-=\frac{\Delta_{\rm charge}}{2}$, 
\begin{align}
\label{eq:ekAgap}
{\bar \varepsilon}^{\tilde {\cal A}}_{{\bm k}_{\parallel}}-\frac{\Delta_{\rm spin}}{2}&
\overset{\text{Eq. (\ref{eq:ekAmin})}}{\ge}\frac{\Delta_{\rm charge}}{2}-\frac{\Delta_{\rm spin}}{2}\overset{\text{Eq. (\ref{eq:CSgaps})}}{>}0,\\
\label{eq:ekRgap}
-{\bar \varepsilon}^{\tilde {\cal R}}_{-{\bm k}_{\parallel}}+\frac{\Delta_{\rm spin}}{2}&
\overset{\text{Eq. (\ref{eq:ekRmin})}}{\le}-\frac{\Delta_{\rm charge}}{2}+\frac{\Delta_{\rm spin}}{2}\overset{\text{Eq. (\ref{eq:CSgaps})}}{<}0
\end{align}
at $\mu=0$. Here, ${\bar \varepsilon}^{\cal A}_{{\bm k}_{\parallel{\rm F}}^+}$ and ${\bar \varepsilon}^{\cal R}_{-{\bm k}_{\parallel{\rm F}}^-}$ denote 
$\varepsilon^{{\cal A}_{\sigma}}_{{\bm k}_{\parallel{\rm F}}^+}$ and $\varepsilon^{{\cal R}_{\sigma}}_{-{\bm k}_{\parallel{\rm F}}^-}$ at $H=0$, respectively. 
\par
Thus, the Fermi level ($\omega=0$) is located within the band gap 
between the emergent mode [Eq. (\ref{eq:ekAdTuH}) for $A^{\downarrow}_{{\cal T}^{1}_{{\bm q}_{\parallel}}}({\bm k},\omega)$ and 
Eq. (\ref{eq:ekRuTuH}) for $A^{\uparrow}_{{\cal T}^{1}_{{\bm q}_{\parallel}}}({\bm k},\omega)$] and conventional mode 
[Eq. (\ref{eq:ekRuHcv}) of $\sigma=\downarrow$ for $A^{\downarrow}_{{\cal T}^{1}_{{\bm q}_{\parallel}}}({\bm k},\omega)$ and 
Eq. (\ref{eq:ekAuHcv}) of $\sigma=\uparrow$ for $A^{\uparrow}_{{\cal T}^{1}_{{\bm q}_{\parallel}}}({\bm k},\omega)$]; 
the system remains insulating even if the magnetization-induced modes emerge under the magnetic field at $\mu=0$. 
\par
The gap between the emergent mode [Eqs. (\ref{eq:ekAdTuH}) and (\ref{eq:ekRuTuH})] and conventional mode [Eq. (\ref{eq:ekRuHcv}) of $\sigma=\downarrow$ and Eq. (\ref{eq:ekAuHcv}) of $\sigma=\uparrow$] 
is obtained as $\Delta_{\rm charge}-\Delta_{\rm spin}$ in $A^{\downarrow}_{{\cal T}^{1}_{{\bm q}_{\parallel}}}({\bm k},\omega)$ and $A^{\uparrow}_{{\cal T}^{1}_{{\bm q}_{\parallel}}}({\bm k},\omega)$ 
[Eqs. (\ref{eq:ekAgap}) and (\ref{eq:ekRgap})] in spin-gapped Mott and Kondo insulators. 
\par
For $\mu\ne0$, because
\begin{align}
\label{eq:ekwAmu}
&\varepsilon^{{\tilde {\cal A}}_{\sigma}}_{{\bm k}_{\parallel}}(\mu)=\varepsilon^{{\tilde {\cal A}}_{\sigma}}_{{\bm k}_{\parallel}}(\mu=0)-\mu,\\
\label{eq:ekwRmu}
&\varepsilon^{{\tilde {\cal R}}_{\sigma}}_{{\bm k}_{\parallel}}(\mu)=\varepsilon^{{\tilde {\cal R}}_{\sigma}}_{{\bm k}_{\parallel}}(\mu=0)+\mu
\end{align}
[Eq. (\ref{eq:eXkFermion}); Tables \ref{tbl:symbols}--\ref{tbl:KLM}; Appendix \ref{sec:effTheory}], 
all dispersion relations of electronic excitations shift by $-\mu$. 
\par
Both the magnetization-induced mode [Eq. (\ref{eq:ekAdTuH}) or (\ref{eq:ekRuTuH})] and 
conventional mode [Eq. (\ref{eq:ekAuHcv}) for $\sigma=\uparrow$ or Eq. (\ref{eq:ekRuHcv}) for $\sigma=\downarrow$] cross the Fermi level ($\omega=0$) 
for $|\mu|\ge\frac{1}{2}(\Delta_{\rm charge}-\Delta_{\rm spin})$ [Eqs. (\ref{eq:ekAgap}) and (\ref{eq:ekRgap})]. 
\par
Thus, the system becomes metallic for $|\mu|\ge\frac{1}{2}(\Delta_{\rm charge}-\Delta_{\rm spin})$ and remains insulating for $|\mu|<\frac{1}{2}(\Delta_{\rm charge}-\Delta_{\rm spin})$ 
under the magnetic field of $|H|=\Delta_{\rm spin}$ regardless of the appearance of the magnetization-induced modes. 
\subsection{Absence of magnetization-induced in-gap modes in spin-gapless Mott and Kondo insulators} 
\label{sec:noMISgaplessSpin}
Although electronic modes are induced within the band gap by magnetizing spin-gapped Mott and Kondo insulators, 
electronic modes are not induced inside the band gap by magnetizing insulators with gapless spin excitation. 
\par
In the case of gapless spin excitation, because the lowest spin-excitation energy at $H=0$ is essentially zero, the magnetic field for $S^z/S_{\max}\rightarrow\pm0$ is also essentially zero 
[$e^{{\cal T}^{2s^z}}_{\bm q}\rightarrow0$ and $H\rightarrow0$ in Eqs. (\ref{eq:ekeMIS}) and (\ref{eq:ekhMIS})]. 
Hence, the dispersion relations of the emergent modes satisfy the following relations: 
\begin{align}
\label{eq:ekAgaplessSpin}
&\omega\overset{\text {Eq. (\ref{eq:ekeMIS})}}{=}\varepsilon^{{\tilde {\cal A}}_{\sigma}}_{{\bm k}+{\bm q}}-e^{{\cal T}^{2s^z}}_{\bm q}\rightarrow
{\bar \varepsilon}^{\tilde {\cal A}}_{{\bm k}+{\bm q}}\ge{\bar \varepsilon}^{\cal A}_{{\bm k}^+_{\rm F}}\overset{\text{Eq. (\ref{eq:mu+})}}{=}\mu_+,\\
\label{eq:ekRgaplessSpin}
&\omega\overset{\text {Eq. (\ref{eq:ekhMIS})}}{=}-\varepsilon^{{\tilde {\cal R}}_{\sigma}}_{-{\bm k}+{\bm q}}+e^{{\cal T}^{2s^z}}_{\bm q}\rightarrow
-{\bar \varepsilon}^{\tilde {\cal R}}_{-{\bm k}+{\bm q}}\le -{\bar \varepsilon}^{\cal R}_{-{\bm k}^-_{\rm F}}\overset{\text{Eq. (\ref{eq:mu-})}}{=}\mu_-,
\end{align}
where ${\bar \varepsilon}^{{\tilde {\cal A}}_{\sigma}}_{{\bm k}+{\bm q}}$ and ${\bar \varepsilon}^{\tilde {\cal R}}_{-{\bm k}+{\bm q}}$ denote 
$\varepsilon^{{\tilde {\cal A}}_{\sigma}}_{{\bm k}+{\bm q}}$ and $\varepsilon^{\tilde {\cal R}}_{-{\bm k}+{\bm q}}$ at $H=0$, respectively, and 
${\bar \varepsilon}^{\cal A}_{{\bm k}^+_{\rm F}}$ and ${\bar \varepsilon}^{\cal R}_{-{\bm k}^-_{\rm F}}$ 
represent the excitation energies of the low-$|\omega|$ electron-addition and electron-removal conventional modes at the Fermi momentum for $H=0$, respectively. 
Equations (\ref{eq:ekAgaplessSpin}) and (\ref{eq:ekRgaplessSpin}) imply that the emergent modes appear in the $\omega$ regimes of $\omega\ge\mu_+$ and $\omega\le\mu_-$; 
electronic modes are not induced inside the band gap by magnetizing insulators with gapless spin excitation. 
\par
\begin{figure} 
\includegraphics[width=\linewidth]{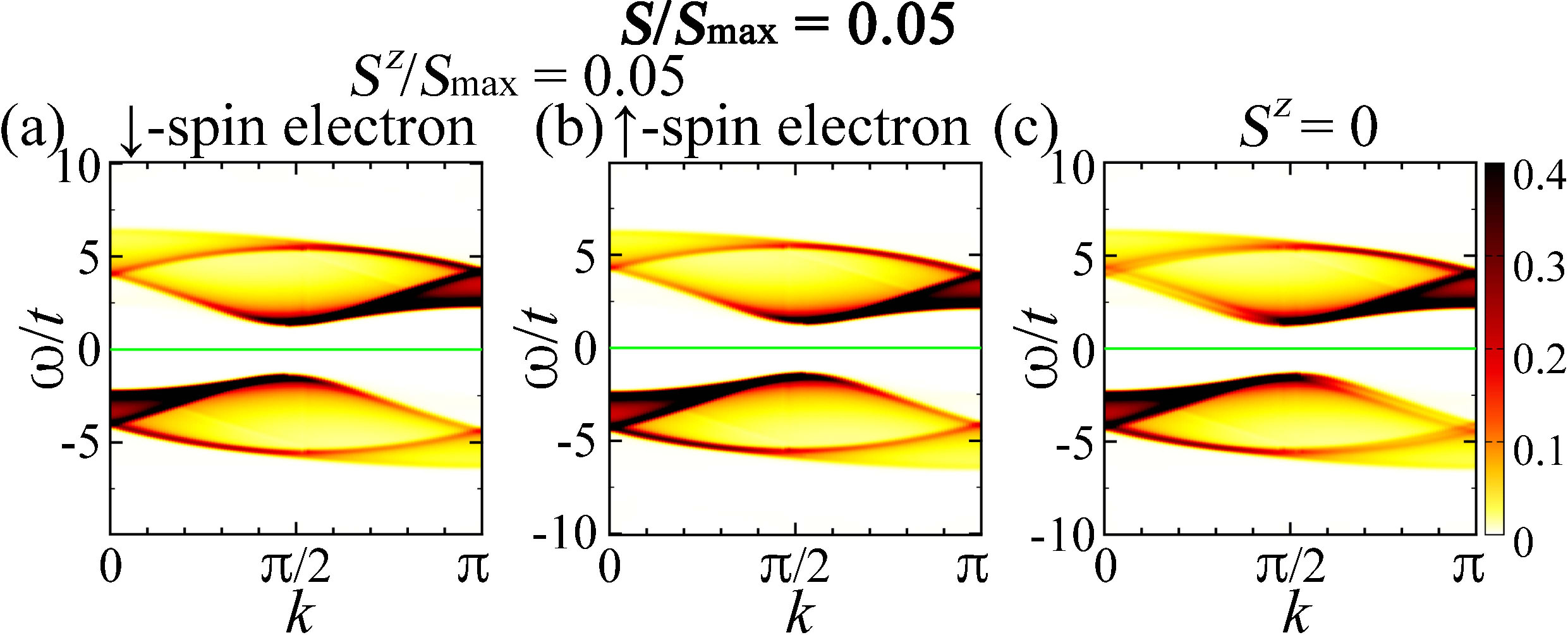}
\caption{Electronic excitation from the lowest-energy spin-perturbed states of $S/S_{\rm max}=0.05$ in the 1D HM. 
(a), (b) $A^\sigma_X(k,\omega)t$ for $|X\rangle$ of the magnetized ground state of $S^z=S$ for $\sigma=\downarrow$ [(a)] and $\uparrow$ [(b)]. 
(c) $A_X(k,\omega)t$ for $|X\rangle$ of the lowest-energy state of $S/S_{\rm max}=0.05$ and $S^z=0$. 
The solid green lines indicate $\omega=0$.}
\label{fig:SzHub1d}
\end{figure}

In fact, in the 1D HM, the spectral functions of the magnetized ground state of $S^z/S_{\rm max}=0.05$ [Figs. \ref{fig:SzHub1d}(a) and \ref{fig:SzHub1d}(b)] 
are almost the same as that of the unmagnetized ground state [Fig. \ref{fig:Hub1d}(a)]; magnetization-induced modes do not appear inside the band gap, 
reflecting the gapless spin excitation. 
\subsection{Spectral function of magnetized states via Shiba transformation} 
\label{sec:AkwM}
Under the transformation of 
\begin{equation}
c_{i,\downarrow}\rightarrow(-)^ic^{\dagger}_{i,\downarrow},\quad f_{i,\downarrow}\rightarrow-(-)^if^{\dagger}_{i,\downarrow}, 
\label{eq:ShibaTrans}
\end{equation}
which is called the Shiba transformation \cite{ShibaTrans,ShibaTrans_Heilmann,LiebWu}, 
the Hamiltonians of the HM and symmetric ($U=2\Delta$) PAM [Eqs. (\ref{eq:HamHub}) and (\ref{eq:HamPAM})] on a bipartite lattice are transformed \cite{ShibaTrans,ShibaTrans_Heilmann,LiebWu,TakahashiBook,Essler} as 
\begin{align}
\label{eq:ShibaHamHub}
{\cal H}_{\rm HM}(t,U,\mu,H)\rightarrow&{\cal H}_{\rm HM}(t,-U,\frac{H}{2},2\mu)+(\frac{H}{2}-\mu)N_{\rm s},\\
\label{eq:ShibaHamPAM}
{\cal H}_{\rm PAM}(t,t_{\rm K},U,\mu,H)\rightarrow&{\cal H}_{\rm PAM}(t,t_{\rm K},-U,\frac{H}{2},2\mu)\nonumber\\
&+(H-2\mu-\frac{U}{2})N_{\rm u}.
\end{align}
\par
The spin operators at a site $i$ are transformed \cite{ShibaTrans_Heilmann,TakahashiBook,Essler} as 
\begin{equation}
\label{eq:defSetai}
\begin{array}{ccc}
S^{c,+}_i=c^{\dagger}_{i,\uparrow}c_{i,\downarrow}&\rightarrow&\eta^{c,+}_i=(-)^ic^{\dagger}_{i,\uparrow}c^{\dagger}_{i,\downarrow},\\
S^{f,+}_i=f^{\dagger}_{i,\uparrow}f_{i,\downarrow}&\rightarrow&\eta^{f,+}_i=-(-)^if^{\dagger}_{i,\uparrow}f^{\dagger}_{i,\downarrow},\\
S^{c,-}_i=c^{\dagger}_{i,\downarrow}c_{i,\uparrow}&\rightarrow&\eta^{c,-}_i=(-)^ic_{i,\downarrow}c_{i,\uparrow},\\
S^{f,-}_i=f^{\dagger}_{i,\downarrow}f_{i,\uparrow}&\rightarrow&\eta^{f,-}_i=-(-)^if_{i,\downarrow}f_{i,\uparrow},\\
S^{c,z}_i=\frac{1}{2}(n^c_{i,\uparrow}-n^c_{i,\downarrow})&\rightarrow&\eta^{c,z}_i=\frac{1}{2}(n^c_{i,\uparrow}+n^c_{i,\downarrow}-1),\\
S^{f,z}_i=\frac{1}{2}(n^f_{i,\uparrow}-n^f_{i,\downarrow})&\rightarrow&\eta^{f,z}_i=\frac{1}{2}(n^f_{i,\uparrow}+n^f_{i,\downarrow}-1).
\end{array}
\end{equation}
Hence, the spin operators are transformed into the $\eta$ operators, 
\begin{equation}
\label{eq:SetaShiba}
{\hat {\bm S}}\rightarrow{\hat {\bm \eta}}.
\end{equation}
Here, 
\begin{equation}
\label{eq:defSeta}
\begin{array}{ccc}
{\hat S}^x=\frac{1}{2}({\hat S}^++{\hat S}^-),&{\hat S}^y=\frac{1}{2i}({\hat S}^+-{\hat S}^-),&{\hat S}^z=\sum_{i,\lambda} S^{\lambda,z}_i,\\
{\hat \eta}^x=\frac{1}{2}({\hat \eta}^++{\hat \eta}^-),&{\hat \eta}^y=\frac{1}{2i}({\hat \eta}^+-{\hat \eta}^-),&{\hat \eta}^z=\sum_{i,\lambda}\eta^{\lambda,z}_i
\end{array}
\end{equation}
with
\begin{equation}
\begin{array}{cc}
{\hat S}^\pm=\sum_{i,\lambda}S^{\lambda,\pm}_i,&{\hat \eta}^\pm=\sum_{i,\lambda}\eta^{\lambda,\pm}_i,\\
\end{array}
\end{equation}
where $\lambda=c$ for the HM and $\lambda=c$ and $f$ for the PAM and KLM [Eqs. (\ref{eq:eta}) and (\ref{eq:etai})]. 
\par
The $\downarrow$-spin-electron operators in the momentum space are transformed [Eq. (\ref{eq:ShibaTrans})] as 
\begin{equation}
\begin{array}{ll}
c_{{\bm k},\downarrow}\rightarrow c^{\dagger}_{{\bm \pi}-{\bm k},\downarrow},&f_{{\bm k},\downarrow}\rightarrow -f^{\dagger}_{{\bm \pi}-{\bm k},\downarrow}.
\label{eq:ckShiba}
\end{array}
\end{equation}
\par
Accordingly, under the Shiba transformation, a magnetized state of $S=j$ and $S^z=m$ at $H=2{\bar \mu}$ and $\mu=\frac{\bar H}{2}$ for ${\cal H}(U)$ is transformed into 
the corresponding doped state of $\eta=j$ and $\eta^z=m$ at $\mu={\bar \mu}$ and $H={\bar H}$ for ${\cal H}(-U)$, 
where ${\cal H}(U)$ and ${\cal H}(-U)$ represent the Hamiltonians before and after the Shiba transformation, respectively [Eqs. (\ref{eq:ShibaHamHub}) and (\ref{eq:ShibaHamPAM})]. 
\par
The spectral function of a magnetized state of $S=j$, $S^z=m$, and $\eta=0$ for $\sigma$-spin electrons, $A^{\lambda,\sigma}_{S=j, S^z=m}({\bm k},\omega)$, at $H=2{\bar \mu}$ and $\mu=0$ for ${\cal H}(U)$ 
can be expressed using those of the doped states of $\eta=j$, $\eta^z=\pm m$, and $S=0$ at $\mu=\pm{\bar \mu}$ and $H=0$ for ${\cal H}(-U)$ 
[Appendix \ref{sec:AkwSnU}; Eqs. (\ref{eq:AkwudShiba}) and (\ref{eq:AkwudShiba2})] as 
\begin{align}
\label{eq:AkwuEta}
&\begin{array}{clclcl}
&A^{\lambda,\uparrow}_{S=j,S^z=m}({\bm k},\omega)&\text{at}&H=2{\bar \mu}&\text{for}&{\cal H}(U)\\
=&{\tilde A}^{\lambda}_{\eta=j,\eta^z=m}({\bm k},\omega)&\text{at}&\mu={\bar \mu}&\text{for}&{\cal H}(-U)\\
\overset{\text{Eq. (\ref{eq:AkwphX})}}{=}&{\tilde A}^{\lambda}_{\eta=j,\eta^z=-m}({\bm \pi}-{\bm k},-\omega)&\text{at}&\mu=-{\bar \mu}&\text{for}&{\cal H}(-U),
\end{array}\\
\label{eq:AkwdEta}
&\begin{array}{clclcl}
&A^{\lambda,\downarrow}_{S=j,S^z=m}({\bm k},\omega)&\text{at}&H=2{\bar \mu}&\text{for}&{\cal H}(U)\\
\overset{\text{ Eq. (\ref{eq:ckShiba})}}{=}&{\tilde A}^{\lambda}_{\eta=j,\eta^z=m}({\bm \pi}-{\bm k},-\omega)&\text{at}&\mu={\bar \mu}&\text{for}&{\cal H}(-U)\\
\overset{\text{Eq. (\ref{eq:AkwphX})}}{=}&{\tilde A}^{\lambda}_{\eta=j,\eta^z=-m}({\bm k},\omega)&\text{at}&\mu=-{\bar \mu}&\text{for}&{\cal H}(-U),
\end{array}
\end{align}
where $\mu=0$ for ${\cal H}(U)$, $H=0$ for ${\cal H}(-U)$, and the spectral function for ${\cal H}(-U)$ is represented by ${\tilde A}^{\lambda}_X({\bm k},\omega)[=\frac{1}{2}\sum_{\sigma}{\tilde A}^{\lambda,\sigma}_X({\bm k},\omega)]$; 
${\tilde A}_X({\bm k},\omega)=\sum_{\lambda}{\tilde A}^{\lambda}_X({\bm k},\omega)$. 
\par
Thus, the magnetization-induced electronic states appear from the ground state of $S=|S^z|>0$ for ${\cal H}(U)$ in spin-gapped strongly correlated insulators, 
as the doping-induced electronic states appear from the ground state of $\eta=|\eta^z|(=\frac{|N_{\rm h}|}{2})>0$ for ${\cal H}(-U)$. 
\par
The numerical results for $A^{\uparrow}_X({\bm k},\omega)$ and $A^{\downarrow}_X({\bm k},\omega)$ for $|X\rangle$ of the state corresponding to the magnetized ground state of $S=S^z>0$ for ${\cal H}(U)$ 
[Figs. \ref{fig:SzLadBilPAM}(a), \ref{fig:SzLadBilPAM}(b), \ref{fig:SzLadBilPAM}(d), \ref{fig:SzLadBilPAM}(e), \ref{fig:SzLadBilPAM}(m), \ref{fig:SzLadBilPAM}(n), 
\ref{fig:SzLadBilPAM}(p), \ref{fig:SzLadBilPAM}(q), \ref{fig:SzHub1d}(a), and \ref{fig:SzHub1d}(b)] were obtained via Eqs. (\ref{eq:AkwuEta}) and (\ref{eq:AkwdEta}) 
using those of ${\tilde A}_X({\bm k},\omega)$ for $|X\rangle$ of the state corresponding to the doped ground state for ${\cal H}(-U)$ in the HM and PAM. 
\subsection{Emergent states corresponding to magnetization-induced states in unmagnetized spin-perturbed systems} 
\label{sec:Sz0MIS}
The state of $|S=j,S^z=0\rangle=({\hat S}^+)^j|{\rm GS}\rangle^{S^z=-j}=({\hat S}^-)^j|{\rm GS}\rangle^{S^z=j}$ is the lowest-energy state of $S=j$ for $H=S^z=0$, 
where $|{\rm GS}\rangle^{S^z=m}$ denotes the magnetized ground state of $S^z=m$, 
because the magnetized ground states of $S^z=\pm j$ are the lowest-energy states of $S=j$ and $S^z=\pm j$, 
provided that phase separation in the magnetization density does not occur at $S^z=\pm j$ for $\mu=\delta=0$. 
\par
The spectral function of the lowest-energy state of $S=j$ and $S^z=0$ for $\mu=H=0$ can be expressed using those of $S^z=\pm j$ as 
\begin{align}
&A^{\lambda}_{S=j,S^z=0}({\bm k},\omega)\overset{\text{Eq. (\ref{eq:Akwspin})}}{=}\frac{1}{2}\sum_{\sigma}A^{\lambda,\sigma}_{S=j,S^z=0}({\bm k},\omega)\nonumber\\
&\overset{\text{Eqs. (\ref{eq:AkwuEta}) and (\ref{eq:AkwdEta})}}{=}{\tilde A}^{\lambda}_{\eta=j,\eta^z=0}({\bm k},\omega)\nonumber\\
&\overset{\text{Eq. (\ref{eq:AkwetaGS})}}{=}\frac{1}{2}[{\tilde A}^{\lambda}_{\eta=j,\eta^z=j}({\bm k},\omega)+{\tilde A}^{\lambda}_{\eta=j,\eta^z=-j}({\bm k},\omega)]\nonumber\\
&\overset{\text{Eqs. (\ref{eq:AkwuEta}) and (\ref{eq:AkwdEta})}}{=}\frac{1}{2}\sum_{\sigma}A^{\lambda,\sigma}_{S=j,S^z=\pm j}({\bm k},\omega)\nonumber\\
&\overset{\text{Eq. (\ref{eq:Akwspin})}}{=}A^{\lambda}_{S=j,S^z=\pm j}({\bm k},\omega).
\label{eq:AkwSSz0}
\end{align}
The relation of $A^{\lambda}_{S=j,S^z=0}({\bm k},\omega)=A^{\lambda}_{S=j,S^z=\pm j}({\bm k},\omega)$ at $H=0$ can also be derived using spin-SU(2) symmetry [Eq. (\ref{eq:AkwS}); Appendix \ref{sec:AkwS}]. 
\par
Thus, the spin perturbation that excites spins each having an excitation energy as large as the spin gap without a magnetic field or magnetization in spin-gapped Mott and Kondo insulators can induce electronic modes 
corresponding to the magnetization-induced modes [Eqs. (\ref{eq:ekeMIS}) and (\ref{eq:ekhMIS})] along 
\begin{align}
\label{eq:ekeMISSz0}
&\omega=\varepsilon^{\tilde {\cal A}}_{{\bm k}+{\bm q}}-e^{{\cal T}^{0}}_{\bm q},\\
\label{eq:ekeMISSz-}
&\omega=-\varepsilon^{\tilde {\cal R}}_{-{\bm k}+{\bm q}}+e^{{\cal T}^{0}}_{\bm q},
\end{align}
where $\varepsilon^{\tilde {\cal A}}_{\bm k}$ and $\varepsilon^{\tilde {\cal R}}_{\bm k}$ denote the electron-addition and electron-removal excitation energies of the states in the conventional bands with a momentum ${\bm k}$, respectively, 
and $e^{{\cal T}^{0}}_{\bm q}$ denotes the excitation energy of the lowest-energy spin-excited state of $S^z=0$, which has the momentum of ${\bm q}$, from $|{\rm GS}\rangle$ at $H=0$. 
\par
The dispersion relations of the emergent modes for $S/S_{\rm max}\rightarrow+0$ at $S^z=0$ 
obtained via Eq. (\ref{eq:AkwSSz0}) using those for $S/S_{\rm max}=S^z/S_{\rm max}\rightarrow+0$ in the effective theory are 
shown by the dotted curves in Figs. \ref{fig:SzLadBilPAM}(i), \ref{fig:SzLadBilPAM}(l), \ref{fig:SzLadBilPAM}(u), \ref{fig:SzLadBilPAM}(x), \ref{fig:SzKLM}(i), and \ref{fig:SzKLM}(l). 
The spectral functions of the state corresponding to the lowest-energy state of $S/S_{\rm max}=0.05$ and $S^z=0$
obtained via Eq. (\ref{eq:AkwSSz0}) using the numerical results for $S/S_{\rm max}=S^z/S_{\rm max}=0.05$ are shown in 
Figs. \ref{fig:SzLadBilPAM}(c), \ref{fig:SzLadBilPAM}(f), \ref{fig:SzLadBilPAM}(o), \ref{fig:SzLadBilPAM}(r), \ref{fig:SzKLM}(c), \ref{fig:SzKLM}(f), and \ref{fig:SzHub1d}(c). 
\par
If the emergent modes cross the Fermi level, the band structure can be regarded as metallic. 
\subsection{Remarks on nonzero-$S$ states} 
\label{sec:RemarksMIS}
As in the case of the doping-induced states (Sec. \ref{sec:swDIS}) \cite{KohnoTinduced,KohnoRPP,KohnoDIS,KohnoKLM,Eskes,DagottoDOS}, 
the spectral weights of the magnetization-induced states should be proportional to the density of the excited spins ($|S^z|/S_{\rm max}$) in the small-$|S^z|/S_{\rm max}$ regime. 
To observe the magnetization-induced states, a macroscopic number $\mathcal{O}(N_{\rm u})$ of spins should be excited. 
The magnetized ground state can be obtained by applying a magnetic field larger than the spin gap: $|H|>\Delta_{\rm spin}$. 
\par
Note that the magnetic field causes not only the Zeeman effect but also the cyclotron motion of electrons. 
In a magnetic field, Landau quantization might considerably affect the electronic states. 
Furthermore, in angle-resolved photoemission experiments, the magnetic-field effects on the electron trajectory should be considered. 
These effects of the magnetic field, which are neglected in this study, should be considered for accurate comparison with experimental results. 
\par
To observe the emergent electronic states corresponding to the magnetization-induced states in unmagnetized systems, 
a macroscopic number $\mathcal{O}(N_{\rm u})$ of spins should be kept excited by the spin perturbation without a magnetic field at the time of measurement. 
\par
In the case of neutron radiation, because a single spin can be excited by a neutron, 
a macroscopic number $\mathcal{O}(N_{\rm u})$ of neutrons that excite spins (excluding reflected and transmitted neutrons that are irrelevant to the excitations) 
should be injected every lifetime (relaxation time) of the single-spin excited states 
to keep the number of excited spins macroscopic, as in a steady state, during the measurement, as mentioned in Sec. \ref{sec:Remarksetaz0DIS}. 
Photons can be used instead of neutrons if they can generate spin excitation \cite{spinRIXS}. 
\subsection{Electronic modes induced by spin fluctuation} 
\label{sec:spinFluc}
\subsubsection{Selection rules for electronic excitation from a state perturbed by spin fluctuation} 
\label{sec:selectionRulesSF}
Electronic modes can also be induced by spin fluctuation, as in the case of charge fluctuation (Sec. \ref{sec:chargeFluc}). 
As a spin-perturbed state, we consider $S^{\lambda,z}_{\bm q}|{\rm GS}\rangle$ at half filling at $H=0$ for $\lambda=c$ and $f$. 
The spin-perturbed state $S^{\lambda,z}_{\bm q}|{\rm GS}\rangle$ has the momentum of ${\bm q}$, $N_{\rm h}=0$, and $S^z=0$ 
for the unperturbed ground state $|{\rm GS}\rangle$ with the momentum of ${\bm 0}$, $N_{\rm h}=0$, and $S^z=0$. 
The excitation energy of each eigenstate in $S^{\lambda,z}_{\bm q}|{\rm GS}\rangle$ from $|{\rm GS}\rangle$ is denoted by $e^{{\cal T}^0}_{\bm q}$. 
\par
The state obtained by adding an electron with a momentum ${\bm k}$ to $S^{\lambda,z}_{\bm q}|{\rm GS}\rangle$ has the momentum of ${\bm k}+{\bm q}$ and $N_{\rm h}=-1$. 
Since this state can overlap with $|{\tilde {\cal A}}\rangle_{{\bm k}+{\bm q}}$ [electron-addition excited states in the conventional bands from $|{\rm GS}\rangle$ (Sec. \ref{sec:selectionRulesCF})], 
electronic states can emerge in the electron-addition spectrum, exhibiting 
\begin{equation}
\omega=\varepsilon^{\tilde {\cal A}}_{{\bm k}+{\bm q}}-e^{{\cal T}^0}_{\bm q}, 
\label{eq:ekAS}
\end{equation}
where $\varepsilon^{\tilde {\cal A}}_{{\bm k}+{\bm q}}$ denotes the excitation energy of $|{\tilde {\cal A}}\rangle_{{\bm k}+{\bm q}}$ from $|{\rm GS}\rangle$. 
\par
Similarly, the state obtained by removing an electron with a momentum ${\bm k}$ from $S^{\lambda,z}_{\bm q}|{\rm GS}\rangle$ has the momentum of $-{\bm k}+{\bm q}$ and $N_{\rm h}=1$. 
Since this state can overlap with $|{\tilde {\cal R}}\rangle_{-{\bm k}+{\bm q}}$ [electron-removal excited states in the conventional bands from $|{\rm GS}\rangle$ (Sec. \ref{sec:selectionRulesCF})], 
electronic states can emerge in the electron-removal spectrum, exhibiting 
\begin{equation}
\omega=-\varepsilon^{\tilde {\cal R}}_{-{\bm k}+{\bm q}}+e^{{\cal T}^0}_{\bm q}, 
\label{eq:ekRS}
\end{equation}
where $\varepsilon^{\tilde {\cal R}}_{-{\bm k}+{\bm q}}$ denotes the excitation energy of $|{\tilde {\cal R}}\rangle_{-{\bm k}+{\bm q}}$ from $|{\rm GS}\rangle$. 
\par
Thus, electronic modes exhibiting the dispersion relations of the conventional bands ($\varepsilon^{\tilde {\cal A}}_{\bm k}$ and $-\varepsilon^{\tilde {\cal R}}_{-{\bm k}}$) 
shifted by the spin-excitation energy ($-e^{{\cal T}^0}_{\bm q}$ and $e^{{\cal T}^0}_{\bm q}$) and momentum (${\bm q}$) 
[Eqs. (\ref{eq:ekAS}) and (\ref{eq:ekRS})] can emerge in the electronic spectrum from the spin-perturbed state $S^{\lambda,z}_{\bm q}|{\rm GS}\rangle$. 
\par
The spin-perturbed state $S^{\lambda,z}_{\bm q}|{\rm GS}\rangle$ includes all the eigenstates of $S^z=0$ at the momentum of ${\bm q}$ caused by $S^{\lambda,z}_{\bm q}$, 
i.e., $\forall|m\rangle$ which satisfies $\langle m|S^{\lambda,z}_{\bm q}|{\rm GS}\rangle\ne 0$ [Eqs. (\ref{eq:SkwHub}) and (\ref{eq:SkwPAMKLM})]. 
However, if a single state dominates at the momentum of ${\bm q}$ in the spin excitation, 
$S^{\lambda,z}_{\bm q}|{\rm GS}\rangle$ can effectively be regarded as a single state, and the emergent states can be well identified as the emergent modes. 
\par
To observe the emergent modes, a macroscopic number $\mathcal{O}(N_{\rm u})$ of spins ($\prod_i S^{\lambda,z}_{{\bm q}_i}$) should be 
kept excited at the time of measurement, as mentioned in Sec. \ref{sec:RemarksMIS}. 
\par
If the emergent modes cross the Fermi level, the band structure can be regarded as metallic. 
\par
\subsubsection{Electronic modes induced by spin fluctuation in spin-gapped Mott and Kondo insulators} 
\label{sec:effectiveTheorySF}
\begin{figure*} 
\includegraphics[width=\linewidth]{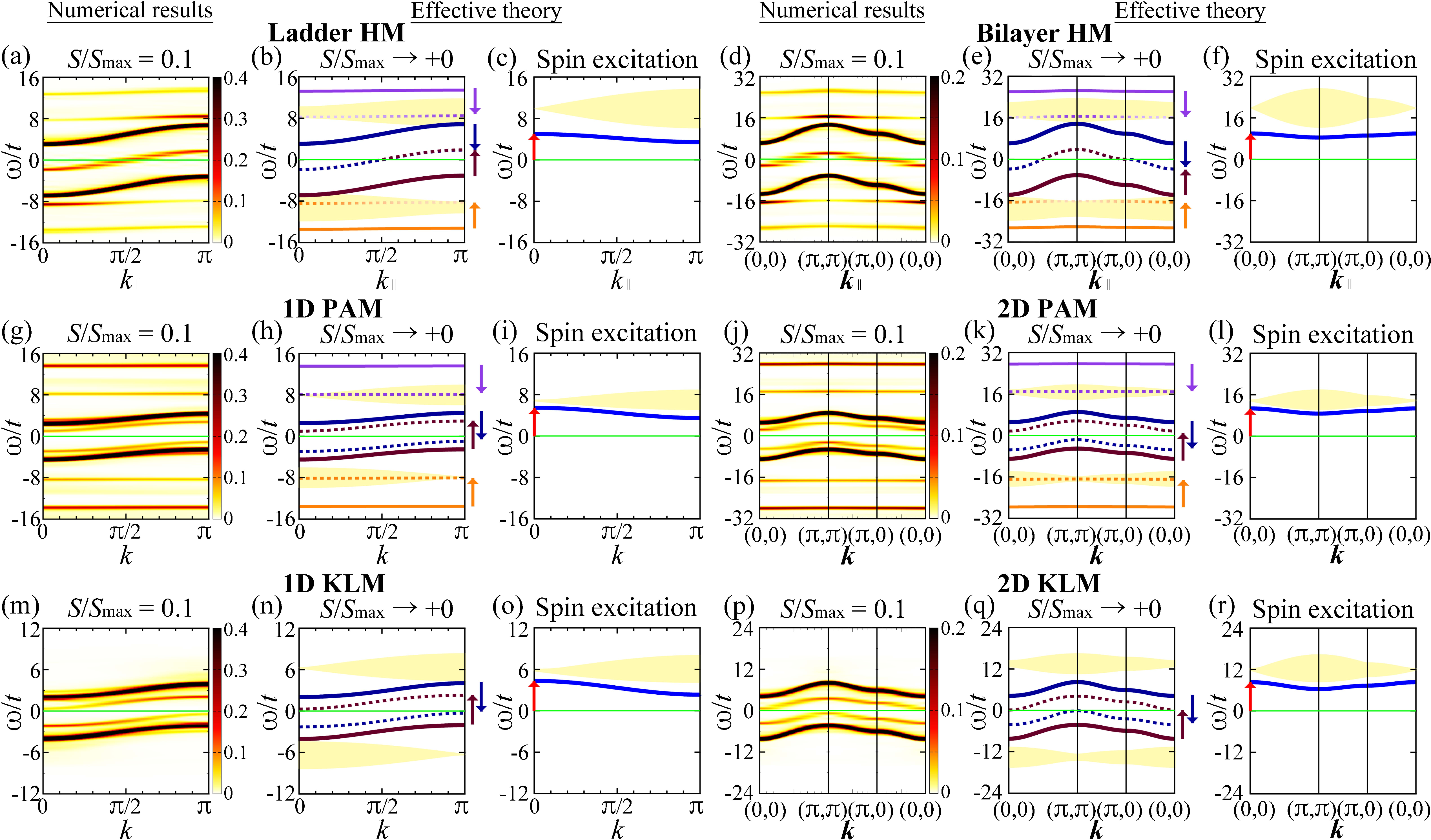}
\caption{Electronic excitation from the spin-perturbed states of small $S/S_{\rm max}$ caused by spin fluctuation 
in the ladder HM [(a), (b)], bilayer HM [(d), (e)], 1D PAM [(g), (h)], 2D PAM [(j), (k)], 1D KLM [(m), (n)], and 2D KLM [(p), (q)]. 
(a), (d), (g), (j), (m), (p) $A_X(k_\parallel,0,\omega)t+A_X(k_\parallel,\pi,\omega)t$ [(a)], 
${\bar A}_X({\bm k}_\parallel,0,\omega)t+{\bar A}_X({\bm k}_\parallel,\pi,\omega)t$ [(d)], 
$A_X({\bm k},\omega)t$ [(g), (m), (p)], and 
${\bar A}_X({\bm k},\omega)t$ [(j)] for 
$|X\rangle$ of the state corresponding to 
$\prod_i S^{c,z}_{{\bm q}_i}|{\rm GS}\rangle$ at ${\bm q}_i=({\bm 0},\pi)$ in (a) and (d); 
$\prod_i(S^{c,z}_{{\bm q}_i}-S^{f,z}_{{\bm q}_i})|{\rm GS}\rangle$ at ${\bm q}_i={\bm 0}$ in (g), (j), (m), and (p) for $S/S_{\rm max}=0.1$. 
(b), (e), (h), (k), (n), (q) Dispersion relations of electronic excitations from $S^{c,z}_{({\bm 0},\pi)}|{\rm GS}\rangle$ [(b), (e)] and 
$(S^{c,z}_{\bm 0}-S^{f,z}_{\bm 0})|{\rm GS}\rangle$ [(h), (k), (n), (q)] in the effective theory. 
The dispersion relations of the emergent modes are 
$\omega=-\varepsilon^{\cal R}_{-{\bm k}_\parallel}+e^{{\cal T}^0}_{\bm 0}$ (dotted brown curves), 
$\omega=\varepsilon^{\cal A}_{{\bm k}_\parallel}-e^{{\cal T}^0}_{\bm 0}$ (dotted blue curves), 
$\omega=-\varepsilon^{\bar {\cal R}}_{-{\bm k}_\parallel}+e^{{\cal T}^0}_{\bm 0}$ (dotted orange curves), and 
$\omega=\varepsilon^{\bar {\cal A}}_{{\bm k}_\parallel}-e^{{\cal T}^0}_{\bm 0}$ (dotted purple curves), 
which are shown in the relevant momentum regimes for comparison with the numerical results. 
Here, $e^{{\cal T}^0}_{\bm 0}$ denotes the excitation energy of the spin mode at ${\bm k}_\parallel={\bm 0}$ [red arrows in (c), (f), (i), (l), (o), and (r)]. 
The dispersion relations of the other modes and continua in (b), (e), (h), (k), (n), and (q) are the same as 
those in Figs. \ref{fig:AkwLadBil}(g), \ref{fig:AkwLadBil}(j), \ref{fig:AkwLadBil}(h), \ref{fig:AkwLadBil}(k), \ref{fig:AkwLadBil}(i), and \ref{fig:AkwLadBil}(l), respectively. 
The up and down arrows on the right sides of (b), (e), (h), (k), (n), and (q) indicate $\omega$-shifts from the unperturbed modes 
by $e^{{\cal T}^0}_{\bm 0}$ and $-e^{{\cal T}^0}_{\bm 0}$, respectively. 
(c), (f), (i), (l), (o), (r) Dispersion relations of spin excitations in the effective theory in the ladder HM [(c)], bilayer HM [(f)], 1D PAM [(i)], 2D PAM [(l)], 1D KLM [(o)], and 2D KLM [(r)]. 
The red arrows indicate the excitation energy of the spin mode at ${\bm k}_\parallel={\bm 0}$, i.e., $e^{{\cal T}^0}_{\bm 0}$. 
The dispersion relations of the modes and continua in (c), (f), (i), (l), (o), and (r) are the same as 
those in Figs. \ref{fig:SkwLadBil}(g), \ref{fig:SkwLadBil}(j), \ref{fig:SkwLadBil}(h), \ref{fig:SkwLadBil}(k), \ref{fig:SkwLadBil}(i), and \ref{fig:SkwLadBil}(l), respectively. 
The solid green lines indicate $\omega=0$.}
\label{fig:Kz0PLadBil}
\end{figure*}
In the ladder and bilayer HMs and 1D and 2D PAMs and KLMs, 
\begin{equation}
\label{eq:T0k}
\frac{{\hat S}^{z}_{{\bm q}_{\parallel}}|{\rm GS}\rangle}
{\sqrt{\langle{\rm GS}|{\hat S}^{z}_{-{\bm q}_{\parallel}}{\hat S}^{z}_{{\bm q}_{\parallel}}|{\rm GS}\rangle}}
=|{\cal T}^0\rangle_{{\bm q}_{\parallel}}
\end{equation}
in the effective theory, where ${\hat S}^{z}_{{\bm q}_{\parallel}}=S^{c,z}_{({\bm q}_{\parallel},\pi)}$ for the ladder and bilayer HMs and 
$S^{c,z}_{{\bm q}_{\parallel}}-S^{f,z}_{{\bm q}_{\parallel}}$ for the 1D and 2D PAMs and KLMs 
[Eqs. (\ref{eq:SkwHub}), (\ref{eq:SkwPAMKLM}), (\ref{eq:GS}), and (\ref{eq:Xk}); Tables \ref{tbl:symbols}--\ref{tbl:KLM}]. 
The electron-addition and electron-removal excited states from this state are obtained as Eqs. (\ref{eq:adaggerX}) and (\ref{eq:aX}) for $|X\rangle_{{\bm q}_{\parallel}}=|{\cal T}^0\rangle_{{\bm q}_{\parallel}}$. 
The first terms on the right-hand sides of Eqs. (\ref{eq:adaggerX}) and (\ref{eq:aX}) exhibit essentially the same band structure as that in the unperturbed case 
[Eqs. (\ref{eq:ekAtilde}), (\ref{eq:ekRtilde}), (\ref{eq:AbX}), and (\ref{eq:RbX})] \cite{KohnoKLM,KohnoTinduced}. 
\par
In addition, spectral weights emerge along 
\begin{align}
\label{eq:ekAT0}
\omega&=\varepsilon^{\tilde {\cal A}}_{{\bm k}_{\parallel}+{\bm q}_{\parallel}}-e^{{\cal T}^0}_{{\bm q}_{\parallel}},\\
\label{eq:ekRT0}
\omega&=-\varepsilon^{\tilde {\cal R}}_{-{\bm k}_{\parallel}+{\bm q}_{\parallel}}+e^{{\cal T}^0}_{{\bm q}_{\parallel}}
\end{align}
(for ${\tilde {\cal A}}_{\sigma}={\rm F}_{\sigma}$ and ${\tilde {\cal R}}_{\bar \sigma}={\rm A}_{\bar \sigma}$ at $k_{\perp}=0$; 
${\tilde {\cal A}}_{\sigma}={\rm G}_{\sigma}$ and ${\tilde {\cal R}}_{\bar \sigma}={\rm B}_{\bar \sigma}$ at $k_{\perp}=\pi$ in the HM), because 
$_{{\bm k}_{\parallel}+{\bm q}_{\parallel}}\langle{\tilde {\cal A}}_{\sigma}|a^{\dagger}_{{\bm k}_{\parallel},\sigma}|{{\cal T}^0}\rangle_{{\bm q}_{\parallel}}\ne 0$ and 
$_{-{\bm k}_{\parallel}+{\bm q}_{\parallel}}\langle{\tilde {\cal R}}_{\bar \sigma}|a_{{\bm k}_{\parallel},\sigma}|{{\cal T}^0}\rangle_{{\bm q}_{\parallel}}\ne 0$ 
(for ${\tilde {\cal A}}_{\sigma}={\rm F}_{\sigma}$ and ${\tilde {\cal R}}_{\bar \sigma}={\rm A}_{\bar \sigma}$ at $k_{\perp}=0$; 
${\tilde {\cal A}}_{\sigma}={\rm G}_{\sigma}$ and ${\tilde {\cal R}}_{\bar \sigma}={\rm B}_{\bar \sigma}$ at $k_{\perp}=\pi$ in the HM) 
between $|{\tilde {\cal A}}_{\sigma}\rangle_{{\bm k}_{\parallel}+{\bm q}_{\parallel}}$ and $|{\tilde {\cal R}}_{\bar \sigma}\rangle_{-{\bm k}_{\parallel}+{\bm q}_{\parallel}}$ [Eq. (\ref{eq:Xk})] and 
the second terms on the right-hand sides of Eqs. (\ref{eq:adaggerX}) and (\ref{eq:aX}) for $|X\rangle_{{\bm q}_{\parallel}}=|{\cal T}^0\rangle_{{\bm q}_{\parallel}}$ 
(Tables \ref{tbl:symbols}--\ref{tbl:KLM}; Appendix \ref{sec:effTheory}). 
Equations (\ref{eq:ekAT0}) and (\ref{eq:ekRT0}) correspond to Eqs. (\ref{eq:ekAS}) and (\ref{eq:ekRS}), respectively. 
\par
The dispersion relations of the emergent modes are those of the unperturbed modes ($\varepsilon^{\tilde {\cal A}}_{{\bm k}_{\parallel}}$ and $-\varepsilon^{\tilde {\cal R}}_{-{\bm k}_{\parallel}}$) 
shifted by the spin-excitation energy ($-e^{{\cal T}^0}_{{\bm q}_{\parallel}}$ and $e^{{\cal T}^0}_{{\bm q}_{\parallel}}$) and momentum (${\bm q}_{\parallel}$) 
[red arrows at ${\bm k}_{\parallel}={\bm 0}(={\bm q}_{\parallel})$ in Figs. \ref{fig:Kz0PLadBil}(c), \ref{fig:Kz0PLadBil}(f), \ref{fig:Kz0PLadBil}(i), \ref{fig:Kz0PLadBil}(l), \ref{fig:Kz0PLadBil}(o), and \ref{fig:Kz0PLadBil}(r)] 
as Eqs. (\ref{eq:ekAT0}) and (\ref{eq:ekRT0}) [dotted curves in Figs. \ref{fig:Kz0PLadBil}(b), \ref{fig:Kz0PLadBil}(e), \ref{fig:Kz0PLadBil}(h), \ref{fig:Kz0PLadBil}(k), \ref{fig:Kz0PLadBil}(n), and \ref{fig:Kz0PLadBil}(q)], 
which are consistent with the numerical results [Figs. \ref{fig:Kz0PLadBil}(a), \ref{fig:Kz0PLadBil}(d), \ref{fig:Kz0PLadBil}(g), \ref{fig:Kz0PLadBil}(j), \ref{fig:Kz0PLadBil}(m), and \ref{fig:Kz0PLadBil}(p)]. 
\subsubsection{Electronic modes induced by spin fluctuation in the 1D and 2D HMs} 
\label{sec:Hub1d2dSF}
\begin{figure} 
\includegraphics[width=\linewidth]{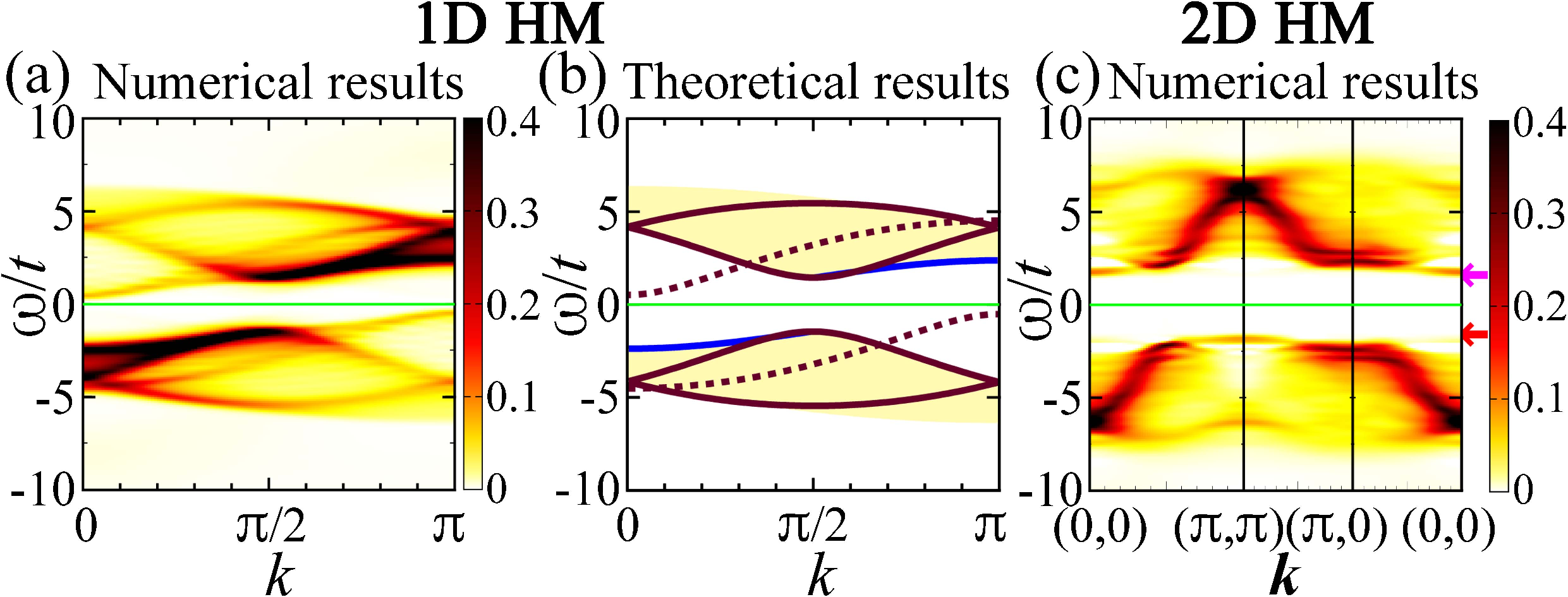}
\caption{Electronic excitation from the spin-perturbed states of small $S/S_{\rm max}$ caused by spin fluctuation in the 1D HM [(a), (b)] and 2D HM [(c)]. 
(a), (c) $A_X({\bm k},\omega)t$ for 
$|X\rangle$ of the state corresponding to $\prod_i S^{c,z}_{{\bm q}_i}|{\rm GS}\rangle$ of $S/S_{\rm max}=0.1$ for ${\bm q}_i=\frac{\bm \pi}{2}$. 
The magenta and red arrows in (c) indicate $\omega=-e^{{\cal T}^0}_{\frac{\bm \pi}{2}}+\mu_+$ and $\omega=e^{{\cal T}^0}_{\frac{\bm \pi}{2}}+\mu_-$, respectively, 
where $e^{{\cal T}^0}_{\frac{\bm \pi}{2}}$ denotes the excitation energy of the spin mode at ${\bm k}=\frac{\bm \pi}{2}$ in the 2D HM. 
(b) Dispersion relations from $S^{c,z}_{\frac{\pi}{2}}|{\rm GS}\rangle$ in the 1D HM. 
Dotted brown curves indicate 
$\omega=\varepsilon^{\rm holon}_{k-\frac{\pi}{2}}-e^{{\cal T}^0}_{\frac{\pi}{2}}$ (upper curve) and 
$\omega=-\varepsilon^{\rm holon}_{-k+\frac{\pi}{2}}+e^{{\cal T}^0}_{\frac{\pi}{2}}$ (lower curve), 
where $\varepsilon^{\rm holon}_k$ and $e^{{\cal T}^0}_{\frac{\pi}{2}}$ denote the excitation energy of the holon mode with a momentum $k$ and that of the spin mode at $k=\frac{\pi}{2}$, respectively. 
Only the relevant holon modes induced by the spin fluctuation for comparison with the numerical results are shown. 
The dispersion relations of the other modes and continua in (b) are the same as those in Fig. \ref{fig:Hub1d}(d). 
The solid green lines indicate $\omega=0$.}
\label{fig:KzHub1d2d}
\end{figure}
In the 1D and 2D HMs, electronic modes can be induced by spin fluctuation. 
According to Eqs. (\ref{eq:ekAS}) and (\ref{eq:ekRS}), electronic modes induced by the spin fluctuation of $S^{c,z}_{\bm q}$ exhibit the dispersion relations of 
\begin{align}
\label{eq:ekAHub1d2d}
\omega&=\varepsilon^{\tilde {\cal A}}_{{\bm k}+{\bm q}}-e^{{\cal T}^0}_{\bm q},\\
\label{eq:ekRHub1d2d}
\omega&=-\varepsilon^{\tilde {\cal R}}_{-{\bm k}+{\bm q}}+e^{{\cal T}^0}_{\bm q},
\end{align}
where $e^{{\cal T}^0}_{\bm q}$ denotes the excitation energy of the spin mode at the momentum of ${\bm q}$ [Figs. \ref{fig:Hub1d}(b), \ref{fig:Hub1d}(e), \ref{fig:Hub2d}(b), and \ref{fig:Hub2d}(c)], 
and the dispersion relations of the electron-addition and electron-removal modes from the unperturbed ground state $|{\rm GS}\rangle$ are represented as 
$\omega=\varepsilon^{\tilde {\cal A}}_{\bm k}$ and $\omega=-\varepsilon^{\tilde {\cal R}}_{-{\bm k}}$, respectively [Figs. \ref{fig:Hub1d}(a), \ref{fig:Hub1d}(d), and \ref{fig:Hub2d}(a)]. 
\par
The numerical results for the spectral function of the spin-perturbed state corresponding to $\prod_i S^{c,z}_{{\bm q}_i}|{\rm GS}\rangle$ for ${\bm q}_i=\frac{\bm \pi}{2}$ 
show that electronic modes are induced by the spin fluctuation [Figs. \ref{fig:KzHub1d2d}(a) and \ref{fig:KzHub1d2d}(c)]. 
\par
The emergent electronic modes in the 1D HM [Fig. \ref{fig:KzHub1d2d}(a)] can be identified as the holon modes shifted by the spin-excitation energy ($\mp e^{{\cal T}^0}_{\frac{\pi}{2}}$) and momentum ($\frac{\pi}{2}$) 
[dotted brown curves in Fig. \ref{fig:KzHub1d2d}(b)], consistent with Eqs. (\ref{eq:ekAHub1d2d}) and (\ref{eq:ekRHub1d2d}). 
\par
The emergent electronic modes in the 2D HM can be identified as the dominant modes around the bottom of the upper band and top of the lower band shifted downward and upward, respectively, 
by the spin-excitation energy ($\mp e^{{\cal T}^0}_{\frac{\bm \pi}{2}}$) and momentum ($\frac{\bm \pi}{2}$), 
as in Eqs. (\ref{eq:ekAHub1d2d}) and (\ref{eq:ekRHub1d2d}) [Fig. \ref{fig:KzHub1d2d}(c)]. 
The magenta and red arrows in Fig. \ref{fig:KzHub1d2d}(c) indicate the $\omega$ values of 
the bottom of the electron-addition emergent mode expected from Eq. (\ref{eq:ekAHub1d2d}) at ${\bm k}={\bm 0}$, 
$\omega=\varepsilon^{\tilde {\cal A}}_{\frac{\bm \pi}{2}}-e^{{\cal T}^0}_{\frac{\bm \pi}{2}}\overset{\text{Eq. (\ref{eq:mu+})}}{=}-e^{{\cal T}^0}_{\frac{\bm \pi}{2}}+\mu_+$, and 
top of the electron-removal emergent mode expected from Eq. (\ref{eq:ekRHub1d2d}) at ${\bm k}={\bm \pi}$, 
$\omega=-\varepsilon^{\tilde {\cal R}}_{-\frac{\bm \pi}{2}}+e^{{\cal T}^0}_{\frac{\bm \pi}{2}}\overset{\text{Eq. (\ref{eq:mu-})}}{=}e^{{\cal T}^0}_{\frac{\bm \pi}{2}}+\mu_-$, respectively. 
\section{Differences from conventional band insulators} 
\label{sec:bandInsulator}
\subsection{Outline} 
\label{sec:outlineBandinsulators}
To elucidate the unconventional nature of the electronic modes induced by spin and charge perturbations in strongly correlated insulators, 
the spectral features of conventional band insulators are reviewed in Sec. \ref{sec:U0HMPAM} for comparison. 
Since the emergence of these electronic modes reflects spin-charge separation in strongly correlated insulators, 
Sec. \ref{sec:SCsep1Dmetal} first reviews spin-charge separation in interacting metals on a chain and points out that it is not well-defined in terms of the spectral-weight distribution of electronic excitation. 
Section \ref{sec:SCinsulator} then explains spin-charge separation in strongly correlated insulators. 
\par
In Sec. \ref{sec:relationDISSCsep}, the relation between doping-induced in-gap electronic modes and spin-charge separation of strongly correlated insulators is clarified. 
Section \ref{sec:relationMISSCsep} discusses the relation between magnetization-induced in-gap electronic modes and spin-charge separation. 
\par
Implications of this study are described in Sec. \ref{sec:implications}. 
\subsection{Conventional band insulators} 
\label{sec:U0HMPAM}
\begin{figure} 
\includegraphics[width=\linewidth]{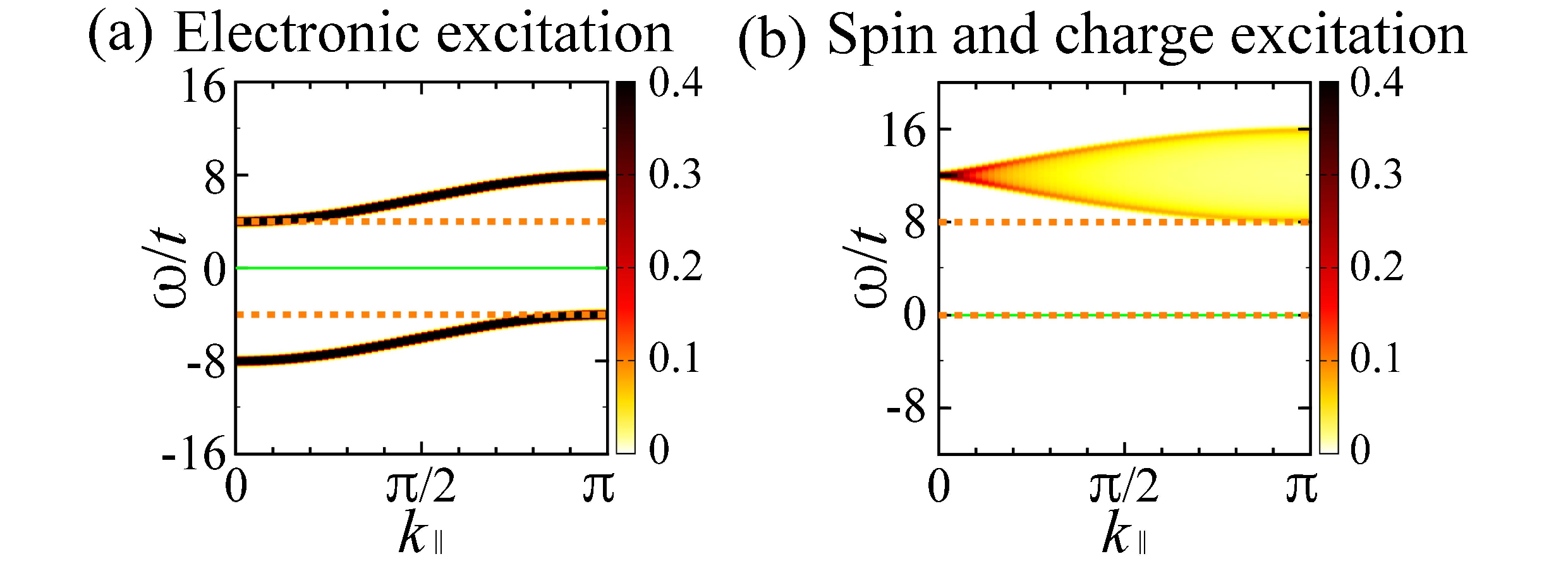}
\caption{Electronic, spin, and charge excitations of the noninteracting ($U=0$) ladder HM for $t_{\perp}/t=6$ at $\mu=H=0$. 
(a) $A_X(k_{\parallel},0,\omega)t+A_X(k_{\parallel},\pi,\omega)t$ for $|X\rangle$ of any eigenstate of the Hamiltonian. 
The dotted orange lines indicate the bottom of the upper band and top of the lower band; the band gap is the region between the dotted orange lines. 
(b) $S(k_{\parallel},\pi,\omega)t/3$ and $N(k_{\parallel},\pi,\omega)t/4$. 
The dotted orange lines indicate the lowest spin- and charge-excitation energies and $\omega=0$; the lowest spin- and charge-excitation energies are the same as the band gap. 
The solid green lines indicate $\omega=0$.}
\label{fig:U0}
\end{figure}
In the noninteracting ($U=0$) HM and PAM, the Hamiltonians [Eqs. (\ref{eq:HamHub}) and (\ref{eq:HamPAM})] can be diagonalized with electronic-quasiparticle operators in the momentum space as 
\begin{align}
\label{eq:HamHubU0}
&{\cal H}_{\rm HM}|_{U=0}=\sum_{{\bm k},\sigma}(E_{\bm k}-\mu-s^zH)c^{\dagger}_{{\bm k},\sigma}c_{{\bm k},\sigma},\\
\label{eq:HamPAMU0}
&{\cal H}_{\rm PAM}|_{U=0}=\sum_{{\bm k},\sigma}\sum_{\zeta=\pm}(E^{\zeta}_{\bm k}-\mu-s^zH)\alpha^{\zeta\dagger}_{{\bm k},\sigma}\alpha^{\zeta}_{{\bm k},\sigma},
\end{align}
where
\begin{align}
&E_{\bm k}=\left\{\begin{array}{ll}
\varepsilon_{\bm k}&\text{for the 1D and 2D HMs},\\
\varepsilon_{{\bm k}_{\parallel}}-t_{\perp}\cos k_{\perp}&\text{for the ladder and bilayer HMs},\end{array}\right.\\
&E^{\pm}_{\bm k}=\frac{\varepsilon_{\bm k}-\Delta}{2}\pm\frac{\sqrt{(\varepsilon_{\bm k}+\Delta)^2+4t_{\rm K}^2}}{2},\\
&\varepsilon_{\bm k}=-2dt\gamma_{\bm k},\\
&\alpha^{\pm}_{{\bm k},\sigma}=u^{\pm}_{\bm k}c_{{\bm k},\sigma}\mp{\rm sgn}(t_{\rm K})u^{\mp}_{\bm k}f_{{\bm k},\sigma},\\
&u^{\pm}_{\bm k}=\sqrt{\frac{1}{2}\left(1\pm\frac{\varepsilon_{\bm k}+\Delta}{\sqrt{(\varepsilon_{\bm k}+\Delta)^2+4t_{\rm K}^2}}\right)},
\end{align}
and $s^z=\frac{1}{2}$ and $-\frac{1}{2}$ for $\sigma=\uparrow$ and $\downarrow$, respectively. 
The ground states for $\delta\ge 0$ are expressed as 
\begin{align}
\label{eq:GSU0HM}
&|{\rm GS}\rangle=\prod_{\sigma}\prod_{{\bm k}\in{\rm FS}}c^{\dagger}_{{\bm k},\sigma}|0\rangle,\\
\label{eq:GSU0PAM}
&|{\rm GS}\rangle=\prod_{\sigma}\prod_{{\bm k}\in{\rm FS}}\alpha^{-\dagger}_{{\bm k},\sigma}|0\rangle
\end{align}
for the noninteracting ($U=0$) HM and PAM, respectively, where $\prod_{{\bm k}\in{\rm FS}}$ indicates the product over the momenta ${\bm k}$ within the Fermi sea. 
\par
The band structure [Fig. \ref{fig:U0}(a)] does not change with the chemical potential $\mu$; 
only the Fermi level ($\omega=0$) shifts [Eqs. (\ref{eq:HamHubU0}) and (\ref{eq:HamPAMU0})], even in doped systems. 
Similarly, the band structure for $\uparrow$-spin and $\downarrow$-spin electrons does not change with the magnetic field $H$; 
only the Fermi levels for $\uparrow$-spin and $\downarrow$-spin electrons shift by $\frac{H}{2}$ and $-\frac{H}{2}$, respectively [Eqs. (\ref{eq:HamHubU0}) and (\ref{eq:HamPAMU0})], even in magnetized systems. 
\par
Thus, doping-induced states and magnetization-induced states do not appear in band insulators where the interaction between electrons can be neglected, 
which is in contrast to the case of strongly correlated insulators (Secs. \ref{sec:DIS}, \ref{sec:DISspingap}, \ref{sec:MIS}, and \ref{sec:MISspingap}; 
Figs. \ref{fig:etaHub1d}--\ref{fig:etaKLM}, \ref{fig:SzLadBilPAM}, and \ref{fig:SzKLM}). 
\par
This implies that the emergence of the electronic modes caused by doping and magnetization within the band gap represents a remarkable strong-correlation effect, which does not occur in band insulators. 
The differences originate from spin-charge separation of strongly correlated insulators (Sec. \ref{sec:SCinsulator}) 
\cite{KohnoMottT,KohnoTinduced,KohnoRPP,Kohno1DHub,Kohno2DHub,Kohno1DtJ,Kohno2DtJ,KohnoDIS,KohnoHubLadder,KohnoSpin,KohnoAF,KohnoGW,KohnoKLM,KohnottpHub,KohnottpJ}. 
\subsection{Spin-charge separation of interacting metals on a chain} 
\label{sec:SCsep1Dmetal}
Spin-charge separation means that the lowest spin-excitation energy differs from the lowest charge-excitation energy. 
In interacting metals on a chain, it is known that the velocity of spin excitation generally differs from that of charge excitation; 
spin-charge separation occurs in the energy scale of $\mathcal{O}(\frac{1}{L})$, 
where $L$ denotes the number of sites on a chain \cite{Essler,HaldaneTLL,TomonagaTLL,LuttingerTLL,MattisLiebTLL,Giamarchi}. 
This characteristic is contrasted with that of a Fermi liquid \cite{LandauFL}, where spin and charge excitations are obtained as particle-hole excitations of electronic quasiparticles. 
In a Fermi liquid, the lowest spin-excitation energy is the same as the lowest charge-excitation energy, and spin-charge separation does not occur. 
\par
In the hole-doped 1D HM, the spectral function has two dominant gapless modes for $\omega<0$ [Figs. \ref{fig:etaHub1d}(a) and \ref{fig:etaHub1d}(d)]. 
The lower-$|\omega|$ and higher-$|\omega|$ gapless modes for $\omega<0$ can be identified as the spinon and holon modes [blue and brown curves in Fig. \ref{fig:etaHub1d}(d)], respectively. 
\par
In the low-energy effective theory (bosonization) \cite{Essler,HaldaneTLL,TomonagaTLL,LuttingerTLL,MattisLiebTLL,Giamarchi}, 
the dispersion relations for $\omega>0$ and $\omega<0$ are symmetric with respect to $\omega=0$ and $k=k_{\rm F}$ (Fermi momentum). 
This is correct in the $|\omega|\rightarrow 0$ limit, which can be verified using the Bethe ansatz as follows: 
The spinon and holon (antiholon) modes smoothly cross the Fermi level ($\omega=0$). 
The slope (velocity) is the same between $\omega\rightarrow -0$ and $\omega\rightarrow +0$ in the $|\omega|$ regime of $\mathcal{O}(\frac{1}{L})$ for both the spinon and holon (antiholon) modes. 
Hence, the eigenstates of the Hamiltonian for the spinon and antiholon modes exist for $\omega>0$, which correspond to those of the spinon and holon modes for $\omega<0$, respectively, 
in the lower band of a hole-doped system; this is consistent with the low-energy effective theory. 
\par
However, for $\omega>0$ in the lower band of a hole-doped system, a significant amount of spectral weight is concentrated along the upper edge of the spinon-antiholon continuum \cite{Kohno1DHub} 
rather than along the spinon and antiholon modes, 
and a single dominant gapless mode exists, i.e., the doping-induced mode [Fig. \ref{fig:etaHub1d}(a); red curve in Fig. \ref{fig:etaHub1d}(d)]. 
The spectral-weight distribution does not exhibit the band structure of two dominant gapless modes for $\omega>0$ in the lower band although spin-charge separation occurs. 
This shows that spin-charge separation can occur even if the spectral-weight distribution of electronic excitation exhibits the band structure consisting primarily of a single dominant gapless mode as in a noninteracting metal or Fermi liquid. 
\par
In addition, the spectral-weight distribution continuously changes with increasing interchain hopping towards higher-dimensional systems; 
in the Fermi-liquid theory, the higher-$|\omega|$ mode should be interpreted as a contribution of the incoherent part. 
In metals, the spin and charge degrees of freedom are generally coupled with each other in the $\omega$ regime away from the $\omega\rightarrow 0$ limit in any spatial dimension. 
This implies that even if the overall spectral-weight distribution seems to exhibit a band structure consisting primarily of two dominant gapless modes in a metal, 
the occurrence of spin-charge separation remains undetermined unless the nature of the excitations in the $\omega\rightarrow 0$ limit (i.e., spin and charge velocities) is clarified. 
\par
Thus, spin-charge separation is not well-defined in terms of the spectral-weight distribution 
but is well-defined only through the excitations in the $\omega\rightarrow 0$ limit (i.e., spin and charge velocities) in a metal. 
\subsection{Spin-charge separation of strongly correlated insulators} 
\label{sec:SCinsulator}
Although spin-charge separation is difficult to verify in metals unless information regarding the excitations in the $\omega\rightarrow 0$ limit can be obtained, it is usually clear in strongly correlated insulators. 
\par
Insulators generally have a nonzero charge gap. 
In strongly correlated insulators, spin excitation exists in the energy regime lower than the charge gap. 
If the charge gap is much larger than the spin-excitation energies, the spin degrees of freedom can be effectively described using spin models such as the Heisenberg model \cite{Anderson}. 
According to its definition (Sec. \ref{sec:SCsep1Dmetal}), spin-charge separation generally occurs in strongly correlated insulators in any spatial dimension 
because the lowest charge-excitation energy (= charge gap) differs from the lowest spin-excitation energy. 
\par
In contrast, spin-charge separation does not occur in band insulators in which the interaction between electrons can be neglected. 
This is because spin and charge excitations are described as particle-hole excitations of noninteracting electrons, as in a Fermi liquid. 
The spin-excited state for $S^{c,+}_{\bm k}$ in a band insulator is obtained by removing a $\downarrow$-spin electron with a momentum ${\bm q}$ ($c_{{\bm q},\downarrow}$) from the filled lower band ($\omega\le\mu_-$) 
and adding an $\uparrow$-spin electron with the momentum of ${\bm k}+{\bm q}$ ($c^{\dagger}_{{\bm k}+{\bm q},\uparrow}$) to the empty upper band ($\omega\ge\mu_+$). 
The lowest-energy spin-excited state is obtained as the particle-hole excited state of 
\begin{equation}
\label{eq:PHS+}
|{\rm PH}\rangle_{S^+}=c^{\dagger}_{{\bm k}^+_{\rm F},\uparrow}c_{{\bm k}^-_{\rm F},\downarrow}|{\rm GS}\rangle
\end{equation}
in band insulators, such as the noninteracting ($U=0$) ladder and bilayer HMs at half filling. 
The lowest spin-excitation energy is equal to the band gap ($\mu_+-\mu_-$), which is equal to the charge gap (Fig. \ref{fig:U0}); 
spin-charge separation (Sec. \ref{sec:SCsep1Dmetal}) does not occur in noninteracting systems. 
\par
Thus, spin-charge separation is a general characteristic of strongly correlated insulators in any spatial dimension and is a strong-correlation effect that does not occur in band insulators. 
\subsection{Relation between doping-induced in-gap electronic modes and spin-charge separation of strongly correlated insulators} 
\label{sec:relationDISSCsep}
According to the quantum-number analysis (selection rules) for doping-induced modes in Sec. \ref{sec:DIS}, 
the dispersion relations of the doping-induced in-gap modes are expressed [Eqs. (\ref{eq:ekhDISspin}) and (\ref{eq:ekeDISspin})] as 
\begin{align}
\label{eq:ekhDISspin2}
&\omega=e^{\rm spin}_{{\bm k}-{\bm k}_{\rm F}^-}+\mu_-,\\
\label{eq:ekeDISspin2}
&\omega=-e^{\rm spin}_{-{\bm k}+{\bm k}_{\rm F}^+}+\mu_+,
\end{align}
which implies that the emergent in-gap electronic modes exhibit the spin-excitation dispersion relation ($\pm e^{\rm spin}_{\pm{\bm k}}$) 
shifted by the Fermi momentum ($\mp{\bm k}_{\rm F}^{\mp}$) from the band edges ($\mu_{\mp}$). 
\par
If the lowest spin-excitation energy, $e^{\rm spin}_{\bm q}$, is lower than the charge gap, $\mu_+-\mu_-$, i.e., 
\begin{equation}
\label{eq:ekSCsep}
0<e^{\rm spin}_{\bm q}<\mu_+-\mu_-,
\end{equation}
the lowest-energy spin-excited state emerges as an electronic excited state [Eqs. (\ref{eq:ekhDISspin2}) and (\ref{eq:ekeDISspin2})] within the band gap, as follows: 
\begin{align}
&\mu_-<e^{\rm spin}_{\bm q}+\mu_-<\mu_+,\\
&\mu_-<-e^{\rm spin}_{\bm q}+\mu_+<\mu_+,
\end{align}
which are obtained by rewriting Eq. (\ref{eq:ekSCsep}). Thus, electronic modes can emerge within the band gap if spin-charge separation [Eq. (\ref{eq:ekSCsep})] occurs 
\cite{KohnoMottT,KohnoTinduced,KohnoDIS,KohnoHubLadder,KohnoGW,KohnoKLM}. 
\par
In the case of gapless spin excitation ($e^{\rm spin}_{\bm q}\rightarrow 0$), because $e^{\rm spin}_{{\bm k}-{\bm k}_{\rm F}^-}$ and $e^{\rm spin}_{-{\bm k}+{\bm k}_{\rm F}^+}\ge e^{\rm spin}_{\bm q}$, 
gapless electronic modes emerge in the band gap from the band edges 
($\omega=e^{\rm spin}_{{\bm k}-{\bm k}_{\rm F}^-}+\mu_-\ge e^{\rm spin}_{\bm q}+\mu_-\rightarrow\mu_-$ and 
$\omega=-e^{\rm spin}_{-{\bm k}+{\bm k}_{\rm F}^+}+\mu_+\le -e^{\rm spin}_{\bm q}+\mu_+\rightarrow\mu_+$) [Eqs. (\ref{eq:ekhDISspin2}) and (\ref{eq:ekeDISspin2})] 
\cite{KohnoMottT,KohnoRPP,Kohno1DHub,Kohno2DHub,Kohno1DtJ,Kohno2DtJ,KohnoDIS,KohnoSpin,KohnoAF,KohnoGW,KohnottpHub,KohnottpJ}. 
\par
On the other hand, if spin-charge separation does not occur, i.e., 
\begin{equation}
\label{eq:eknonSCsep}
e^{\rm spin}_{\bm q}=\mu_+-\mu_-,
\end{equation}
according to the selection rules, the lowest-energy spin-excited state can appear as an electronic excited state [Eqs. (\ref{eq:ekhDISspin2}) and (\ref{eq:ekeDISspin2})] at the band edges, as follows: 
\begin{align}
&\omega=e^{\rm spin}_{\bm q}+\mu_-\overset{\text{Eq. (\ref{eq:eknonSCsep})}}{=}\mu_+,\\
&\omega=-e^{\rm spin}_{\bm q}+\mu_+\overset{\text{Eq. (\ref{eq:eknonSCsep})}}{=}\mu_-.
\end{align}
Because $e^{\rm spin}_{{\bm k}-{\bm k}_{\rm F}^-}$ and $e^{\rm spin}_{-{\bm k}+{\bm k}_{\rm F}^+}\ge e^{\rm spin}_{\bm q}$, 
the emergent electronic modes [Eqs. (\ref{eq:ekhDISspin2}) and (\ref{eq:ekeDISspin2})] can appear outside the band gap 
($\omega=e^{\rm spin}_{{\bm k}-{\bm k}_{\rm F}^-}+\mu_-\ge e^{\rm spin}_{\bm q}+\mu_-=\mu_+$ and 
$\omega=-e^{\rm spin}_{-{\bm k}+{\bm k}_{\rm F}^+}+\mu_+\le -e^{\rm spin}_{\bm q}+\mu_+=\mu_-$); 
in the absence of spin-charge separation, the electronic modes do not emerge within the band gap after doping. 
\par
Thus, the doping-induced in-gap modes reflect spin-charge separation of strongly correlated insulators 
\cite{KohnoMottT,KohnoTinduced,KohnoRPP,Kohno1DHub,Kohno2DHub,Kohno1DtJ,Kohno2DtJ,KohnoDIS,KohnoHubLadder,KohnoSpin,KohnoAF,KohnoGW,KohnoKLM,KohnottpHub,KohnottpJ}; 
these do not appear in conventional band insulators without spin-charge separation. 
\par
The emergence of in-gap modes in the electronic excitation from the lowest-energy state of a nonzero value of $\eta$ at half filling ($\eta^z=0$) 
[Sec. \ref{sec:etaz0DIS}; Figs. \ref{fig:etaHub1d}(c), \ref{fig:etaHub1d}(f), \ref{fig:etaHub2d}(c), \ref{fig:etaLadBilPAM}(c), \ref{fig:etaLadBilPAM}(f), \ref{fig:etaLadBilPAM}(i), \ref{fig:etaLadBilPAM}(l), 
\ref{fig:etaLadBilPAM}(o), \ref{fig:etaLadBilPAM}(r), \ref{fig:etaLadBilPAM}(u), \ref{fig:etaLadBilPAM}(x), \ref{fig:etaKLM}(c), \ref{fig:etaKLM}(f), \ref{fig:etaKLM}(i), and \ref{fig:etaKLM}(l)] also reflects spin-charge separation. 
\subsection{Relation between magnetization-induced in-gap electronic modes and spin-charge separation of strongly correlated insulators} 
\label{sec:relationMISSCsep}
\subsubsection{Emergence of magnetization-induced modes within the band gap in spin-gapped Mott and Kondo insulators} 
\label{sec:ingapMISspingap}
Similarly, according to the quantum-number analysis (selection rules) for magnetization-induced modes in Sec. \ref{sec:MIS}, 
the dispersion relations of the magnetization-induced in-gap modes are expressed [Eqs. (\ref{eq:ekeMIS}) and (\ref{eq:ekhMIS})] as 
\begin{align}
\label{eq:ekeMIS2}
&\omega=\varepsilon^{{\cal A}_{\sigma}}_{{\bm k}+{\bm q}}-e^{{\cal T}^{2s^z}}_{\bm q}\quad\text{in}\quad A^{\bar \sigma}_{{\cal T}^{2s^z}_{\bm q}}({\bm k},\omega),\\
\label{eq:ekhMIS2}
&\omega=-\varepsilon^{{\cal R}_{\sigma}}_{-{\bm k}+{\bm q}}+e^{{\cal T}^{2s^z}}_{\bm q}\quad\text{in}\quad A^{\sigma}_{{\cal T}^{2s^z}_{\bm q}}({\bm k},\omega),
\end{align}
where $\varepsilon^{{\cal A}_{\sigma}}_{\bm k}$ and $\varepsilon^{{\cal R}_{\sigma}}_{-{\bm k}}$ 
represent the excitation energies of the low-$|\omega|$ electron-addition and electron-removal modes of the conventional bands under the magnetic field, respectively, from the unperturbed ground state of $S^z=0$ at half filling. 
Equations (\ref{eq:ekeMIS2}) and (\ref{eq:ekhMIS2}) imply that the emergent electronic modes exhibit 
the conventional electronic dispersion relations ($\varepsilon^{{\cal A}_{\sigma}}_{\bm k}$ and $-\varepsilon^{{\cal R}_{\sigma}}_{-{\bm k}}$) 
shifted by the excitation energy ($-e^{{\cal T}^{2s^z}}_{\bm q}$ and $e^{{\cal T}^{2s^z}}_{\bm q}$) and momentum (${\bm q}$) of the lowest-energy spin-excited state. 
\par
Equations (\ref{eq:ekeMIS2}) and (\ref{eq:ekhMIS2}) for $\sigma=\uparrow$ can be rewritten using the excitation energies at $H=0$ 
(${\bar \varepsilon}^{\cal A}_{{\bm k}+{\bm q}}$, ${\bar \varepsilon}^{\cal R}_{-{\bm k}+{\bm q}}$, and $\Delta_{\rm spin}$ [Eqs. (\ref{eq:ebarkA2})--(\ref{eq:HDelta2})]) [Eqs. (\ref{eq:ekAdTuH})--(\ref{eq:HDelta})] as 
\begin{align}
\label{eq:ekAdTuH2}
&\omega={\bar \varepsilon}^{\cal A}_{{\bm k}+{\bm q}}-\frac{\Delta_{\rm spin}}{2}\quad\text{in}\quad A^{\downarrow}_{{\cal T}^{1}_{\bm q}}({\bm k},\omega),\\
\label{eq:ekRuTuH2}
&\omega=-{\bar \varepsilon}^{\cal R}_{-{\bm k}+{\bm q}}+\frac{\Delta_{\rm spin}}{2}\quad\text{in}\quad A^{\uparrow}_{{\cal T}^{1}_{\bm q}}({\bm k},\omega),
\end{align}
where 
\begin{align}
\label{eq:ebarkA2}
&\varepsilon^{{\cal A}_{\sigma}}_{{\bm k}+{\bm q}}={\bar \varepsilon}^{\cal A}_{{\bm k}+{\bm q}}-s^zH,\\
\label{eq:ebarkR2}
&\varepsilon^{{\cal R}_{\sigma}}_{-{\bm k}+{\bm q}}={\bar \varepsilon}^{\cal R}_{-{\bm k}+{\bm q}}-s^zH,\\
\label{eq:ebarkT2}
&e^{{\cal T}^{2s^z}}_{\bm q}=\Delta_{\rm spin}-2s^zH,\\
\label{eq:HDelta2}
&H\rightarrow\Delta_{\rm spin}\quad\text{for}\quad S^z/S_{\rm max}\rightarrow+0.
\end{align}
The spin gap at $H=0$ is denoted by $\Delta_{\rm spin}$. 
\par
If the lowest spin-excitation energy, $\Delta_{\rm spin}$, is lower than the charge gap, $\mu_+-\mu_-$, at $H=0$, i.e., 
\begin{equation}
\label{eq:spingapSCsep}
0<\Delta_{\rm spin}<\mu_+-\mu_-,
\end{equation}
the lower-$|\omega|$ band edges of the emergent electronic modes (${\bar \varepsilon}^{\cal A}_{{\bm k}_{\rm F}^+}-\frac{\Delta_{\rm spin}}{2}$ [Eq. (\ref{eq:ekAdTuH2}) at ${\bm k}+{\bm q}={\bm k}_{\rm F}^+$] and 
$-{\bar \varepsilon}^{\cal R}_{-{\bm k}_{\rm F}^-}+\frac{\Delta_{\rm spin}}{2}$ [Eq. (\ref{eq:ekRuTuH2}) at $-{\bm k}+{\bm q}=-{\bm k}_{\rm F}^-$]) 
are located within the band gap [($\mu_--s^zH$)--($\mu_+-s^zH$) for $\sigma$-spin electrons], as follows: 
\begin{align}
&\mu_-+\frac{H}{2}<{\bar \varepsilon}^{\cal A}_{{\bm k}_{\rm F}^+}-\frac{\Delta_{\rm spin}}{2}<\mu_++\frac{H}{2}\quad\text{in}\quad A^{\downarrow}_{{\cal T}^{1}_{\bm q}}({\bm k},\omega),\\
&\mu_--\frac{H}{2}<-{\bar \varepsilon}^{\cal R}_{-{\bm k}_{\rm F}^-}+\frac{\Delta_{\rm spin}}{2}<\mu_+-\frac{H}{2}\quad\text{in}\quad A^{\uparrow}_{{\cal T}^{1}_{\bm q}}({\bm k},\omega),
\end{align}
which are obtained by rewriting Eq. (\ref{eq:spingapSCsep}) using $H\rightarrow\Delta_{\rm spin}$, ${\bar \varepsilon}^{\cal A}_{{\bm k}_{\rm F}^+}=\mu_+$, and $-{\bar \varepsilon}^{\cal R}_{-{\bm k}_{\rm F}^-}=\mu_-$. 
\par
Thus, electronic modes can emerge within the band gap by magnetizing spin-gapped insulators 
if the lowest spin-excitation energy is lower than the charge gap [Eq. (\ref{eq:spingapSCsep})] (spin-charge separation occurs). 
\par
In the case of gapless spin excitation, as mentioned in Sec. \ref{sec:noMISgaplessSpin} (Fig. \ref{fig:SzHub1d}), 
the magnetization-induced modes do not emerge inside the band gap even though spin-charge separation occurs. 
This is because the emergent modes are essentially degenerate in $\omega$ with the conventional bands 
[Eqs. (\ref{eq:ekeMIS2}) and (\ref{eq:ekhMIS2}) with $e^{{\cal T}^{\pm1}}_{\bm q}\rightarrow0$ and $H\rightarrow\pm0$ for $S^z/S_{\rm max}\rightarrow\pm0$]. 
\subsubsection{Absence of magnetization-induced modes within the band gap in insulators without spin-charge separation} 
\label{sec:ingapMISnoSCsep}
If spin-charge separation does not occur, i.e., 
\begin{equation}
\label{eq:spingapnoSCsep}
\Delta_{\rm spin}=\mu_+-\mu_-,
\end{equation}
the bottom of the emergent upper band for $\downarrow$-spin electrons (${\bar \varepsilon}^{\cal A}_{{\bm k}_{\rm F}^+}-\frac{\Delta_{\rm spin}}{2}$ [Eq. (\ref{eq:ekAdTuH2}) at ${\bm k}+{\bm q}={\bm k}_{\rm F}^+$]) and 
top of the emergent lower band for $\uparrow$-spin electrons ($-{\bar \varepsilon}^{\cal R}_{-{\bm k}_{\rm F}^-}+\frac{\Delta_{\rm spin}}{2}$ [Eq. (\ref{eq:ekRuTuH2}) at $-{\bm k}+{\bm q}=-{\bm k}_{\rm F}^-$]) 
are located at the top of the conventional lower band for $\downarrow$-spin electrons ($\mu_-+\frac{H}{2}$) and bottom of the conventional upper band for $\uparrow$-spin electrons ($\mu_+-\frac{H}{2}$), respectively, as follows: 
\begin{align}
&{\bar \varepsilon}^{\cal A}_{{\bm k}_{\rm F}^+}-\frac{\Delta_{\rm spin}}{2}\overset{\text{Eq. (\ref{eq:spingapnoSCsep})}}{=}\mu_-+\frac{H}{2}\quad\text{in}\quad A^{\downarrow}_{{\cal T}^{1}_{\bm q}}({\bm k},\omega),\\
&-{\bar \varepsilon}^{\cal R}_{-{\bm k}_{\rm F}^-}+\frac{\Delta_{\rm spin}}{2}\overset{\text{Eq. (\ref{eq:spingapnoSCsep})}}{=}\mu_+-\frac{H}{2}\quad\text{in}\quad A^{\uparrow}_{{\cal T}^{1}_{\bm q}}({\bm k},\omega)
\end{align}
using $H\rightarrow\Delta_{\rm spin}$, ${\bar \varepsilon}^{\cal A}_{{\bm k}_{\rm F}^+}=\mu_+$, and $-{\bar \varepsilon}^{\cal R}_{-{\bm k}_{\rm F}^-}=\mu_-$. 
\par
Because ${\bar \varepsilon}^{\cal A}_{{\bm k}+{\bm q}}\ge{\bar \varepsilon}^{\cal A}_{{\bm k}_{\rm F}^+}$ and 
${\bar \varepsilon}^{\cal R}_{-{\bm k}+{\bm q}}\ge{\bar \varepsilon}^{\cal R}_{-{\bm k}_{\rm F}^-}$, 
the emergent electronic modes [Eqs. (\ref{eq:ekAdTuH2}) and (\ref{eq:ekRuTuH2})] can appear within the band gap 
[$\omega={\bar \varepsilon}^{\cal A}_{{\bm k}+{\bm q}}-\frac{\Delta_{\rm spin}}{2}\ge \mu_-+\frac{H}{2}$ in $A^{\downarrow}_{{\cal T}^{1}_{\bm q}}({\bm k},\omega)$ and 
$\omega=-{\bar \varepsilon}^{\cal R}_{-{\bm k}+{\bm q}}+\frac{\Delta_{\rm spin}}{2}\le \mu_+-\frac{H}{2}$ in $A^{\uparrow}_{{\cal T}^{1}_{\bm q}}({\bm k},\omega)$] 
from the viewpoint of the selection rules. 
\par
However, if spin-charge separation does not occur, the lowest-energy spin-excited state and lowest-energy charge-excited state are 
the particle-hole-excited states at the bottom of the particle-hole continuum [Fig. \ref{fig:U0}(b)]. 
The lowest-energy spin-excited state of $S^z=1$ is expressed as $|{\rm PH}\rangle_{S^+}$ of Eq. (\ref{eq:PHS+}). 
The electronic states of the emergent modes obtained by adding and removing an electron on $|{\rm PH}\rangle_{S^+}$ are as follows: 
\begin{align}
\label{eq:cdaggerPH}
&c^{\dagger}_{{\bm k}^-_{\rm F},\downarrow}|{\rm PH}\rangle_{S^+}=-c^{\dagger}_{{\bm k}^+_{\rm F},\uparrow}|{\rm GS}\rangle,\\
\label{eq:cPH}
&c_{{\bm k}^+_{\rm F},\uparrow}|{\rm PH}\rangle_{S^+}=c_{{\bm k}^-_{\rm F},\downarrow}|{\rm GS}\rangle,
\end{align}
and $c^{(\dagger)}_{{\bm k},\sigma}|{\rm PH}\rangle_{S^+}$ of the other momenta or spin contribute to the same band structure as that in the unperturbed case 
[similar to the first terms on the right-hand sides of Eqs. (\ref{eq:adaggerX}) and (\ref{eq:aX}); Sec. \ref{sec:eleEff}]. 
\par
The momentum and spin of the added and removed electrons in Eqs. (\ref{eq:cdaggerPH}) and (\ref{eq:cPH}) are 
the same as those of the removed and added electrons at the band edges for the lowest-energy particle-hole excitation, respectively [Eq. (\ref{eq:PHS+})]; 
the emergent electronic states in Eqs. (\ref{eq:cdaggerPH}) and (\ref{eq:cPH}) also contribute to the same band structure as that in the unperturbed case at the band edges. 
Thus, if the lowest-energy spin-excited state is a particle-hole excited state, electronic modes are not considered to emerge with magnetization. 
\subsubsection{Spin-mode formation as a manifestation of spin-charge separation in spin-gapped Mott and Kondo insulators} 
\label{sec:spinmodeSCsep}
From the above consideration, the formation of the spin mode (triplon mode) separated from the particle-hole continuum can be considered the key characteristic for the emergence of electronic modes with magnetization. 
As shown in Figs. \ref{fig:SkwLadBil} and \ref{fig:charge}, the spin mode (triplon mode) appears below the particle-hole continuum, and the charge mode appears above the particle-hole continuum in spin-gapped Mott and Kondo insulators. 
Because the lowest excitation energy of the spin mode is lower than that of the particle-hole continuum, which is equal to the lowest charge-excitation energy, 
the formation of the spin mode below the particle-hole continuum is a manifestation of spin-charge separation in spin-gapped Mott and Kondo insulators. 
\par
Thus, the emergence of magnetization-induced modes within the band gap reflects spin-charge separation, which is realized as 
the formation of the spin mode below the particle-hole continuum, in spin-gapped Mott and Kondo insulators. 
\par
The emergence of in-gap modes in the electronic excitation from the lowest-energy state of a nonzero value of $S$ without magnetization at half filling 
[Sec. \ref{sec:Sz0MIS}; Figs. \ref{fig:SzLadBilPAM}(c), \ref{fig:SzLadBilPAM}(f), \ref{fig:SzLadBilPAM}(i), \ref{fig:SzLadBilPAM}(l), 
\ref{fig:SzLadBilPAM}(o), \ref{fig:SzLadBilPAM}(r), \ref{fig:SzLadBilPAM}(u), \ref{fig:SzLadBilPAM}(x), \ref{fig:SzKLM}(c), \ref{fig:SzKLM}(f), \ref{fig:SzKLM}(i), and \ref{fig:SzKLM}(l)] also reflects spin-charge separation. 
\subsubsection{Strongly correlated insulators of attractively interacting electrons} 
\label{sec:attInsulators}
In the insulators of attractively interacting electrons ($U<0$) which are obtained by applying the Shiba transformation (Sec. \ref{sec:AkwM}) to spin-gapped Mott and Kondo insulators with repulsive interactions ($U>0$), 
because the roles of spin and charge are interchanged by the Shiba transformation [Eqs. (\ref{eq:defSetai}) and (\ref{eq:SetaShiba})], 
the charge mode appears below the particle-hole continuum. 
In this case, the doping-induced modes appear within the band gap, reflecting spin-charge separation 
(the lowest charge-excitation energy is lower than the lowest spin-excitation energy, which is opposite to the repulsive case). 
\par
Similarly, the results for the doping-induced states in repulsive-interaction systems ($U>0$) (Sec. \ref{sec:chargePerturbed}) 
\cite{KohnoMottT,KohnoTinduced,KohnoRPP,Kohno1DHub,Kohno2DHub,KohnoHubLadder,KohnoSpin,KohnoAF,KohnoKLM} can be used straightforwardly 
for the magnetization-induced states in attractive-interaction systems ($U<0$) via the Shiba transformation (Sec. \ref{sec:AkwM}). 
\subsection{Implications} 
\label{sec:implications}
Although various interpretations of electronic states in strongly correlated systems have been considered 
\cite{KohnoMottT,KohnoTinduced,KohnoRPP,Kohno1DHub,Kohno2DHub,Kohno1DtJ,Kohno2DtJ,KohnoDIS,KohnoHubLadder,KohnoSpin,KohnoAF,KohnoGW,KohnoKLM,KohnottpHub,KohnottpJ,
Eskes,DagottoDOS,CPTPRL,CPTPRB,CPTHanke,TsunetsuguRMP,localRungHubLadder,
lightPump1dHub,pumpProbe1dEHM,photoSubbands,FKFloquetDMFT,noneqDMFTRMP,DCAkw1dHub,entropyCoolingDMFT,etaPairingSuper,nonthermalSuper,KanekoEtaPairing,etaPairingHubLad,
Essler,HaldaneTLL,TomonagaTLL,LuttingerTLL,MattisLiebTLL,Giamarchi,DDMRGAkw,
HubbardI,LandauFL,BrinkmanRice,SlaterAF,PennAF,
ImadaRMP,SakaiImadaPRL,SakaiImadaPRB,SakaiJPSJ,ImadaCofermionPRL,ImadaCofermionPRB,PhillipsRMP,PhillipsRPP,EderOhta2DHub,EderOhtaIPES,
PreussQP,GroberQMC,NoceraDMRT_T,Akw2dtJ_T}, 
the view presented in this paper along with 
Refs. [\onlinecite{KohnoTinduced,KohnoMottT,KohnoRPP,Kohno1DHub,Kohno2DHub,Kohno1DtJ,Kohno2DtJ,KohnoDIS,KohnoHubLadder,KohnoSpin,KohnoAF,KohnoGW,KohnoKLM,
KohnottpHub,KohnottpJ}] can consistently explain unconventional features of the electronic states induced not only by temperature increase, doping, and magnetization but also 
by spin and charge perturbations that cause nonequilibrium states around Mott and Kondo insulators (Secs. \ref{sec:chargePerturbed} and \ref{sec:spinPerturbed}). 
Furthermore, this view can clarify the differences from the conventional features of the temperature-, doping-, and magnetization-independent band structure of noninteracting band insulators (Sec. \ref{sec:bandInsulator}). 
\par
According to this view, electronic modes, whose dispersion relations can be expressed as combinations of the electronic and either the spin or the charge dispersion relations of unperturbed systems, 
can be induced by various spin and charge perturbations as well as by temperature increase in strongly correlated insulators 
\cite{KohnoMottT,KohnoTinduced,KohnoRPP,Kohno1DHub,Kohno2DHub,KohnoDIS,Kohno1DtJ,Kohno2DtJ,KohnoSpin,KohnoAF,KohnoGW,KohnottpHub,KohnottpJ,KohnoHubLadder,KohnoKLM}. 
This characteristic can be applied to band-structure engineering, which makes it possible to increase the number of electronic bands, control the band gap, and 
even change the insulating band structure to a metallic one at half filling by appropriately tuning and selecting perturbations and strongly correlated materials. 
\par
Although the particle--hole-symmetric case has been considered an unperturbed system in this study, 
electronic modes should be induced by spin and charge perturbations even in the nonsymmetric case. 
In particular, the dispersion relations obtained through the quantum-number analysis (selection rules) 
[Secs. \ref{sec:DIS}, \ref{sec:selectionRulesCF}, \ref{sec:MIS}, and \ref{sec:selectionRulesSF}; 
Eqs. (\ref{eq:ekhDISspin}), (\ref{eq:ekhDISchargeNh0}), (\ref{eq:ekhDISchargeNh2}), (\ref{eq:ekeDISspin}), (\ref{eq:ekeDISchargeNh0}), (\ref{eq:ekeDISchargeNh-2}), (\ref{eq:ekAN}), (\ref{eq:ekRN}), 
(\ref{eq:ekeMIS}), (\ref{eq:ekhMIS}), (\ref{eq:ekAS}), and (\ref{eq:ekRS})] 
should remain valid, 
and the experimentally obtained spin- and charge-mode dispersion relations can also be used to interpret the emergent electronic modes in strongly correlated materials. 
\par
In this study, only the low-energy or dynamically dominant characteristic spin- and charge-excited states have been considered perturbed states. 
Nevertheless, the results would be useful for interpreting essential spectral features, which are represented by the dominant part of the spectral-weight distribution, for more complicated nonequilibrium or perturbed states. 
\section{Summary} 
\label{sec:summary}
Building upon the understanding of doping- or temperature-induced electronic modes exhibiting momentum-shifted magnetic dispersion relations, 
this study generalizes the underlying mechanism and demonstrates that various spin and charge perturbations can induce electronic modes and alter the band structure in strongly correlated insulators, 
provided that a macroscopic number $\mathcal{O}(N_{\rm u})$ of spins or charges are excited. 
The origins and dispersion relations of these emergent electronic modes, particularly why and how the dispersion relations depend on the momentum and energy of the perturbations, 
are elucidated by investigating the selection rules and using the Bethe ansatz and the effective theory for weak inter-unit-cell hopping. 
The validity and generality of the theoretical results across different models and spatial dimensions are verified by numerical calculations for the 1D and 2D HMs, PAMs, KLMs, and ladder and bilayer HMs. 
\par
For charge perturbations, not only the chemical-potential shift that causes hole or electron doping 
but also the perturbation that generates the lowest-energy state of a nonzero value (${\bar \eta}$) of $\eta$ at half filling ($N_{\rm h}=0$) 
can induce electronic modes. The induced electronic modes at half filling exhibit the same dispersion relations as those of the doping-induced modes at $|N_{\rm h}|=2{\bar \eta}$ for $\mu=0$, 
reflecting $\eta$-SU(2) symmetry. 
In addition, the charge fluctuation that leads to effectively a single dominant state at a given momentum in the charge excitation can induce electronic modes. 
The charge perturbations at half filling can be realized by photon (light) radiation. 
\par
For spin perturbations, the magnetic field that magnetizes spin-gapped Mott and Kondo insulators can induce electronic modes within the band gap. 
The spin perturbation that generates the lowest-energy state of a nonzero value (${\bar S}$) of $S$ without magnetization ($S^z=0$) 
can also induce electronic modes, which exhibit the same dispersion relations as those of the magnetization-induced modes at $|S^z|={\bar S}$ for $H=0$, reflecting spin-SU(2) symmetry. 
In addition, the spin fluctuation that leads to effectively a single dominant state at a given momentum in the spin excitation can induce electronic modes. 
The spin perturbations without a magnetic field can be realized through neutron or photon radiation. 
\par
The dispersion relations of the emergent electronic modes can be simply expressed as combinations of the electronic and either the spin or the charge dispersion relations of unperturbed systems 
because the emergent modes exhibit the dispersion relations of unperturbed systems shifted by the momentum and energy (i.e., a point in the dispersion relations) of the perturbations. 
The results can be used to interpret changes in the band structure caused by perturbations, including the emergence of electronic states in nonequilibrium systems. 
Furthermore, the energy and momentum regimes of the emergent electronic states can be controlled by tuning and selecting perturbations and strongly correlated materials based on these results. 
This approach can be applied to the basic principle of band-structure engineering for strong-correlation electronics, utilizing the unconventional characteristics of strongly correlated insulators. 
\begin{acknowledgments} 
This work was supported by JSPS KAKENHI Grants No. JP22K03477 and No. JP25K07160 and World Premier International Research Center Initiative (WPI), MEXT, Japan. 
Numerical calculations were partly performed on the Numerical Materials Simulator at NIMS. 
\end{acknowledgments}
\section*{Data availability} 
The data that support the findings of this article are openly available \cite{dataAvailability}. 
\appendix
\section{Symmetry of spectral function} 
\label{sec:AkwSym}
\subsection{Particle--hole symmetry} 
\label{sec:phSym}
Under the transformation of 
\begin{equation}
c_{i,\sigma}\rightarrow(-)^ic^{\dagger}_{i,\sigma},\quad f_{i,\sigma}\rightarrow-(-)^if^{\dagger}_{i,\sigma}
\label{eq:phTrans}
\end{equation}
for $\sigma=\uparrow$ and $\downarrow$, where 
\begin{equation}
(-)^i=\left\{\begin{array}{rl}
1&\text{for a site $i$ on the A sublattice},\\
-1&\text{for a site $i$ on the B sublattice}
\end{array}\right.
\end{equation}
on a bipartite lattice, the Hamiltonians of the HM, symmetric ($U=2\Delta$) PAM, and KLM [Eqs. (\ref{eq:HamHub})--(\ref{eq:HamKLM})] are transformed as 
\begin{align}
\label{eq:phHamHub}
{\cal H}_{\rm HM}(t,U,\mu,H)&\rightarrow{\cal H}_{\rm HM}(t,U,-\mu,-H)-2N_{\rm s}\mu,\\
\label{eq:phHamPAM}
{\cal H}_{\rm PAM}(t,t_{\rm K},U,\mu,H)&\rightarrow{\cal H}_{\rm PAM}(t,t_{\rm K},U,-\mu,-H)-4N_{\rm u}\mu,\\
\label{eq:phHamKLM}
{\cal H}_{\rm KLM}(t,J_{\rm K},\mu,H)&\rightarrow{\cal H}_{\rm KLM}(t,J_{\rm K},-\mu,-H)-4N_{\rm u}\mu. 
\end{align}
This transformation is called the particle--hole transformation. 
At $\mu=H=0$, the Hamiltonians are invariant under the transformation \cite{Essler,TakahashiBook}. 
\par
Under the particle--hole transformation [Eq. (\ref{eq:phTrans})], 
\begin{align}
\label{eq:phck}
&c_{{\bm k},\sigma}\rightarrow c^{\dagger}_{{\bm \pi}-{\bm k},\sigma},\quad f_{{\bm k},\sigma}\rightarrow -f^{\dagger}_{{\bm \pi-{\bm k}},\sigma},\\
\label{eq:phdelta}
&|X\rangle\text{ of }\delta=\delta_X \rightarrow|X^\prime\rangle\text{ of }\delta=-\delta_X,
\end{align}
and the spectral function [Eq. (\ref{eq:Akwdef})] is transformed as 
\begin{equation}
A^{\lambda,\sigma}_X({\bm k},\omega)\rightarrow A^{\lambda,\sigma}_{X^\prime}({\bm \pi}-{\bm k},-\omega)
\label{eq:AkwphTrans}
\end{equation}
for $\lambda=c$ and $f$. 
\par
Thus, $A_{X^\prime}({\bm k},\omega)$ of the electron-doped eigenstate $|X^\prime\rangle$ is obtained as $A_X({\bm \pi}-{\bm k},-\omega)$ of the hole-doped eigenstate $|X\rangle$, 
and vice versa. 
According to Eqs. (\ref{eq:phHamHub})--(\ref{eq:AkwphTrans}), 
\begin{equation}
\label{eq:AkwphX}
\begin{array}{llllll}
&A_X({\bm k},\omega)&\text{at}&\mu={\bar \mu},&H={\bar H},&\delta=\delta_X\\
=&A_{X^\prime}({\bm \pi}-{\bm k},-\omega)&\text{at}&\mu=-{\bar \mu},&H=-{\bar H},&\delta=-\delta_X
\end{array}
\end{equation}
in the HM, symmetric PAM, and KLM on a bipartite lattice. 
\subsection{Spectral function of nonzero-$\eta$ states} 
\label{sec:Akweta}
In the HM, symmetric PAM, and KLM on a bipartite lattice, 
by applying $({\hat \eta}^+)^m$ to a $2m$-hole-doped state ($\eta^z=-m$) or by applying $({\hat \eta}^-)^m$ to a $2m$-electron-doped state ($\eta^z=m$), 
an undoped state ($\eta^z=0$) can be obtained (Sec. \ref{sec:etaSU2}). 
The spectral function of an undoped eigenstate of the Hamiltonian with $\eta=n(>0)$ for $\mu=H=0$, $A_{n,0}({\bm k},\omega)$, 
can be obtained from those of the $2m$-electron-doped and $2m$-hole-doped eigenstates with $\eta=n$ for $\mu=H=0$, $A_{n,\pm m}({\bm k},\omega)$, as 
\begin{align}
\label{eq:Akwetaz0}
A_{n,0}({\bm k},{\omega})
&=\frac{1}{2}[A_{n,m}({\bm k},\omega)+A_{n,-m}({\bm k},\omega)]\nonumber\\
&=\frac{1}{2}[A_{n,m}({\bm k},\omega)+A_{n,m}({\bm \pi}-{\bm k},-\omega)]
\end{align}
for $-n\le m\le n$, where $A_{n,m}({\bm k},\omega)$ denotes the spectral function of the state with $\eta=n$ and $\eta^z=m$ in an $\eta$-SU(2) multiplet. 
Equation (\ref{eq:Akwetaz0}) is obtained as follows: 
\par
The Wigner-Eckart theorem states that the matrix element of a rank-$J$ spherical tensor operator $T^J_M$ between states $|j,m\rangle$ and $|j^\prime,m^\prime\rangle$ 
(total angular-momentum quantum number $j^{(\prime)}$ and $z$ component of the angular momentum $m^{(\prime)}$) can be expressed as 
\begin{equation}
\langle j^\prime,m^\prime|T^J_M|j,m\rangle=(-1)^{j^\prime-m^\prime}
\left(\begin{array}{ccc}
j^\prime&J&j\\
-m^\prime&M&m
\end{array}\right)
\langle j^\prime||T^J||j\rangle,
\label{eq:WEtheorem}
\end{equation}
where $\langle j^\prime||T^J||j\rangle$ is a quantity that does not depend on $m$, $m^\prime$, or $M$, and the Wigner 3-$j$ symbol 
$\left(\begin{array}{ccc}
j_1&j_2&j_3\\
m_1&m_2&m_3
\end{array}\right)$ has the following properties: 
\begin{align}
\label{eq:3jSymbol1}
&\begin{array}{ll}
\left(\begin{array}{ccc}
j_1&j_2&j_3\\
m_1&m_2&m_3
\end{array}\right)&=
\left(\begin{array}{ccc}
j_2&j_3&j_1\\
m_2&m_3&m_1
\end{array}\right)\\
&=(-1)^{s}\left(\begin{array}{ccc}
j_1&j_2&j_3\\
-m_1&-m_2&-m_3
\end{array}\right),
\end{array}\\
\label{eq:3jSymbol2}
&\left(\begin{array}{ccc}
j_1&j_2&j_3\\
m_1&-m_1-j_3&j_3
\end{array}\right)
=(-1)^{j_1+m_1+2j_2}\nonumber\\
&\quad\times\sqrt{\frac{(2j_3)!(s-2j_3)!(j_1-m_1)!(j_2+m_1+j_3)!}{(s+1)!(s-2j_1)!(s-2j_2)!(j_1+m_1)!(j_2-m_1-j_3)!}},
\end{align}
where $s=j_1+j_2+j_3$.
\par
Using Eqs. (\ref{eq:WEtheorem})--(\ref{eq:3jSymbol2}), the following relation is obtained: 
\begin{equation}
\label{eq:WEnorm}
\sum_{M=\pm\frac{1}{2}}|\langle j\pm\frac{1}{2},m+M|T^{\frac{1}{2}}_M|j,m\rangle|^2=\frac{1}{2j+1}|\langle j\pm\frac{1}{2}||T^{\frac{1}{2}}||j\rangle|^2,
\end{equation}
which does not depend on $m$. 
\par
The electronic operators $(c^{\dagger}_{{\bm k},\sigma},2s^z c_{{\bm \pi}-{\bm k},{\bar \sigma}})$, 
where $s^z=\frac{1}{2}$ and $-\frac{1}{2}$ for $\sigma=\uparrow$ and $\downarrow$, respectively, behave 
as rank-$\frac{1}{2}$ spherical tensor operators $(T^\frac{1}{2}_\frac{1}{2},T^\frac{1}{2}_{-\frac{1}{2}})$ for ${\hat {\bm \eta}}$, as follows: 
\begin{equation}
\begin{array}{ll}
&[{\hat \eta}^{\pm},T^J_M]=\sqrt{(J\mp M)(J\pm M+1)}T^J_{M\pm 1},\\
&[{\hat \eta}^z,T^J_M]=MT^J_M.
\end{array}
\label{eq:tensorOp}
\end{equation}
\par
The sum of the spectral functions $A^{c}_{n,m}({\bm k},\omega)$ and $A^{c}_{n,m}({\bm \pi}-{\bm k},-\omega)$ for $\mu=0$ can be expressed as 
\begin{align}
\label{eq:Akwsum}
&A^{c}_{n,m}({\bm k},\omega)+A^{c}_{n,m}({\bm \pi}-{\bm k},-\omega)\nonumber\\
&=\frac{1}{2Z_{n,m}}\sum_{i,\sigma}\left[\begin{array}{l}
|\langle i|c^{\dagger}_{{\bm k},\sigma}|n,m\rangle|^2\delta(\omega-\epsilon_i)\\
+|\langle i|c_{{\bm k},{\bar \sigma}}|n,m\rangle|^2\delta(\omega+\epsilon_i)\\
+|\langle i|c^{\dagger}_{{\bm \pi}-{\bm k},\sigma}|n,m\rangle|^2\delta(-\omega-\epsilon_i)\\
+|\langle i|c_{{\bm \pi}-{\bm k},{\bar \sigma}}|n,m\rangle|^2\delta(-\omega+\epsilon_i)\end{array}\right]\nonumber\\
&=\frac{1}{2Z_{n,m}}\sum_{\sigma}\sum_{\zeta=\pm\frac{1}{2}}\sum_{i\in(n+\zeta)}\left[\begin{array}{l}
|\langle i|T^{\frac{1}{2};{\bm k},\sigma}_{\frac{1}{2}}|n,m\rangle|^2\delta(\omega-\epsilon_i)\\
+|\langle i|T^{\frac{1}{2};{\bm \pi}-{\bm k},\sigma}_{-\frac{1}{2}}|n,m\rangle|^2\delta(\omega+\epsilon_i)\\
+|\langle i|T^{\frac{1}{2};{\bm \pi}-{\bm k},\sigma}_{\frac{1}{2}}|n,m\rangle|^2\delta(\omega+\epsilon_i)\\
+|\langle i|T^{\frac{1}{2};{\bm k},\sigma}_{-\frac{1}{2}}|n,m\rangle|^2\delta(\omega-\epsilon_i)\end{array}\right]\nonumber\\
&\overset{\text{Eq. (\ref{eq:WEnorm})}}{=}\frac{1}{2(2n+1)Z_{n}}\nonumber\\
&\quad\times\sum_{\sigma}\sum_{\zeta=\pm\frac{1}{2}}\sum_{i\in(n+\zeta)}\left[\begin{array}{l}
|_i\langle n+\zeta||T^{\frac{1}{2};{\bm k},\sigma}||n\rangle|^2\delta(\omega-\epsilon_i)\\
+|_i\langle n+\zeta||T^{\frac{1}{2};{\bm \pi}-{\bm k},\sigma}||n\rangle|^2\delta(\omega+\epsilon_i)\end{array}\right],
\end{align}
where $(T^{\frac{1}{2};{\bm k},\sigma}_\frac{1}{2},T^{\frac{1}{2};{\bm k},\sigma}_{-\frac{1}{2}})=(c^{\dagger}_{{\bm k},\sigma},2s^z c_{{\bm \pi}-{\bm k},{\bar \sigma}})$, 
$Z_{n,m}=\langle n,m|n,m\rangle=Z_n$, and $i\in(l)$ means that $|i\rangle$ has $\eta=l$. 
The quantities with subscript $i$ indicate those for $|i\rangle$, and the excitation energy of $|i\rangle$ from $|n,m\rangle$ is denoted by $\epsilon_i$. 
\par
In Eq. (\ref{eq:Akwsum}), because the energies of $|n\pm\frac{1}{2},m^{\prime}\rangle$ for $|m^{\prime}|\le n\pm\frac{1}{2}$ are the same regardless of $m^{\prime}$ in an SU(2) multiplet, 
$\epsilon_i$ does not depend on $\eta^z$ of $|i\rangle$ in an $\eta$-SU(2) multiplet, 
and $\langle n^{\prime},m^{\prime}|T^{\frac{1}{2}}_{M}|n,m\rangle$ can be nonzero only for $n^{\prime}=n\pm\frac{1}{2}$ and $m^{\prime}=m+M$. 
\par
Equation (\ref{eq:Akwsum}) means that $A^{c}_{n,m}({\bm k},\omega)+A^{c}_{n,m}({\bm \pi}-{\bm k},-\omega)$ does not depend on $m$ at $\mu=0$, 
\begin{align}
\label{eq:Akweta}
&A^{c}_{n,n}({\bm k},\omega)+A^{c}_{n,n}({\bm \pi}-{\bm k},-\omega)\nonumber\\
&=A^{c}_{n,n-1}({\bm k},\omega)+A^{c}_{n,n-1}({\bm \pi}-{\bm k},-\omega)=\cdots\nonumber\\
&=A^{c}_{n,-n}({\bm k},\omega)+A^{c}_{n,-n}({\bm \pi}-{\bm k},-\omega).
\end{align}
\par
At $\mu=H=0$, using particle--hole symmetry, 
\begin{align}
\label{eq:An0kw}
A^{c}_{n,0}({\bm k},\omega)&\overset{\text{Eq. (\ref{eq:AkwphX})}}{=}\frac{1}{2}[A^{c}_{n,0}({\bm k},\omega)+A^{c}_{n,0}({\bm \pi}-{\bm k},-\omega)]\nonumber\\
&\overset{\text{Eq. (\ref{eq:Akweta})}}{=}\frac{1}{2}[A^{c}_{n,m}({\bm k},\omega)+A^{c}_{n,m}({\bm \pi}-{\bm k},-\omega)]\nonumber\\
&\overset{\text{Eq. (\ref{eq:AkwphX})}}{=}\frac{1}{2}[A^{c}_{n,m}({\bm k},\omega)+A^{c}_{n,-m}({\bm k},\omega)].
\end{align}
\par
In the symmetric PAM for $\mu=H=0$, because the electronic operators $(f^{\dagger}_{{\bm k},\sigma},-2s^z f_{{\bm \pi}-{\bm k},{\bar \sigma}})$ also behave 
as rank-$\frac{1}{2}$ spherical tensor operators $(T^\frac{1}{2}_\frac{1}{2},T^\frac{1}{2}_{-\frac{1}{2}})$ for ${\hat {\bm \eta}}$, 
\begin{align}
\label{eq:An0kwf}
A^{f}_{n,0}({\bm k},\omega)&=\frac{1}{2}[A^{f}_{n,m}({\bm k},\omega)+A^{f}_{n,m}({\bm \pi}-{\bm k},-\omega)]\nonumber\\
&=\frac{1}{2}[A^{f}_{n,m}({\bm k},\omega)+A^{f}_{n,-m}({\bm k},\omega)].
\end{align}
\par
By combining Eqs. (\ref{eq:An0kw}) and (\ref{eq:An0kwf}), Eq. (\ref{eq:Akwetaz0}) is obtained [Eq. (\ref{eq:Akwcf})]. 
\subsection{Spectral function of nonzero-$S$ states} 
\label{sec:AkwS}
The electron creation operators $(c^{\dagger}_{{\bm k},\uparrow}, c^{\dagger}_{{\bm k},\downarrow})$ and 
annihilation operators $(-c_{{\bm k},\downarrow}, c_{{\bm k},\uparrow})$ behave as rank-$\frac{1}{2}$ spherical tensor operators 
$(T^\frac{1}{2}_\frac{1}{2}, T^\frac{1}{2}_{-\frac{1}{2}})$ for ${\hat {\bm S}}$. 
\par
The electron-addition and electron-removal spectral functions of an eigenstate of the Hamiltonian with $S=n$ and $S^z=m$ for $H=0$ 
can be expressed as 
\begin{align}
&A^{c,+}_{n,m}({\bm k},\omega)=\frac{1}{2Z_{n,m}}\sum_{i,\sigma}|\langle i|c^{\dagger}_{{\bm k},\sigma}|n,m\rangle|^2\delta(\omega-\epsilon_i)\nonumber\\
&=\frac{1}{2Z_{n,m}}\sum_{s^z=\pm\frac{1}{2}}\left[\begin{array}{r}
\sum_{i\in (n+\frac{1}{2})}|\langle i|T^\frac{1}{2}_{s^z}|n,m\rangle|^2\delta(\omega-\epsilon_i)\\
+\sum_{i\in (n-\frac{1}{2})}|\langle i|T^\frac{1}{2}_{s^z}|n,m\rangle|^2\delta(\omega-\epsilon_i)\end{array}\right]\nonumber\\
&\overset{\text{Eq. (\ref{eq:WEnorm})}}{=}\frac{1}{2(2n+1)Z_{n}}\nonumber\\
&\quad\times\left[\begin{array}{r}
\sum_{i\in (n+\frac{1}{2})}|_i\langle n+\frac{1}{2}||T^\frac{1}{2}||n\rangle|^2\delta(\omega-\epsilon_i)\\
+\sum_{i\in (n-\frac{1}{2})}|_i\langle n-\frac{1}{2}||T^\frac{1}{2}||n\rangle|^2\delta(\omega-\epsilon_i)\end{array}\right]
\label{eq:ApkwSz}
\end{align}
with $(T^\frac{1}{2}_\frac{1}{2}, T^\frac{1}{2}_{-\frac{1}{2}})=(c^{\dagger}_{{\bm k},\uparrow}, c^{\dagger}_{{\bm k},\downarrow})$ and $Z_n=Z_{n,m}$, and 
\begin{align}
&A^{c,-}_{n,m}({\bm k},\omega)=\frac{1}{2Z_{n,m}}\sum_{i,\sigma}|\langle i|c_{{\bm k},\sigma}|n,m\rangle|^2\delta(\omega+\epsilon_i)\nonumber\\
&=\frac{1}{2Z_{n,m}}\sum_{s^z=\pm\frac{1}{2}}\left[\begin{array}{r}
\sum_{i\in (n+\frac{1}{2})}|\langle i|T^\frac{1}{2}_{s^z}|n,m\rangle|^2\delta(\omega+\epsilon_i)\\
+\sum_{i\in (n-\frac{1}{2})}|\langle i|T^\frac{1}{2}_{s^z}|n,m\rangle|^2\delta(\omega+\epsilon_i)\end{array}\right]\nonumber\\
&\overset{\text{Eq. (\ref{eq:WEnorm})}}{=}\frac{1}{2(2n+1)Z_{n}}\nonumber\\
&\quad\times\left[\begin{array}{r}
\sum_{i\in (n+\frac{1}{2})}|_i\langle n+\frac{1}{2}||T^\frac{1}{2}||n\rangle|^2\delta(\omega+\epsilon_i)\\
+\sum_{i\in (n-\frac{1}{2})}|_i\langle n-\frac{1}{2}||T^\frac{1}{2}||n\rangle|^2\delta(\omega+\epsilon_i)\end{array}\right]
\label{eq:AnkwSz}
\end{align}
with $(T^\frac{1}{2}_\frac{1}{2}, T^\frac{1}{2}_{-\frac{1}{2}})=(-c_{{\bm k},\downarrow}, c_{{\bm k},\uparrow})$ and $Z_n=Z_{n,m}$, 
where $i\in (n\pm\frac{1}{2})$ means that $|i\rangle$ has $S=n\pm\frac{1}{2}$, and the quantities with subscript $i$ indicate those for $|i\rangle$; 
the excitation energy of $|i\rangle$ from $|n,m\rangle$ is denoted by $\epsilon_i$. 
Here, $\epsilon_i$ does not depend on $S^z$ of $|i\rangle$ in a spin-SU(2) multiplet, and 
$\langle n^{\prime},m^{\prime}|T^\frac{1}{2}_{s^z}|n,m\rangle$ can be nonzero only for $n^{\prime}=n\pm\frac{1}{2}$ and $m^{\prime}=m+s^z$. 
\par
Equations (\ref{eq:ApkwSz}) and (\ref{eq:AnkwSz}) mean that $A^{c,\pm}_{n,m}({\bm k},\omega)$ do not depend on $m$ at $H=0$: 
\begin{equation}
A^{c,\pm}_{S,S}({\bm k},\omega)=A^{c,\pm}_{S,S-1}({\bm k},\omega)
=\cdots=A^{c,\pm}_{S,-S}({\bm k},\omega).
\label{eq:AkwS}
\end{equation}
This relation generally holds true for spin-SU(2)-symmetric systems. 
In the PAM for $H=0$, $A^{f,\pm}_{S,S^z}({\bm k},\omega)$ are similarly shown to be independent of $S^z$. 
\subsection{Relation between spectral functions before and after Shiba transformation} 
\label{sec:AkwSnU}
The spectral function of an eigenstate of the Hamiltonian $|X\rangle$ for $\lambda(=c,f)$ electrons of spin $\sigma(=\uparrow,\downarrow)$ can be expressed [Eq. (\ref{eq:Akwdef})] as 
\begin{align}
\label{eq:Akwud2}
A^{\lambda,\sigma}_X({\bm k},\omega)&=\frac{1}{Z_X}\sum_{n}
\left[\begin{array}{r}
|\langle n|\lambda^{\dagger}_{{\bm k},\sigma}|X\rangle|^2\delta(\omega-E_n+E_X)\\
+|\langle n|\lambda_{{\bm k},\sigma}|X\rangle|^2\delta(\omega+E_n-E_X)\end{array}\right].
\end{align}
Under the Shiba transformation [Eq. (\ref{eq:ShibaTrans})], the $\downarrow$-spin-electron operators and spin operators are transformed [Eqs. (\ref{eq:SetaShiba}) and (\ref{eq:ckShiba})] as 
\begin{equation}
\label{eq:ckSetaShiba2}
c_{{\bm k},\downarrow}\rightarrow c^{\dagger}_{{\bm \pi}-{\bm k},\downarrow},\quad f_{{\bm k},\downarrow}\rightarrow -f^{\dagger}_{{\bm \pi-{\bm k}},\downarrow},\quad 
{\hat {\bm S}}\rightarrow{\hat {\bm \eta}}.
\end{equation}
Accordingly, the $\downarrow$-spin-electron spectral function of a state $|X\rangle$ of $S=j$ and $S^z=m$ 
for ${\cal H}(U)$, 
\begin{equation}
A^{\lambda,\downarrow}_X({\bm k},\omega)=\frac{1}{Z_X}\sum_{i}\left[
\begin{array}{r}|\langle i|\lambda^{\dagger}_{{\bm k},\downarrow}|X\rangle|^2\delta(\omega-\epsilon_i)\\
+|\langle i|\lambda_{{\bm k},\downarrow}|X\rangle|^2\delta(\omega+\epsilon_i)\end{array}\right],
\end{equation}
is transformed to 
\begin{equation}
{\tilde A}^{\lambda,\downarrow}_{X^\prime}({\bm \pi}-{\bm k},-\omega)=
\frac{1}{Z_{X^\prime}}\sum_{i}\left[
\begin{array}{r}|\langle i|\lambda_{{\bm \pi}-{\bm k},\downarrow}|{X^\prime}\rangle|^2\delta(\omega-\epsilon_i)\\
+|\langle i|\lambda^{\dagger}_{{\bm \pi}-{\bm k},\downarrow}|{X^\prime}\rangle|^2\delta(\omega+\epsilon_i)\end{array}\right]\\
\label{eq:AkwdXprime2}
\end{equation}
for ${\cal H}(-U)$, where $|X^{\prime}\rangle$ is the state of $\eta=j$ and $\eta^z=m$. 
Here, the excitation energy of $|i\rangle$ from $|X\rangle$ or $|X^\prime\rangle$ is denoted by $\epsilon_i$, 
and ${\cal H}(U)$ and ${\cal H}(-U)$ denote the Hamiltonians before and after the Shiba transformation, respectively [Eqs. (\ref{eq:ShibaHamHub}) and (\ref{eq:ShibaHamPAM})]. 
To distinguish the spectral functions before and after the transformation, the spectral function after the Shiba transformation is denoted by ${\tilde A}^{\lambda,\sigma}_{X^\prime}({\bm k},\omega)$. 
The $\uparrow$-spin electron spectral function is transformed as 
\begin{align}
A^{\lambda,\uparrow}_X({\bm k},\omega)\rightarrow&\frac{1}{Z_{X^\prime}}\sum_{i}\left[
\begin{array}{r}|\langle i|\lambda^{\dagger}_{{\bm k},\uparrow}|{X^\prime}\rangle|^2\delta(\omega-\epsilon_i)\\
+|\langle i|\lambda_{{\bm k},\uparrow}|{X^\prime}\rangle|^2\delta(\omega+\epsilon_i)\end{array}\right]\\
&={\tilde A}^{\lambda,\uparrow}_{X^\prime}({\bm k},\omega).
\label{eq:AkwuXprime}
\end{align}
\par
If the state $|X^{\prime}\rangle$ is a spin-singlet ($S=0$) state, 
\begin{equation}
{\tilde A}^{\lambda,\uparrow}_{X^\prime}({\bm k},\omega)={\tilde A}^{\lambda,\downarrow}_{X^\prime}({\bm k},\omega)={\tilde A}^{\lambda}_{X^\prime}({\bm k},\omega)
\label{eq:AkwudXprime}
\end{equation}
at $H=0$ for ${\cal H}(-U)$ [Eq. (\ref{eq:Akwlambda})] because 
\begin{equation}
\begin{array}{c}
\sum_{i}|\langle i|\lambda^{\dagger}_{{\bm k},\uparrow}|{X^\prime}\rangle|^2\delta(\omega-\epsilon_i)
=\sum_{i}|\langle i|\lambda^{\dagger}_{{\bm k},\downarrow}|{X^\prime}\rangle|^2\delta(\omega-\epsilon_i),\\
\sum_{i}|\langle i|\lambda_{{\bm k},\uparrow}|{X^\prime}\rangle|^2\delta(\omega+\epsilon_i)
=\sum_{i}|\langle i|\lambda_{{\bm k},\downarrow}|{X^\prime}\rangle|^2\delta(\omega+\epsilon_i).
\end{array}
\end{equation}
\par
Hence, the spectral functions of an $\eta$-singlet ($\eta=0$) state of $S=j$ and $S^z=m$, $|S=j,S^z=m\rangle$, for $\uparrow$-spin and $\downarrow$-spin electrons at $\mu=0$ for ${\cal H}(U)$ 
can be expressed as that of the corresponding spin-singlet ($S=0$) state of $\eta=j$ and $\eta^z=m$, $|\eta=j,\eta^z=m\rangle$, at $H=0$ for ${\cal H}(-U)$ as 
\begin{align}
\label{eq:AkwudShiba}
&\left(\begin{array}{l}A^{\lambda,\uparrow}_{S=j,S^z=m}({\bm k},\omega)\\
A^{\lambda,\downarrow}_{S=j,S^z=m}({\bm k},\omega)\end{array}\right)\text{ at }\mu=0\text{ for }{\cal H}(U)\nonumber\\
=&\left(\begin{array}{l}{\tilde A}^{\lambda}_{\eta=j,\eta^z=m}({\bm k},\omega)\\
{\tilde A}^{\lambda}_{\eta=j,\eta^z=m}({\bm \pi}-{\bm k},-\omega)\end{array}\right)\text{ at }H=0\text{ for }{\cal H}(-U),
\end{align}
which can further be rewritten using the relation of the particle--hole transformation [Eq. (\ref{eq:AkwphX})] as 
\begin{equation}
\left(\begin{array}{l}{\tilde A}^{\lambda}_{\eta=j,\eta^z=-m}({\bm \pi}-{\bm k},-\omega)\\
{\tilde A}^{\lambda}_{\eta=j,\eta^z=-m}({\bm k},\omega)\end{array}\right)\text{ at }H=0\text{ for }{\cal H}(-U),
\label{eq:AkwudShiba2}
\end{equation}
where the sign of $\mu$ for ${\cal H}(-U)$ is reversed by the particle--hole transformation. 
\par
At $\mu=H=0$, using Eqs. (\ref{eq:An0kw}), (\ref{eq:An0kwf}), (\ref{eq:AkwudShiba}), and (\ref{eq:AkwudShiba2}), 
\begin{align}
\label{eq:AkwSz0}
&A^{\lambda}_{S=j,S^z=0}({\bm k},\omega)\text{ for }{\cal H}(U)\nonumber\\
=&\frac{1}{2}[{\tilde A}^{\lambda}_{\eta=j,\eta^z=m}({\bm k},\omega)+{\tilde A}^{\lambda}_{\eta=j,\eta^z=m}({\bm \pi}-{\bm k},-\omega)]\text{ for }{\cal H}(-U)\nonumber\\
=&\frac{1}{2}[{\tilde A}^{\lambda}_{\eta=j,\eta^z=m}({\bm k},\omega)+{\tilde A}^{\lambda}_{\eta=j,\eta^z=-m}({\bm k},\omega)]\text{ for }{\cal H}(-U)
\end{align}
for $-j\le m\le j$. 
\par
Thus, the spectral functions of an eigenstate with $S=j$, $S^z=m$, and $\eta=0$ at half filling ($\eta^z=0$) at $\mu=0$ for ${\cal H}(U)$ 
can be expressed using those of the eigenstates with $\eta=j$, $\eta^z=\pm m$, and $S=0$ at $H=0$ for ${\cal H}(-U)$ as Eqs. (\ref{eq:AkwudShiba})--(\ref{eq:AkwSz0}); 
in particular, the spectral functions of the magnetized ground state at half filling for ${\cal H}(U)$ can be expressed using those of the doped ground states for ${\cal H}(-U)$. 
\section{Effective theory for weak inter-unit-cell hopping in spin-gapped Mott and Kondo insulators} 
\label{sec:effTheory}
\subsection{Effective states and energies in systems of weakly coupled unit cells} 
\label{sec:effStates}
Without inter-unit-cell hopping ($t=0$), the system is decoupled into unit cells. 
The eigenstates in a unit cell for the ladder and bilayer HMs and 1D and 2D symmetric ($U=2\Delta$) PAMs and KLMs are obtained 
as shown in Tables \ref{tbl:HM}--\ref{tbl:KLM} \cite{KohnoTinduced,KohnoHubLadder,KohnoKLM,OneSiteKondo,localRungHubLadder,TsunetsuguRMP}. 
For shorthand notation, the spin indices are omitted for spin-independent energies, 
e.g., $\varepsilon^{\rm F}_{{\bm k}_{\parallel}}=\varepsilon^{{\rm F}_{\sigma}}_{{\bm k}_{\parallel}}$ for $H=0$, 
$t_{\rm eff}^{\rm T}=t_{\rm eff}^{{\rm T}^\gamma}$, $t_{\rm eff}^{\rm P}=t_{\rm eff}^{{\rm P}_{\sigma}}$, and $\xi_{\psi^-{\rm B}}=\xi_{\psi^-{{\rm B}_{\sigma}}}$. 
\begin{table*} 
\caption{Eigenstates of the dimer HM.}
\begin{center}
\begin{tabular}{cccccccc}
\hline\hline
$N_{\rm e}$&$S$&$S^z$&$\eta$&$\eta^z$&$k_{\perp}$&Eigenstates\tnote&Energies\\\hline
4&0&0&1&1&0&$|{\rm Z}\rangle=-|\uparrow\downarrow,\uparrow\downarrow\rangle$&${\cal E}_{\rm Z}=\frac{U}{2}-4\mu$\\ 
2&0&0&1&0&$\pi$&$|{\rm D}^-\rangle$&${\cal E}_{{\rm D}^-}=\frac{U}{2}-2\mu$\\ 
0&0&0&1&$-1$&0&$|{\rm V}\rangle=|0,0\rangle$&${\cal E}_{\rm V}=\frac{U}{2}$\\
2&1&$2s^z$&0&0&$\pi$&$|{\rm T}^{2s^z}\rangle=|\sigma,\sigma\rangle$&${\cal E}_{{\rm T}^{2s^z}}=-\frac{U}{2}-2\mu-2s^zH$\\ 
2&1&$0$&0&0&$\pi$&$|{\rm T}^0\rangle=\frac{1}{\sqrt{2}}\left(|\uparrow,\downarrow\rangle+|\downarrow,\uparrow\rangle\right)$&${\cal E}_{{\rm T}^0}=-\frac{U}{2}-2\mu$\\
2&0&0&0&0&0&$|\psi^{\pm}\rangle=u^{\mp}|{\rm S}\rangle\mp{\rm sgn}(t_{\perp})u^{\pm}|{\rm D}^+\rangle$&${\cal E}_{\psi^{\pm}}=E^{\pm}-2\mu$\\
3&$\frac{1}{2}$&$s^z$&$\frac{1}{2}$&$\frac{1}{2}$&$\pi$&$|{\rm F}_{\sigma}\rangle=\frac{1}{\sqrt{2}}\left(|\uparrow\downarrow,\sigma\rangle-|\sigma,\uparrow\downarrow\rangle\right)$
&${\cal E}_{{\rm F}_{\sigma}}=-t_{\perp}-3\mu-s^zH$\\ 
1&$\frac{1}{2}$&$s^z$&$\frac{1}{2}$&$-\frac{1}{2}$&0&$|{\rm B}_{\sigma}\rangle=\frac{1}{\sqrt{2}}\left(|0,\sigma\rangle+|\sigma,0\rangle\right)$&${\cal E}_{{\rm B}_{\sigma}}=-t_{\perp}-\mu-s^zH$\\
3&$\frac{1}{2}$&$s^z$&$\frac{1}{2}$&$\frac{1}{2}$&0&$|{\rm G}_{\sigma}\rangle=\frac{1}{\sqrt{2}}\left(|\uparrow\downarrow,\sigma\rangle+|\sigma,\uparrow\downarrow\rangle\right)$
&${\cal E}_{{\rm G}_{\sigma}}=t_{\perp}-3\mu-s^zH$\\ 
1&$\frac{1}{2}$&$s^z$&$\frac{1}{2}$&$-\frac{1}{2}$&$\pi$&$|{\rm A}_{\sigma}\rangle=\frac{1}{\sqrt{2}}\left(|0,\sigma\rangle-|\sigma,0\rangle\right)$&${\cal E}_{{\rm A}_{\sigma}}=t_{\perp}-\mu-s^zH$\\\hline\hline 
\end{tabular}
\begin{tablenotes}
\item{\begin{center}\begin{tabular}{cccc}
\multicolumn{4}{l}{The state with $|\alpha_1\rangle$ and $|\alpha_2\rangle$ at sites 1 and 2, respectively, in a dimer is denoted by $|\alpha_1,\alpha_2\rangle$.}\\
$|{\rm S}\rangle=\frac{1}{\sqrt{2}}\left(|\uparrow,\downarrow\rangle-|\downarrow,\uparrow\rangle\right)$,
&$|{\rm D}^{\pm}\rangle=\frac{1}{\sqrt{2}}\left(|\uparrow\downarrow,0\rangle\pm|0,\uparrow\downarrow\rangle\right)$,&
$E^{\pm}=\pm\frac{\sqrt{U^2+16t_{\perp}^2}}{2}$, 
&$u^{\pm}=\sqrt{\frac{1}{2}\left(1\pm\frac{U}{\sqrt{U^2+16t_{\perp}^2}}\right)}$.\\
\multicolumn{4}{l}{$s^z=\frac{1}{2}$ and $-\frac{1}{2}$ for $\sigma=\uparrow$ and $\downarrow$, respectively.}
\end{tabular}\end{center}}
\end{tablenotes}
\end{center}\label{tbl:HM}
\end{table*}
\begin{table*} 
\caption{Eigenstates of the single-site PAM for $U=2\Delta$.}
\begin{center}
\begin{tabular}{ccccccc}
\hline\hline
$N_{\rm e}$&$S$&$S^z$&$\eta$&$\eta^z$&Eigenstates\tnote&Energies\\\hline
4&0&0&1&1&$|{\rm Z}\rangle=-|\uparrow\downarrow,\uparrow\downarrow\rangle$&${\cal E}_{\rm Z}=-4\mu$\\ 
2&0&0&1&0&$|{\rm D}^-\rangle$&${\cal E}_{{\rm D}^-}=-2\mu$\\ 
0&0&0&1&$-1$&$|{\rm V}\rangle=|0,0\rangle$&${\cal E}_{\rm V}=0$\\
2&1&$2s^z$&0&0&$|{\rm T}^{2s^z}\rangle=|\sigma,\sigma\rangle$&${\cal E}_{{\rm T}^{2s^z}}=-\Delta-2\mu-2s^zH$\\ 
2&1&$0$&0&0&$|{\rm T}^0\rangle=\frac{1}{\sqrt{2}}\left(|\uparrow,\downarrow\rangle+|\downarrow,\uparrow\rangle\right)$&${\cal E}_{{\rm T}^0}=-\Delta-2\mu$\\
2&0&0&0&0&$|\psi^{\pm}\rangle=u^{\mp}_{2t_{\rm K}}|{\rm S}\rangle\mp{\rm sgn}(t_{\rm K})u^{\pm}_{2t_{\rm K}}|{\rm D}^+\rangle$&${\cal E}_{\psi^{\pm}}=E^{\pm}_{2t_{\rm K}}-2\mu$\\
3&$\frac{1}{2}$&$s^z$&$\frac{1}{2}$&$\frac{1}{2}$&
$|{\rm P}^{\pm}_{\sigma}\rangle=u^{\mp}_{t_{\rm K}}|\uparrow\downarrow,\sigma\rangle\pm{\rm sgn}(t_{\rm K})u^{\pm}_{t_{\rm K}}|\sigma,\uparrow\downarrow\rangle$&
${\cal E}_{{\rm P}^{\pm}_{\sigma}}=E^{\pm}_{t_{\rm K}}-3\mu-s^zH$\\ 
1&$\frac{1}{2}$&$s^z$&$\frac{1}{2}$&$-\frac{1}{2}$&
$|{\rm R}^{\pm}_{\sigma}\rangle=u^{\mp}_{t_{\rm K}}|0,\sigma\rangle\mp{\rm sgn}(t_{\rm K})u^{\pm}_{t_{\rm K}}|\sigma,0\rangle$&
${\cal E}_{{\rm R}^{\pm}_{\sigma}}=E^{\pm}_{t_{\rm K}}-\mu-s^zH$\\\hline\hline 
\end{tabular}
\begin{tablenotes}
\item{\begin{center}\begin{tabular}{cccc}
\multicolumn{4}{l}{The state with $|\alpha_c\rangle$ in the conduction orbital and $|\alpha_f\rangle$ in the localized orbital is denoted by $|\alpha_c,\alpha_f\rangle$.}\\
$|{\rm S}\rangle=\frac{1}{\sqrt{2}}\left(|\uparrow,\downarrow\rangle-|\downarrow,\uparrow\rangle\right)$,
&$|{\rm D}^{\pm}\rangle=\frac{1}{\sqrt{2}}\left(|\uparrow\downarrow,0\rangle\pm|0,\uparrow\downarrow\rangle\right)$,
&$E^{\pm}_{t_{\rm K}}=-\frac{\Delta}{2}\pm\frac{\sqrt{\Delta^2+4t_{\rm K}^2}}{2}$, 
&$u^{\pm}_{t_{\rm K}}=\sqrt{\frac{1}{2}\left(1\pm\frac{\Delta}{\sqrt{\Delta^2+4t_{\rm K}^2}}\right)}$.\\
\multicolumn{2}{l}{$s^z=\frac{1}{2}$ and $-\frac{1}{2}$ for $\sigma=\uparrow$ and $\downarrow$, respectively.}&&
\end{tabular}\end{center}}
\end{tablenotes}
\end{center}\label{tbl:PAM}
\end{table*}
\begin{table*} 
\caption{Eigenstates of the single-site KLM.}
\begin{center}
\begin{tabular}{ccccccc}
\hline\hline
$N_{\rm e}$&$S$&$S^z$&$\eta$&$\eta^z$&Eigenstates\tnote&Energies\\\hline
2&1&$2s^z$&0&0&$|{\rm T}^{2s^z}\rangle=|\sigma,\sigma\rangle$&${\cal E}_{{\rm T}^{2s^z}}=\frac{J_{\rm K}}{4}-2\mu-2s^zH$\\ 
2&1&$0$&0&0&$|{\rm T}^0\rangle=\frac{1}{\sqrt{2}}\left(|\uparrow,\downarrow\rangle+|\downarrow,\uparrow\rangle\right)$&${\cal E}_{{\rm T}^0}=\frac{J_{\rm K}}{4}-2\mu$\\
2&0&0&0&0&$|{\rm S}\rangle=\frac{1}{\sqrt{2}}\left(|\uparrow,\downarrow\rangle-|\downarrow,\uparrow\rangle\right)$&${\cal E}_{\rm S}=-\frac{3J_{\rm K}}{4}-2\mu$\\
3&$\frac{1}{2}$&$s^z$&$\frac{1}{2}$&$\frac{1}{2}$&$|{\rm P}_{\sigma}\rangle=|\uparrow\downarrow,\sigma\rangle$&${\cal E}_{{\rm P}_{\sigma}}=-3\mu-s^zH$\\ 
1&$\frac{1}{2}$&$s^z$&$\frac{1}{2}$&$-\frac{1}{2}$&$|{\rm R}_{\sigma}\rangle=|0,\sigma\rangle$&${\cal E}_{{\rm R}_{\sigma}}=-\mu-s^zH$\\\hline\hline 
\end{tabular}
\begin{tablenotes}
\item{\begin{center}\begin{tabular}{l}
The state with $|\alpha_c\rangle$ in the conduction orbital and $|\alpha_f\rangle$ in the localized orbital is denoted by $|\alpha_c,\alpha_f\rangle$.\\
$s^z=\frac{1}{2}$ and $-\frac{1}{2}$ for $\sigma=\uparrow$ and $\downarrow$, respectively.
\end{tabular}\end{center}}
\end{tablenotes}
\end{center}
\label{tbl:KLM}
\end{table*}
\par
In the regime of weak inter-unit-cell hopping (small $|t|$), effective states and energies are obtained using the perturbation theory with respect to the inter-unit-cell hopping parameter $t$ 
\cite{KohnoDIS,KohnoHubLadder,KohnoKLM,KohnoTinduced,KohnoUinf,TsunetsuguRMP,OneSiteKondo,localRungHubLadder}. 
The ground state at half filling ($N_{\rm e}=2N_{\rm u}$) and single-particle excited state of $X$ can be effectively expressed [Eqs. (\ref{eq:GS}) and (\ref{eq:Xk})] as
\begin{align}
\label{eq:GS2}
&|{\rm GS}\rangle=\prod_i|{\cal G}\rangle_i,\\
\label{eq:Xk2}
&|X\rangle_{{\bm k}_{\parallel}}=\frac{1}{\sqrt{N_{\rm u}}}\sum_ie^{i{\bm k}_{\parallel}\cdot{\bm r}_i}|X\rangle_i\prod_{j(\ne i)}|{\cal G}\rangle_j,
\end{align}
where ${\cal G}$ represents the ground state in a unit cell (Table \ref{tbl:symbols}). 
The ground-state energy ($E_{\rm GS}$) and single-particle excitation energies at a momentum ${{\bm k}_{\parallel}}$ for a boson $X$ ($e^X_{{\bm k}_{\parallel}}$) and fermion $X$ ($\varepsilon^X_{{\bm k}_{\parallel}}$) 
are obtained [Eqs. (\ref{eq:eGS})--(\ref{eq:eXkFermion})] \cite{KohnoTinduced,KohnoHubLadder,KohnoKLM,TsunetsuguRMP} as
\begin{align}
\label{eq:eGS2}
E_{\rm GS}=&({\cal E}_{\cal G}+d\xi_{{\cal G}{\cal G}})N_{\rm u},\\
\label{eq:eXkBoson2}
e^X_{{\bm k}_{\parallel}}=&-2t_{\rm eff}^Xd\gamma_{{\bm k}_{\parallel}}+{\cal E}_X-{\cal E}_{\cal G}+2d(\xi_{{\cal G}X}-\xi_{{\cal G}{\cal G}}),\\
\label{eq:eXkFermion2}
\varepsilon^X_{{\bm k}_{\parallel}}=&-2t_{\rm eff}^Xd\gamma_{{\bm k}_{\parallel}}+{\cal E}_X-{\cal E}_{\cal G}+2d(\xi_{{\cal G}X}-\xi_{{\cal G}{\cal G}})\nonumber\\
&-2t_{3{\rm uc}}^Xd(2d\gamma_{{\bm k}_{\parallel}}^2-1)
\end{align}
in $d$ dimensions; $d=1$ for the ladder HM and 1D PAM and KLM, and $d=2$ for the bilayer HM and 2D PAM and KLM, where 
\begin{equation}
\label{eq:gammak2}
\gamma_{{\bm k}_{\parallel}}=\frac{1}{d}\sum_{i=1}^d\cos k_{\parallel i} 
\end{equation}
with $k_{\parallel 1}=k_x$ and $k_{\parallel 2}=k_y$ in $d$ dimensions [Eq. (\ref{eq:gammak})]. 
\par
The bond energies of $\mathcal{O}(t^2)$ between states $X$ and $Y$ on neighboring unit cells ($\xi_{XY}$) are obtained as 
\begin{align}
&\xi_{\psi^-\psi^-}=-\frac{2t^2U^2}{(U^2+16t_{\perp}^2)^{\frac{3}{2}}},\quad\xi_{\psi^-{\rm T}}=-\frac{4t^2}{U}+\frac{2t^2}{\sqrt{U^2+16t_{\perp}^2}},\nonumber\\
&\xi_{\psi^-{\rm A}}=\xi_{\psi^-{\rm G}}=\frac{\xi_{\psi^-\psi^-}}{8}+\frac{3t^2}{8t_{\perp}}+\frac{3t^2}{2\sqrt{U^2+16t_{\perp}^2}},\nonumber\\
&\xi_{\psi^-{\rm B}}=\xi_{\psi^-{\rm F}}=\frac{\xi_{\psi^-\psi^-}}{8}-\frac{3t^2}{8t_{\perp}}+\frac{3t^2}{2\sqrt{U^2+16t_{\perp}^2}},\nonumber\\
&\xi_{\psi^-{\rm V}}=\xi_{\psi^-{\rm Z}}=\xi_{\psi^-{\rm D}^-}=\frac{4t^2}{U}+\frac{2t^2}{\sqrt{U^2+16t_{\perp}^2}}
\label{eq:xiHub}
\end{align}
in the HM \cite{KohnoTinduced,KohnoHubLadder}, 
\begin{align}
&\xi_{\psi^-\psi^-}=-\frac{t^2(\Delta^4+27t_{\rm K}^2\Delta^2+192t_{\rm K}^4)}{6t_{\rm K}^2(\Delta^2+16t_{\rm K}^2)^{\frac{3}{2}}},\nonumber\\
&\xi_{\psi^-{\rm T}}=-\frac{t^2}{8t_{\rm K}^2}\left[\frac{\Delta^2+8t_{\rm K}^2}{\Delta}+\frac{3(\Delta^2+12t_{\rm K}^2)}{\sqrt{\Delta^2+16t_{\rm K}^2}}\right],\nonumber\\
&\xi_{\psi^-{\rm P}^{\pm}}=\xi_{\psi^-{\rm R}^{\pm}}
=-\frac{t^2}{96t_{\rm K}^2}\left[
\begin{array}{l}
\frac{29\Delta^4+810t_{\rm K}^2\Delta^2+5568t_{\rm K}^4}{(\Delta^2+16t_{\rm K}^2)^\frac{3}{2}}\\
\pm\frac{11\Delta^4+144t_{\rm K}^2\Delta^2+408t_{\rm K}^4}{(\Delta^2+4t_{\rm K}^2)^\frac{3}{2}}
\end{array}
\right],\nonumber\\
&\xi_{\psi^-{\rm V}}=\xi_{\psi^-{\rm Z}}=\xi_{\psi^-{\rm D}^-}=\frac{t^2}{8t_{\rm K}^2}\left[\frac{\Delta^2+8t_{\rm K}^2}{\Delta}-\frac{3(\Delta^2+12t_{\rm K}^2)}{\sqrt{\Delta^2+16t_{\rm K}^2}}\right]
\label{eq:xiPAM}
\end{align}
in the PAM, and 
\begin{equation}
\begin{array}{ccc}
\xi_{\rm SS}=-\frac{2t^2}{3J_{\rm K}},&\xi_{\rm ST}=-\frac{2t^2}{J_{\rm K}},&\xi_{\rm SP}=\xi_{\rm SR}=-\frac{3t^2}{4J_{\rm K}}
\end{array}
\label{eq:xiKLM}
\end{equation}
in the KLM \cite{TsunetsuguRMP}. 
\par
The effective inter-unit-cell hopping parameters of $\mathcal{O}(t^2)$ for bosons $X$ ($t_{\rm eff}^X$), those of $\mathcal{O}(t)$ for fermions $Y$ ($t_{\rm eff}^Y$), 
and the effective three-unit-cell hopping parameters of $\mathcal{O}(t^2)$ for fermions $Y$ ($t_{3{\rm uc}}^Y$) are obtained \cite{KohnoTinduced,KohnoHubLadder,KohnoKLM,TsunetsuguRMP} as 
\begin{align}
&t_{\rm eff}^{\rm T}=\xi_{\psi^-{\rm T}},\quad t_{\rm eff}^{{\rm D}^-}=-t_{\rm eff}^{\rm V}=-t_{\rm eff}^{\rm Z}=\xi_{\psi^-{\rm V}},\nonumber\\
&t_{\rm eff}^{\rm F}=-t_{\rm eff}^{\rm B}=t_{\rm eff}^+,\quad t_{\rm eff}^{\rm G}=-t_{\rm eff}^{\rm A}=t_{\rm eff}^-,\nonumber\\
&t_{3{\rm uc}}^{\rm A}=t_{3{\rm uc}}^{\rm B}=t_{3{\rm uc}}^{\rm F}=t_{3{\rm uc}}^{\rm G}=\frac{\xi_{\psi^-\psi^-}}{8}
\label{eq:teffHub}
\end{align}
with $t_{\rm eff}^{\pm}=\frac{t}{2}\left(1\pm\frac{4t_\perp}{\sqrt{U^2+16t_{\perp}^2}}\right)$ in the HM \cite{KohnoTinduced,KohnoHubLadder}, 
\begin{align}
&t_{\rm eff}^{\rm T}=-\frac{t^2}{4t_{\rm K}^2}\left[\frac{\Delta^2+4t_{\rm K}^2}{\Delta}+\frac{\Delta^2+14t_{\rm K}^2}{\sqrt{\Delta^2+16t_{\rm K}^2}}\right],\nonumber\\
&t_{\rm eff}^{{\rm D}^-}=-t_{\rm eff}^{\rm V}=-t_{\rm eff}^{\rm Z}=\frac{t^2}{4t_{\rm K}^2}\left[\frac{\Delta^2+4t_{\rm K}^2}{\Delta}-\frac{\Delta^2+14t_{\rm K}^2}{\sqrt{\Delta^2+16t_{\rm K}^2}}\right],\nonumber\\
&t_{\rm eff}^{{\rm p}^{\pm}}=-t_{\rm eff}^{{\rm R}^{\pm}}=\frac{t}{4}\left[1\mp\frac{\Delta^2+8t_{\rm K}^2}{\sqrt{(\Delta^2+4t_{\rm K}^2)(\Delta^2+16t_{\rm K}^2)}}\right],\nonumber\\
&t_{3{\rm uc}}^{{\rm P}^{\pm}}=t_{3{\rm uc}}^{{\rm R}^{\pm}}=-\frac{t^2}{48t_{\rm K}^2}\left[
\begin{array}{l}
\frac{\Delta^4+27t_{\rm K}^2\Delta^2+192t_{\rm K}^4}{(\Delta^2+16t_{\rm K}^2)^\frac{3}{2}}\\
\mp\frac{\Delta^4+9t_{\rm K}^2\Delta^2+24t_{\rm K}^4}{(\Delta^2+4t_{\rm K}^2)^\frac{3}{2}}
\end{array}\right]
\label{eq:teffPAM}
\end{align}
in the PAM, and 
\begin{equation}
t_{\rm eff}^{\rm T}=\xi_{\rm ST},\quad t_{\rm eff}^{\rm P}=-t_{\rm eff}^{\rm R}=\frac{t}{2},\quad 
t_{3{\rm uc}}^{\rm P}=t_{3{\rm uc}}^{\rm R}=-\frac{t^2}{6J_{\rm K}}
\label{eq:teffKLM}
\end{equation}
in the KLM \cite{TsunetsuguRMP}. 
\subsection{Symmetry of effective parameters} 
\label{sec:symParam}
Under the particle--hole transformation [Eq. (\ref{eq:phTrans})], 
a fermion of $N_{\rm e}=3$ with a momentum ${\bm k}$ is transformed to a fermion of $N_{\rm e}=1$ with the momentum of ${\bm \pi}-{\bm k}$ [Eq. (\ref{eq:phck})]. 
Thus, particle--hole symmetry at $\mu=H=0$ [Eq. (\ref{eq:AkwphTransHF})] requires 
\begin{equation}
\begin{array}{ll}
\varepsilon^{\rm F}_{{\bm k}_\parallel}=\varepsilon^{\rm B}_{{\bm \pi}-{\bm k}_\parallel},&
\varepsilon^{\rm G}_{{\bm k}_\parallel}=\varepsilon^{\rm A}_{{\bm \pi}-{\bm k}_\parallel},\\
\varepsilon^{{\rm P}^{\pm}}_{{\bm k}_\parallel}=\varepsilon^{{\rm R}^{\pm}}_{{\bm \pi}-{\bm k}_\parallel},&
\varepsilon^{\rm P}_{{\bm k}_\parallel}=\varepsilon^{\rm R}_{{\bm \pi}-{\bm k}_\parallel},
\end{array}
\end{equation}
which implies 
\begin{equation}
\begin{array}{ll}
\xi_{\psi^-{\rm F}}=\xi_{\psi^-{\rm B}},&\xi_{\psi^-{\rm G}}=\xi_{\psi^-{\rm A}},\\
\xi_{\psi^-{\rm P}^{\pm}}=\xi_{\psi^-{\rm R}^{\pm}},&\xi_{\psi^-{\rm P}}=\xi_{\psi^-{\rm R}},\\
t_{\rm eff}^{\rm F}=-t_{\rm eff}^{\rm B},&t_{\rm eff}^{\rm G}=-t_{\rm eff}^{\rm A},\\
t_{\rm eff}^{{\rm P}^{\pm}}=-t_{\rm eff}^{{\rm R}^{\pm}},&t_{\rm eff}^{\rm P}=-t_{\rm eff}^{\rm R},\\
t_{3{\rm uc}}^{\rm F}=t_{3{\rm uc}}^{\rm B},&t_{3{\rm uc}}^{\rm G}=t_{3{\rm uc}}^{\rm A},\\
t_{3{\rm uc}}^{{\rm P}^{\pm}}=t_{3{\rm uc}}^{{\rm R}^{\pm}},&t_{3{\rm uc}}^{\rm P}=t_{3{\rm uc}}^{\rm R}
\end{array}
\label{eq:symtFB}
\end{equation}
in the HM, symmetric PAM, and KLM, because $\gamma_{{\bm \pi}-{\bm k}_\parallel}=-\gamma_{{\bm k}_\parallel}$ [Eq. (\ref{eq:gammak2})] and Eq. (\ref{eq:eXkFermion2}). 
The relations in Eq. (\ref{eq:symtFB}) are satisfied in Eqs. (\ref{eq:xiHub})--(\ref{eq:teffKLM}). 
\par
In the HM, the energies of the bonding electronic state with $k_\perp=0$ and antibonding electronic state with $k_\perp=\pi$ are 
interchanged by the transformation of $t_\perp\rightarrow -t_\perp$, 
\begin{equation}
\begin{array}{ll}
{\cal E}_{{\rm B}_{\sigma}}(t_\perp)={\cal E}_{{\rm A}_{\sigma}}(-t_\perp),&
{\cal E}_{{\rm G}_{\sigma}}(t_\perp)={\cal E}_{{\rm F}_{\sigma}}(-t_\perp),\\
\varepsilon^{{\rm B}_{\sigma}}_{{\bm k}_\parallel}(t_\perp)=\varepsilon^{{\rm A}_{\sigma}}_{{\bm k}_\parallel}(-t_\perp),&
\varepsilon^{{\rm G}_{\sigma}}_{{\bm k}_\parallel}(t_\perp)=\varepsilon^{{\rm F}_{\sigma}}_{{\bm k}_\parallel}(-t_\perp),
\end{array}
\end{equation}
which implies [Eq. (\ref{eq:eXkFermion2})] 
\begin{equation}
\begin{array}{ll}
\xi_{\psi^-{\rm B}}(t_\perp)=\xi_{\psi^-{\rm A}}(-t_\perp),&\xi_{\psi^-{\rm G}}(t_\perp)=\xi_{\psi^-{\rm F}}(-t_\perp),\\
t_{\rm eff}^{\rm B}(t_\perp)=t_{\rm eff}^{\rm A}(-t_\perp),&t_{\rm eff}^{\rm G}(t_\perp)=t_{\rm eff}^{\rm F}(-t_\perp),\\
t_{3{\rm uc}}^{\rm B}(t_\perp)=t_{3{\rm uc}}^{\rm A}(-t_\perp),&t_{3{\rm uc}}^{\rm G}(t_\perp)=t_{3{\rm uc}}^{\rm F}(-t_\perp).
\end{array}
\end{equation}
These relations are satisfied in Eqs. (\ref{eq:xiHub}) and (\ref{eq:teffHub}). 
\par
In the HM and symmetric PAM at $\mu=0$, 
${\hat \eta}^{\pm}$ shifts a momentum ${\bm k}$ by ${\bm \pi}$ because of $(-)^i$ [Eqs. (\ref{eq:eta}) and (\ref{eq:etai})], 
\begin{equation}
e^{{\rm D}^-}_{{\bm k}_\parallel}=e^{\rm V}_{{\bm k}_\parallel+{\bm \pi}}=e^{\rm Z}_{{\bm k}_\parallel+{\bm \pi}},
\end{equation}
which implies 
\begin{equation}
\begin{array}{l}
\xi_{\psi^-{\rm D}^-}=\xi_{\psi^-{\rm V}}=\xi_{\psi^-{\rm Z}},\\
t_{\rm eff}^{{\rm D}^-}=-t_{\rm eff}^{\rm V}=-t_{\rm eff}^{\rm Z},
\end{array}
\end{equation}
because $\gamma_{{\bm k}_\parallel+{\bm \pi}}=-\gamma_{{\bm k}_\parallel}$ [Eq. (\ref{eq:gammak2})] and Eq. (\ref{eq:eXkBoson2}). 
These relations are satisfied in Eqs. (\ref{eq:xiHub}), (\ref{eq:xiPAM}), (\ref{eq:teffHub}), and (\ref{eq:teffPAM}). 
\par
Under the Shiba transformation [Eq. (\ref{eq:ShibaTrans})], the spin-excited states for ${\cal H}(U)$ correspond to the $\eta$ (charge) excited states for $H(-U)$ 
in the HM and symmetric ($U=2\Delta$) PAM [Eqs. (\ref{eq:defSetai}) and (\ref{eq:SetaShiba})], 
where ${\cal H}(U)$ and ${\cal H}(-U)$ denote the Hamiltonians before and after the Shiba transformation, respectively [Eqs. (\ref{eq:ShibaHamHub}) and (\ref{eq:ShibaHamPAM})]. 
This is fulfilled as 
\begin{equation}
\begin{array}{l}
\xi_{\psi^-{\rm T}}(U)=\xi_{\psi^-{\rm D}^-}(-U)=\xi_{\psi^-{\rm V}}(-U)=\xi_{\psi^-{\rm Z}}(-U),\\
t_{\rm eff}^{\rm T}(U)=t_{\rm eff}^{{\rm D}^-}(-U)=-t_{\rm eff}^{\rm V}(-U)=-t_{\rm eff}^{\rm Z}(-U)
\end{array}
\end{equation}
in the HM [Eqs. (\ref{eq:xiHub}) and (\ref{eq:teffHub})] and 
symmetric ($U=2\Delta$) PAM [Eqs. (\ref{eq:xiPAM}) and (\ref{eq:teffPAM})]. 
\par
In the limit of $U=2\Delta\gg |t_{\rm K}|$, the symmetric PAM reduces to the KLM with $J_{\rm K}=\frac{4t_{\rm K}^2}{\Delta}$ \cite{KohnoKLM}. 
The effective parameters of the symmetric PAM [Eqs. (\ref{eq:xiPAM}) and (\ref{eq:teffPAM})] in this limit 
reduce to those of the KLM [Eqs. (\ref{eq:xiKLM}) and (\ref{eq:teffKLM})] as 
\begin{equation}
\begin{array}{ll}
\multicolumn{2}{l}{t_{\rm eff}^{\rm T}\text{ of the PAM}\rightarrow t_{\rm eff}^{\rm T}}\text{ of the KLM},\\
t_{\rm eff}^{{\rm P}^-}\rightarrow t_{\rm eff}^{\rm P},&t_{\rm eff}^{{\rm R}^-}\rightarrow t_{\rm eff}^{\rm R},\\
t_{3{\rm uc}}^{{\rm P}^-}\rightarrow t_{3{\rm uc}}^{\rm P},&t_{3{\rm uc}}^{{\rm R}^-}\rightarrow t_{3{\rm uc}}^{\rm R},\\
\xi_{\psi^-\psi^-}\rightarrow \xi_{\rm SS},&\xi_{\psi^-{\rm T}}\rightarrow \xi_{\rm ST},\\
\xi_{\psi^-{\rm P}^-}\rightarrow \xi_{\rm SP},&\xi_{\psi^-{\rm R}^-}\rightarrow \xi_{\rm SR}.
\end{array}
\end{equation}

\end{document}